\documentclass[11pt]{article}
\usepackage{axodraw2,pix}
\usepackage{epsfig}
\usepackage{amsfonts}
\usepackage{amsmath,amssymb}
\usepackage{xcolor}
\usepackage{bbm,bm}
\usepackage{cite}
\allowdisplaybreaks[1]

 \usepackage[
   colorlinks=true,
   linkcolor={red!50!black},
   citecolor={blue!50!black},
   filecolor=black,
   urlcolor=black,
   breaklinks=true
   ]{hyperref}
\usepackage{orcidlink}

\usepackage[normalem]{ulem}

\renewcommand{\rmi}[1]{{\mbox{\scriptsize #1}}}
\newcommand{\rmii}[1]{{\mbox{\tiny\rm{#1}}}}

\newcommand{\Lamd}{\bar\Lambda_{\rmii{3d}}}
\newcommand{\LamD}{\bar\Lambda}

\newcommand{\phibar}{\phi_4}
\newcommand{\Nf}{N_{\rm f}}

\newcommand{\Nc}{N_{\rm c}}

\newcommand{\Tc}{T_{\rm c}}
\newcommand{\Tp}{T_{\rm p}}

\newcommand{\tc}{t_{\rm c}}
\newcommand{\tp}{t_{\rm p}}

\newcommand{\Hp}{H_{\rm p}}

\newcommand{\alphap}{\alpha_{\rm p}}

\newcommand{\yc}{y_{\rm c}}

\newcommand{\bxc}{\bar{x}_{\rm c}}
\newcommand{\byc}{\bar{y}_{\rm c}}

\newcommand{\yp}{y_{\rm p}}
\newcommand{\cs}{c_s}
\newcommand{\vw}{v_w}

\newcommand{\mD}{m_\rmii{D}}
\newcommand{\yD}{y_\rmii{D}}
\newcommand{\mDLO}{m_\rmii{D}^\rmii{LO}}

\newcommand{\mB}{m_\rmii{$B$}}
\newcommand{\mG}{m_\chi}
\newcommand{\mH}{m_h}

\newcommand{\phid}{v}
\newcommand{\phiD}{v_4}
\newcommand{\phib}{v_b}

\newcommand{\alphaFR}{\alpha}
\newcommand{\alphaF}{\widehat\alpha}
\newcommand{\alphaSF}{\widehat{\overline{\alpha}}}
\newcommand{\alphaSFR}{\overline\alpha}

\renewcommand{\Lb}{L_b}

\newcommand{\T}{\rmii{$T$}}

\newcommand\MSbar{$\overline{\rm MS}$}

\renewcommand{\textemdash}{ --- }

\renewcommand{\vec}[1]{{\bf #1}}
\renewcommand{\nn}{\nonumber \\}
\newcommand{\gammaE}{\gamma_\rmii{E}}

\newcommand{\sumint}[1]{{\hbox{$\sum$}\!\!\!\!\!\!\!\int\,}_{\!\!\!\!\raise-0.9ex\hbox{$\scriptstyle{#1}$}}}
\newcommand{\Tint}[1]{{\hbox{$\sum$}\!\!\!\!\!\!\!\int\,}_{\!\!\!\!\raise-0.9ex\hbox{$\scriptstyle{#1}$}}}
\newcommand{\Tinti}[1]{{{\Sigma}\!\!\!\!\raise0.3ex\hbox{$\int$}_\rmii{${#1}$}}}
\newcommand{\Tintip}[1]{{{\Sigma'}\!\!\!\!\!\raise0.3ex\hbox{$\int$}_\rmii{${#1}$}}}

\def\scfc{0.7}  %
\def\phgt{21}   %
\def\pwc{21}    %
\def\pwcb{31.5} %
\def\pwcc{42} %
\newcommand{\PIC}[4]{\;\parbox[c]{#2 pt}{\begin{picture}(#2,#3)(0,0)
\SetWidth{1.0}\SetScale{#4} #1 \end{picture}}\;}
\renewcommand{\pic}[1]{\PIC{#1}{\pwc}{\phgt}{\scfc}}
\renewcommand{\picb}[1]{\PIC{#1}{\pwcb}{\phgt}{\scfc}}
\renewcommand{\picc}[1]{\PIC{#1}{\pwcc}{\phgt}{\scfc}}

\makeatletter \@addtoreset{equation}{section} \makeatother
\renewcommand{\theequation}{\arabic{section}.\arabic{equation}}
\makeatletter
\renewcommand\section{\@startsection{section}{1}{\z@}%
  {-5.5ex \@plus -1ex \@minus -.2ex}%
  {2.3ex \@plus.2ex}%
  {\normalfont\large\bfseries}}
\renewcommand\subsection{\@startsection{subsection}{2}{\z@}%
  {-3.25ex\@plus -1ex \@minus -.2ex}%
  {1.5ex \@plus .2ex}%
  {\normalfont\normalsize\bfseries}}
\renewcommand\thesection{\@arabic\c@section}
\renewcommand\thesubsection{\thesection.\@arabic\c@subsection}
\renewcommand{\@seccntformat}[1]{%
  \csname the#1\endcsname.\hspace{1.0em}}
\makeatother

\begin{document}

\flushbottom

\begin{titlepage}

\begin{flushright}
HIP-2025-6/TH
\end{flushright}
\begin{centering}

\vfill

{\Large{\bf%
Higher-dimensional operators
at finite temperature
\\
affect
gravitational-wave predictions
}}

\vspace{0.8cm}

\renewcommand{\thefootnote}{\fnsymbol{footnote}}
Fabio Bernardo%
\orcidlink{0009-0008-0719-3219},%
$^{\rm a,}$%
\footnotemark[1]
Philipp Klose%
\orcidlink{0000-0003-3702-4738},%
$^{\rm b,c,d,}$%
\footnotemark[2]\\
Philipp Schicho%
\orcidlink{0000-0001-5869-7611},%
$^{\rm a,}$%
\footnotemark[3]
and Tuomas V.~I.~Tenkanen%
\orcidlink{0000-0002-3087-8450}
$^{\rm e,}$%
\footnotemark[4]

\vspace{0.8cm}

$^\rmi{a}$%
{\em
  D\'epartement de Physique Th\'eorique, Universit\'e de Gen\`eve,\\
  24 quai Ernest Ansermet, CH-1211 Gen\`eve 4, Switzerland
}
\vspace{0.3cm}

$^\rmi{b}$%
{\em
AEC, Institute for Theoretical Physics, University of Bern,\\
Sidlerstrasse 5,
CH-3012 Bern,
Switzerland}
\vspace{0.3cm}

$^\rmi{c}$%
{\em
Fakultät für Physik, Bielefeld University,\\
Universitätsstraße 4,
33615 Bielefeld,
Germany}
\vspace{0.3cm}

$^\rmi{d}$%
{\em
Theory group, Dutch National Institute for Subatomic Physics (Nikhef),\\
Science Park 105,
1098 XG Amsterdam,
Netherlands}
\vspace{0.3cm}

$^{\rmi{e}}$%
{\em
Department of Physics and Helsinki Institute of Physics,\\
P.O.~Box 64,
FI-00014 University of Helsinki,
Finland}
\vspace{0.3cm}

\vspace*{0.8cm}

\mbox{\bf Abstract}

\end{centering}

\vspace*{0.3cm}

\noindent
We investigate the effect of higher-dimensional marginal operators on the thermodynamics
of cosmological phase transitions.
Using the Abelian Higgs model as a representative
for radiatively-generated one-step transitions,
we systematically match these operators,
which arise at higher orders in the underlying high-temperature
expansion of thermal effective field theory, and use field redefinitions to construct a complete, minimal, and
gauge-invariant operator basis.
The Abelian Higgs model shares the essential infrared structure of more realistic
gauge-Higgs theories at high temperatures, allowing us to test the validity of
dimensional reduction in a simplified setting.
We argue that for strong transitions, temporal gauge modes, which enhance the transition
strength, should be treated on equal footing with spatial ones.
Marginal operators are found to weaken the transition and introduce significant
uncertainties for strong transitions.
For transitions strong enough to produce gravitational waves detectable by LISA,
our findings suggest that the high-temperature expansion may break down entirely.
This would limit the applicability of effective theory techniques, including
their use in non-perturbative lattice studies.

\vfill
\end{titlepage}

{\hypersetup{hidelinks}
\tableofcontents
}

\renewcommand{\thefootnote}{\fnsymbol{footnote}}
\footnotetext[1]{fabio.bernardo@unige.ch}
\footnotetext[2]{pklose@nikhef.nl}
\footnotetext[3]{philipp.schicho@unige.ch}
\footnotetext[4]{tuomas.tenkanen@helsinki.fi}
\clearpage

\renewcommand{\thefootnote}{\arabic{footnote}}
\setcounter{footnote}{0}

\section{Introduction}

Cosmological phase transitions offer a compelling window into the physics of the early universe,
potentially resulting in observational signatures such as gravitational waves (GWs).
Since the Standard Model does not exhibit a thermal
phase transition~\cite{Kajantie:1996mn,Kajantie:1996qd,Gurtler:1997hr,Csikor:1998eu,DOnofrio:2015gop},
GWs from strong first-order phase transitions
would provide a particularly promising probe for new physics beyond the Standard Model (BSM)~\cite{LISACosmologyWorkingGroup:2022jok}.
Effective field theory (EFT) techniques such as dimensional reduction at high temperatures~\cite{Kajantie:1995dw,Braaten:1995cm}
play a central role in describing these transitions.
However, the effect of higher-dimensional marginal operators
that arise from higher-order matching contributions in the underlying high-temperature expansion,
and especially their impact on non-perturbative phenomena,
remains an open question.

In the context of electroweak theories, this issue was framed more precisely in the seminal reference~\cite{Kajantie:1995dw},
which stated that, apart from simple power-counting estimates,
it remains difficult to comprehensively assess the importance of dimension-six operators.
While there are comprehensive studies that have since examined
the impact of
parity-violating marginal operators in
the electroweak sector of
the SM~\cite{Kajantie:1997ky},
fermion-induced operators in the minimal SM~\cite{Moore:1995jv}, and
marginal operators in effective theories of
hot QCD~\cite{Chapman:1994vk,Landsman:1989be,Laine:2018lgj},
analogous estimates for gauge-Higgs theories beyond the SM have indeed
remained a long-standing challenge.
However, several recent works have reported simple estimates of the impact of
higher-dimensional operators in thermal EFTs for
the Two-Higgs Doublet model~\cite{Gorda:2018hvi,Andersen:2017ika,Kainulainen:2019kyp},
singlet~\cite{Gould:2019qek,Niemi:2021qvp,Niemi:2024axp,Niemi:2024vzw} and
triplet~\cite{Niemi:2018asa} extensions of the SM,
SMEFT~\cite{Camargo-Molina:2021zgz,Camargo-Molina:2024sde},
and other theories~\cite{Croon:2020cgk,Kierkla:2023von,Chakrabortty:2024wto}.

The primary goal of this work is to extend these investigations by providing a
comprehensive study on the impact of higher-dimensional operators in
the Abelian Higgs model.
This simplified toy setup captures many key features of more complex gauge-Higgs
theories and BSM scenarios, including
the role of temporal gauge fields,
the hierarchy between gauge and scalar modes, and
radiative barrier formation,
which result in a qualitatively similar infrared structure at high temperatures.
These structural similarities make the Abelian Higgs model a useful proxy
for studying cosmological phase transitions in generic gauge-Higgs theories.
However, applications to more intricate BSM theories are beyond the scope of this
article and are left for future work.

Notably, \cite{Croon:2020cgk} found that even in the SM,
the matching contribution to the Wilson coefficient of
the sextic $(\phi^\dagger \phi)^3$ operator is gauge-dependent.
This indicates that the chosen operator basis for the matching is incomplete.
Similar gauge-dependence issues, related to the singlet-Higgs portal,
have been reported in the singlet-extended SM~\cite{Schicho:2021gca}.
Using the Abelian Higgs model as a case study,
we explicitly demonstrate how to overcome such problems by employing
field redefinitions to construct a minimal and fully gauge-invariant operator basis.
This approach not only ensures the theoretical consistency of thermal EFTs but also
clarifies the limitations of perturbative expansions in describing strong phase transitions.
We emphasize that an incomplete operator basis can introduce gauge artifacts,
leading to ambiguities in the computation of condensates and other thermodynamic quantities.
These deficiencies may, in turn, propagate into uncertainties in
predictions of the GW spectrum and other observables derived from the EFT.

Our findings directly impact the interpretation of potential future GW signals and
aid in assessing the reliability of non-perturbative studies in regimes where
marginal operators may not be negligible.
Importantly, \cite{Gould:2019qek} demonstrated that any model which maps onto a
SM–like effective theory without marginal operators in the infrared (IR)
cannot produce phase transitions strong enough to be detectable by LISA.
This result clearly
motivates the inclusion of marginal operators in effective theories to accurately describe
the strongest transitions.
Recently, it was also demonstrated how
higher-dimensional operators
can influence
the GW spectrum in
the Yukawa model~\cite{Chala:2024xll} and
the electroweak sector of the SM~EFT~\cite{Chala:2025aiz}
further underscoring the importance of systematically incorporating them into
EFT descriptions.
In particular, they can alter the transition dynamics and, consequently,
the properties of the predicted GWs.
In fact, \cite{Chala:2024xll,Chala:2025aiz} found that phase transitions
strong enough to be observable in LISA-generation GW detectors,
often occur near the boundary of EFT validity.
In this work, we conduct a similar analysis and
arrive at comparable conclusions.
Our results demonstrate that neglecting marginal operators can introduce uncertainties of up to
$\mathcal{O}(5\%)$ even for moderately strong transitions.
These uncertainties can be significant compared to those from other sources
in perturbative approaches and non-perturbative lattice simulations.
For the strongest transitions, which are
characterized by large discontinuities in the scalar background field and the resulting large Higgs mechanism-induced mass contributions,
our results suggest a breakdown of the high-temperature expansion.
This breakdown
would severely limit the reliability of perturbative and non-perturbative methods
that rely on such an expansion.
In this context, our results build on and strengthen
the conclusions of~\cite{Gould:2019qek},
suggesting that one-step phase transitions with radiatively generated barriers,
strong enough to be detectable by LISA,
may face a serious breakdown of the high-temperature expansion.
This breakdown would imply that new methods beyond
the standard effective dimensional reduction framework at
high temperature need to be devised.
While our analysis is performed in a simplified toy model,
its key features,
such as the IR behavior and the structure of the dimensional reduction mapping,
especially the generation of marginal operators,
are shared by more realistic gauge-Higgs theories, including the
SU(2) gauge-Higgs model studied in~\cite{Gould:2019qek}.

It remains unclear how to include marginal operators
in state-of-the-art, lattice simulations of high-temperature EFTs.
While they may be necessary to describe very strong first-order transitions,
they break the super-renormalizability of such theories and
compromise the exact lattice-continuum relations required to interpret
lattice simulation results~\cite{Farakos:1994xh,Laine:1995np,Laine:1997dy}.
In practice, direct lattice simulations of full theories are not feasible
if they contain fermions with non-Abelian chiral gauge couplings~\cite{Luscher:2000hn},
though some approaches treat
the fermionic sector perturbatively~\cite{Csikor:1998eu,Aoki:1999fi,Csikor:2000sq}.
This highlights the need to develop techniques
to compute phase transition thermodynamics entirely
without relying on high-temperature expansions\textemdash
an endeavor we leave for future work.
As shown in~\cite{Niemi:2024vzw},
this is especially relevant for BSM theories with heavy additional scalars and
large portal couplings to the Higgs.

The structure of this paper is as follows.
Section~\ref{sec:model} introduces the Abelian Higgs model,
followed by its dimensional reduction to thermal effective field theories in
section~\ref{sec:3d:eft}, with
a focus on the role of dimension-six operators and the temporal gauge mode.
In Section~\ref{sec:thermodynamics},
we compute thermodynamic properties and discuss constraints on the validity of
the effective theory,
emphasizing the role of higher-dimensional operators in modifying the transition strength.
In section~\ref{sec:GW},
we briefly discuss the computation of
three thermal parameters required for predicting
a GW background using results of hydrodynamic simulations~\cite{Hindmarsh:2015qta}.
Finally, section~\ref{sec:outlook} summarizes our findings and
outlines future directions.
The main text is supplemented by detailed appendices.
Appendix~\ref{sec:field:redefinitions} provides an overview of relevant field redefinitions.
The renormalization,
$\beta$-functions, and operator matching
for integrating out the hard scale are covered in
appendix~\ref{sec:dimred:hard-soft} and for integrating out the soft scale in
appendix~\ref{sec:dimred:soft-softer}.

\section{Model}
\label{sec:model}

We consider the Abelian Higgs model%
\footnote{
Also known as Scalar electrodynamics, Scalar QED or U(1)-Higgs theory.
}
at finite temperature~\cite{Weinberg:1992ds,Metaxas:1995ab,Garny:2012cg}.
The corresponding
four-dimensional (4d)
Euclidean Lagrangian density is
\begin{align}
\label{eq:lag:4d}
\mathcal{L}^{\rmi{4d}} &=
    \frac{1}{4} F_{\mu\nu}F_{\mu\nu}
  + (D_\mu \phibar)^\dagger (D_\mu \phibar)
  + \mu^2 \phibar^\dagger \phibar
  + \lambda (\phibar^\dagger \phibar)^2
  \;,
\end{align}
where
$B_\mu$ is a U(1) gauge field with gauge coupling $g$ and
$\phibar$ a complex scalar field.
The covariant derivative of the complex Higgs field reads
$D_\mu \phibar = \partial_\mu \phibar - ig Y_\phi B_\mu \phibar$,
the field strength tensor
$F_{\mu\nu} =\partial_\mu B_\nu-\partial_\nu B_\mu$, and
the hypercharge for the complex scalar $Y_{\phi} = 1$.
Since our goal is to compute the equilibrium thermodynamics
(as well as the statistical part of the bubble nucleation rate) of this model,
we already define the Lagrangian in a Euclidean (rather than Minkowski) spacetime.
One can express the complex field in terms of real fields,
\begin{equation}
\label{eq:ComplexPhi}
  \phibar = \frac{1}{\sqrt{2}}\left(\phiD+h_4+i \chi_4 \right)
  \;,
\end{equation}
where $\phiD$ is a scalar background field and
$h_4$ and
$\chi_4$ are propagating
mass eigenstates.
We apply general $R_\xi$ (or Fermi) gauge fixing~\cite{Fukuda:1975di,Martin:2018emo},
\begin{align}
\label{eq:Rxi:F}
\mathcal{L}^{R_\xi}_{\rmii{GF}}=
  \frac{1}{2\xi}  \bigl[ F(\phibar,\phibar^\dagger)\bigr]^2
  \;,\quad
F(\phibar,\phibar^\dagger) \equiv
  -\partial_\mu B_\mu
  \;,
\end{align}
with the gauge fixing functional $F(\phibar,\phibar^\dagger)$.
This implies the corresponding ghost Lagrangian
\begin{align}
\label{eq:L:FP}
\mathcal{L}_{\rmii{FP}} &=
\bar c ( -\Box )c
\;,
\end{align}
where
$c,(\bar{c})$ are (anti)ghost fields.
Gauge-fixing choices are comprehensively discussed
in~\cite{Fukuda:1975di,Martin:2018emo}.
In this simple model,
the ghosts completely decouple from other sectors and do not affect our calculations.
The resulting tree-level relations between (pole) masses and Lagrangian parameters are
\begin{align}
\label{eq:masses:4d}
  m_{\rmii{$B$},4}^2 &= g^2 \phid_4^2
  \,, &
  m_{h,4}^2 &= \mu^2 + 3\lambda \phid_4^2
  \,, &
  m_{\chi,4}^2 &= \mu^2 + \lambda \phid_4^2
  \,.
\end{align}

Before proceeding, we briefly discuss our motivation for studying the Abelian Higgs model.
As a Higgs-gauge theory, it shares several key features with realistic models of
electroweak phase transitions, including similar scale hierarchies that
enable EFT constructions at high temperatures.
It also provides a valuable framework for investigating effects related to gauge-fixing,
while its relatively small field content simplifies technical calculations,
rendering the required operator basis more tractable.

Beyond its aforementioned role in emulating electroweak physics,
the Abelian Higgs model has important applications for studying dark-sector phase transitions that are decoupled from the SM~\cite{Croon:2018erz}.
One such example arises from introducing a Yukawa sector at high energies,
which could help to explain the recently observed
pulsar timing array signals~\cite{Han:2023olf,Banik:2024zwj}.
The Abelian Higgs model also has deep connections to condensed matter physics,
where its
three-dimensional
EFT, the Ginzburg-Landau theory~\cite{Kleinert:1982dz},
serves as a mean-field description for
superconductivity~\cite{Dasgupta:1981zz,Halperin:1973jh,Kajantie:1998zn,Kajantie:1997vc}.
The model is also known to exhibit a tricritical endpoint separating regimes of
first-order (type~I super conductors) and
second-order (type~II superconductors) phase transitions.
This point was determined non-perturbatively~\cite{Mo:2001fi} and studied
perturbatively in~\cite{Kleinert:1982dz,Herbut:1996ut}.
In contrast to the second-order behavior of the Abelian case,
${\rm SU}(N)$ models with a scalar in the fundamental representation
exhibit only a smooth crossover beyond their critical endpoint~\cite{Kajantie:1996mn}.
Topological structures in the Abelian Higgs model, such as vortices, were studied
in~\cite{Kajantie:1998bg,Kajantie:1998zn,Kajantie:1999ih}.
Such vortices are often used to model cosmic strings~\cite{Hindmarsh:1994re} studied
by large real-time classical field theory simulations~\cite{Vincent:1997cx,Hindmarsh:2017qff}.
Recently,
the classically conformal Abelian Higgs model has been used to predict
primordial black holes from
supercooled phase transitions~\cite{Baldes:2023rqv,Gouttenoire:2023pxh,Lewicki:2024sfw},
although their production appears to be smaller than
initially expected~\cite{Franciolini:2025ztf,Kierkla:2025vwp}.

\section{Dimensional reduction with marginal operators}
\label{sec:3d:eft}

To study the thermodynamics of a field theory,
one must compute its free-energy density, which corresponds to
the effective potential evaluated at its global minimum.
Our objective, therefore, is to determine the effective potential within perturbation theory.
To achieve this, we begin by reviewing the EFT construction of the effective potential,
while adding as a novel ingredient the leading marginal operators in the high-temperature expansion.

One major challenge in finite-temperature computations is
the emergence of a hierarchy of dynamically generated scales due to in-medium effects.
These scales are defined relative to the hard scale $\pi T$ that
sets the typical energy of on-shell particles at high temperatures:
\begin{itemize}
  \item
    The {\em soft (or screening) scale} $gT$ is the characteristic scale of thermal masses and
    the associated in-medium screening of charged particles.
    This screening regulates a number of apparent IR divergences in
    the full theory by giving masses to temporal gauge boson (or temporal scalar) modes.
  \item
    The {\em supersoft scale} $g^{3/2} T$ is the characteristic scale associated with
    radiatively-induced first-order
    phase transitions~\cite{Gould:2021ccf,Ekstedt:2022zro,Gould:2023ovu},
    which is particularly relevant for perturbative computations of the effective potential.
  \item
    The {\em ultrasoft scale} $g^2 T$ is the characteristic scale of
    the Linde problem~\cite{Linde:1980ts}.
    At this scale, sufficiently light bosonic degrees of freedom become strongly coupled and their dynamics confining, ultimately invalidating perturbative expansions.
\end{itemize}
Fortunately,
the confining ultrasoft scale contributes only subdominantly to
scalar-driven phase transitions.
As a result,
purely perturbative descriptions remain viable for
sufficiently strong transitions~\cite{Gould:2023ovu,Niemi:2024axp},
in which the
thermally-induced
field-dependent mass terms
(cf.~eq.~\eqref{eq:masses:3d})
at the soft scale
are large compared to the ultrasoft scale.
Thus, while handling the thermal scale hierarchy is a challenge,
such a hierarchy is also a prerequisite to generate phase transitions
perturbatively~\cite{Arnold:1992rz,Lofgren:2023sep,Gould:2023ovu}.

One way to systematically handle 
the finite-temperature hierarchy in perturbative computations is to use
effective field theory techniques~\cite{Kajantie:1995dw, Braaten:1995cm, Gould:2023ovu}.
This approach constructs a sequence of effective theories by first integrating out
hard-scale contributions, then
soft-scale contributions, and finally arriving at a
supersoft scale effective theory.
In the imaginary-time formalism,
the full theory is formulated in four Euclidean dimensions, with a compactified time direction.
This compactification leads to a discrete spectrum of allowed energy modes, known as
Matsubara frequencies~\cite{Matsubara:1955ws}:
\begin{align}
  \omega_n^{\rmii{B}} &= 2 n \pi T
    &&
    \text{for bosons}
    \,,
  \\
  \omega_n^{\rmii{F}} &= (2n + 1) \pi T
    &&
    \text{for fermions}
    \,.
\end{align}
At the soft scale and lower,
only bosonic zero modes (with $\omega_n^{\rmii{B}} = 0$) remain dynamical.
Consequently, both the soft and supersoft effective theories are defined in
a three-dimensional (3d) Euclidean spacetime, with no explicit time dimension.

The initial hard-to-soft EFT is constructed following
the standard generic rules used to construct EFTs,
by identifying the relevant degrees of freedom and symmetries, and
establishing a suitable power-counting scheme~\cite{Kajantie:1995dw}.
A key ingredient in this step is the high-temperature expansion,
since the masses of the surviving scalar and gauge bosons modes
must be small compared to the temperature.
Integrating out the soft scale is more intricate,
as there are two main approaches~\cite{Gould:2023ovu},
\begin{align}
\label{eq:DR:A}
\mathcal L^\text{4d} &\rightarrow
  \mathcal L^\text{3d}_{\text{soft}} \rightarrow
  \mathcal L^\text{3d}_{\text{softer}} \rightarrow
  V^{\text{supersoft}}_{\text{eff}}
  \tag{DR-A}
\,, \\
\label{eq:DR:B}
  \mathcal L^\text{4d} &\rightarrow
  \mathcal L^\text{3d}_{\text{soft}} \rightarrow
  V^{\text{supersoft}}_{\text{eff}}
  \tag{DR-B}
\,.
\end{align}
Henceforth,
we will suppress the 3d superscript of the Lagrangians and effective potentials.

In the~\eqref{eq:DR:A} approach,
soft temporal scalar modes are integrated out to obtain an effective description below the soft scale, the {\em softer} scale,
assuming that their Debye mass dominates over field-dependent masses in the Higgs phase.
The effective potential is then computed in the Higgs phase by integrating out the spatial gauge boson modes.
In contrast, the~\eqref{eq:DR:B} approach treats temporal and spatial modes on equal footing,
integrating them out simultaneously to directly construct the supersoft-scale effective potential.
This approach assumes instead that field-dependent masses dominate in the Higgs phase.

In the following, we first construct both the soft and softer EFTs before then detailing how to obtain the resulting EFT-induced supersoft theory.

\subsection{Soft scale}
\label{sec:soft:EFT}

The degrees of freedom in the soft effective theory are the zero modes of
the Higgs field $\phi$,
the spatial gauge field $B_i$, and
the temporal gauge field $B_0$,
all of which have mass dimension
$[\phi]=[B_i]=[B_0]=\frac{1}{2}$ when normalized canonically.
The temporal gauge field,
represented by a Lorentz scalar (gauge) singlet,
is symmetric under the transformation $B_0 \to - B_0$
while the Higgs boson zero mode exhibits a residual gauge symmetry under purely
spatial gauge rotations $\phi \to e^{i \alpha(x)} \phi$,
where $\alpha(x)$ is a complex phase.
The associated 3d covariant derivative is
$D_i = \partial_i - i g_3 \, B_i$.

Assuming that
$g^4 \ll \lambda \ll g^2$~\cite{Arnold:1992rz},
the leading matching contribution to an operator with
$n_\phi$ Higgs field insertions,
$n_\rmii{$B_0$}$ temporal gauge field insertions,
and $m$ covariant derivative insertions is
\begin{align}
\label{eq:soft power counting}
\mathcal L_\text{soft} \supset \text{const.} \times
  \frac{(2\pi T)^3}{8\pi}
  \Bigl( \frac{D_i}{2\pi T} \Bigr)^m
  \Bigl( \frac{g \phi}{2\pi T} \Bigr)^{n_\phi}
  \Bigl( \frac{g B_0}{2\pi T} \Bigr)^{n_\rmii{$B_0$}}
  \ ,
\end{align}
where the constant prefactor is expected to be of $\mathcal{O}(1)$.
This power counting implies that
higher-dimensional operators are suppressed by factors of $g$ according to their mass dimension.
Up to corrections of $\mathcal O (g^8)$,
corresponding to dimension-eight operators,
the effective 3d Lagrangian at the soft-scale is
\begin{align}
\label{eq:lag:3d}
\mathcal L_\text{soft} =
\frac{1}{4}F_{ij}F_{ij}
+ (D_i\phi)^\dagger (D_i\phi)
&
+ \mu_3^2 \phi^\dagger\phi
+ \frac{1}{2} (\partial_i B_0)(\partial_i B_0)
+ \frac{1}{2} \mD^2 B_0^2
\nn[1mm]
&
+ \lambda_3 (\phi^\dagger\phi)^2 
+ h_3 (\phi^\dagger\phi)B_0^2
+ \kappa_3 B_0^4
+ \mathcal L_\text{soft}^{(6)}
+ \mathcal O\bigl( g^8 \bigr)
\,,
\end{align}
where
$\mD$ is the Debye mass and
$\mu_3^2$ the squared effective mass of the scalar field, while 
$\lambda_3$,
$h_3$, and
$\kappa_3$ are effective coupling constants.
In the presence of a 3d background field $\phid \propto v_4 / T^{\frac12}$,
the effective masses of the zero modes are
\begin{align}
\label{eq:masses:3d}
  \mB^2 &= g_{3}^2 \phid^2
  \,, &
  \mH^2 &= \mu_{3}^2 + 3\lambda_{3}^{ }\phid^2
  \,, &
  \mG^2 &= \mu_{3}^2 + \lambda_{3}^{ }\phid^2
  \,, &
  m_{\rmii{$B_0$}}^{2} &=
  \mD^2 + h_{3} \phid^2
  \,.
\end{align}
Finally, the term $\mathcal L_\text{soft}^{(6)}$
collects higher-dimensional operators that contribute to correlation functions at
$\mathcal O (g^6)$.
Table \ref{tab:softop:dim6} in appendix~\ref{sec:dim6:preShift}
lists all sixteen allowed operators.
Using the field redefinitions detailed in appendix~\ref{sec:dim6:preShift},
it is possible to eliminate eight of these operators,
which yields the final Lagrangian
\begin{align}
\label{eq:lag:3d:dim6}
\mathcal L_\text{soft}^{(6)} =
  \alphaFR_{\phi^2 F^2}F_{ij}F_{ij}\phi^\dagger\phi
&
+ \alphaFR_{D^2 \phi^4}\phi^\dagger\phi(D_i\phi)^\dagger(D_i\phi)
+ \alphaFR_{B_0^2 F^2} B_0^2F_{ij}F_{ij}
+ \alphaFR_{D^2 \phi^2 B_0^2}\phi^\dagger\phi(\partial_iB_0)^2
  \nn[1mm] &
+ \alphaFR_{B_0^6} B_0^6
+ \alphaFR_{\phi^2 B_0^4}(\phi^\dagger\phi)B_0^4
+ \alphaFR_{\phi^4 B_0^2}(\phi^\dagger\phi)^2B_0^2
+ \alphaFR_{\phi^6}(\phi^\dagger\phi)^3
\ ,
\end{align}
where
the dimension-six Wilson coefficients
are denoted by $\alphaFR$;
see~\cite{Bernardo:2026nyq} for a recent update
of {\tt DRalgo}~\cite{Ekstedt:2022bff}, which performs operator matching up to dimension six.
The coefficients in eq.~\eqref{eq:lag:3d} are known up to $\mathcal O(g^4)$;
see e.g.~\cite{Farakos:1994kx,Hirvonen:2021zej}.
The matching relations for the mass parameters are~\cite{Hirvonen:2021zej}
\begin{subequations}
\label{eq:soft eft couplings}
\begin{align}
\mu_3^2 &=
    \mu^2
  + \Bigl(\frac14 g^{2} + \frac13 \lambda \Bigr)T^2
  + \frac{\mu^2}{(4\pi)^2} \bigl( 3 g^2 - 4 \lambda \bigr)\Lb 
  + \delta
  \nn &
+ \frac{T^2}{(4\pi)^2} \Bigl[
    - \frac{8 + 39 \Lb}{36} g^4
    + \frac{2(1 + 3 \Lb)}{3} g^2 \lambda
    - \frac{10}{3}\Lb \lambda^2
  \Bigr]
+ \mathcal{O}(g^6)
\,,
\\[2mm]
\mD^2 &=
      \frac13 g^2 T^2
    + \frac{g^2 \mu^2}{4 \pi^2}
    + \frac{g^2 T^2}{(4\pi)^2} \left[
        \frac{7 - \Lb}{9} g^2
      + \frac{4}{3} \lambda
    \right]
  + \mathcal{O}(g^6)
\, .
\end{align}
Notably, the expression for the Debye mass is only valid in the high-temperature expansion
since at low temperatures $\lim_{\T\to 0} \mD^2 = \frac{g^2\mu^2}{4\pi^2} \neq 0$.
Likewise, the matching relations for the coupling constants are
\begin{align}
g_3^2 &= g^2 T -\frac{\Lb T}{48\pi^2} g^4 + \mathcal O (g^6)
\,,\\
\kappa_3 &= \frac{g^4 T}{24\pi^2} + \mathcal{O}(g^6)
\,,\\
h_3 &=
    g^2 T
  + \frac{T}{(4\pi)^2} \Bigl[ \frac{4 - \Lb}{3} g^4 + 8 g^2 \lambda \Bigr]
  + \mathcal{O}(g^6)
\,,\\
\lambda_3 &=
      \lambda T
    + \frac{T}{(4\pi)^2} \left[ (2 - 3\Lb) g^4 + 6\Lb g^2 \lambda - 10 \Lb \lambda^2 \right]
    +\mathcal{O}(g^6)
    \ ,
\end{align}
\end{subequations}
where the quantities
\begin{align}
  \Lb &\equiv 2 \ln \frac{\LamD e^{\gammaE}}{4\pi T}
  \ , &
\delta &\equiv \frac{T^2}{(4\pi)^2} \Bigl( c + \ln\frac{3 T}{\LamD}  \Bigr)
\left[ - 6 g^4 + 8 g^2 \lambda - 8 \lambda^2 \right]
\,,
\end{align}
encode the dependence on the matching scale $\LamD$,
\begin{align}
c &= \frac12 \Bigl(
    \ln \frac{8\pi}{9}
  + (\ln\zeta_2)'
- 2 \gammaE \Bigr) \approx -0.124
\end{align}
is a dimensionless constant,
$\zeta^{ }_n \equiv \zeta(n)$ is the Riemann $\zeta$-function with
$(\ln\zeta_n)' = \zeta'(n)/\zeta(n)$,
and
$\gammaE$ is the Euler-Mascheroni constant.

We determine the coefficients
for the remaining higher dimensional operators in eq.~\eqref{eq:lag:3d:dim6}
by matching to the full theory at one loop;
see again appendix~\ref{sec:dimred:hard-soft} for details.
In the following, we focus in particular on the $(\phi^\dagger \phi)^3 $ operator, which is
the only dimension-six operator contributing to the effective potential at tree-level.
Its Wilson coefficient is
\begin{align}
\label{eq:c6:soft:EFT} 
  c_6^{ } &=
  \alphaFR_{\phi^6}^{ } =
\frac{\zeta_3}{32 \pi^4} \Bigl(
    g^6
  - \frac{6}{5} g^4 \lambda
  + 5 \, g^2 \lambda^2
  + \frac{20}3 \lambda^3 \Bigr)
  + \mathcal{O}(g^8)
  \,.
\end{align}
Notably, this expression is gauge-invariant, as expected.
This resolves the issue with the naive matching result reported in~\cite{Croon:2020cgk}.
Although that work focused on the SM sector,
the gauge-dependence issue it identified can also be reproduced in the Abelian Higgs model.
The resolution lies in performing appropriate field redefinitions and
adopting a physically meaningful operator basis\textemdash
precisely as anticipated in~\cite{Croon:2020cgk}.

None of the Wilson coefficients in
eqs.~\eqref{eq:lag:3d} and~\eqref{eq:lag:3d:dim6} run at leading order,
because the soft EFT exhibits no one-loop ultraviolet (UV) divergences in
dimensional regularization.
However, the mass parameter $\mu_3^2$ does run at two loops.
In fact, $\mu_3^2$ is the only Wilson coefficient that runs in
the super-renormalizable soft theory without marginal operators.
The full effect of its running is captured exactly by replacing $\delta$ according
to~\cite{Niemi:2021qvp},
such that
\begin{align}
\delta \to \delta_3 \equiv \frac1{(4\pi)^2} \Bigl( c + \ln \frac{3 T}{\Lamd} \Bigr)
\Bigl[ - 4 g_3^4 + 8 g_3^2 \lambda_3^{ } - 8 \lambda_3^2 - 2 h_3^2 \Bigr]
\ ,
\end{align}
where $\Lamd$ is the renormalization scale of the soft EFT.

\subsection{Softer scale}
\label{sec:softer:EFT}

Next, we integrate out the temporal gauge field $B_0$,
following the conventional approach outlined in~\cite{Kajantie:1995dw}.
As alluded to in eqs.~\eqref{eq:DR:A} and~\eqref{eq:DR:B},
this procedure is more nuanced than often acknowledged in the recent literature,
with the exception of~\cite{Gould:2023ovu,Kierkla:2023von,Kierkla:2025qyz,Gould:2024jjt}.
We revisit this issue in detail in the next section.
We refer to the resulting theory as a softer-scale EFT,
rather than an ultrasoft-scale EFT, as the latter is a misnomer,
as argued in~\cite{Gould:2023ovu}.
The characteristic mass scale for a first-order,
scalar-driven phase transition lies above the ultrasoft scale,
at the supersoft scale.

One important distinction compared to the soft theory lies in
the somewhat more delicate power counting
of the softer EFT.
Assuming
$g_3^4 \ll \lambda_3 \ll g_3^2$,
the leading matching contribution to an operator with
$n$ Higgs field insertions and
$m$ covariant derivatives is given by%
\footnote{%
  Since $F_{ij} \propto \left[ D_i, D_j \right]$,
  insertions of $F_{ij}$ 
  do not need to be accounted for separately.
  Matching contributions generated at higher-loop level or by diagrams that
  involve insertions of $\lambda_3$, $\kappa_3$, or $h_3$ scale as
  \begin{align}
  \label{eq:full softer power counting}
  \mathcal L_\text{softer} \supset
  \frac{\mD^3}{4\pi}
  \Bigl( \frac{2 g_3^2}{4\pi \mD} \Bigr)^{\ell - 1}
  \Bigl( \frac{2 \lambda_3}{g_3^2} \Bigr)^{n_\lambda}
  \Bigl( \frac{h_3}{g_3^2} \Bigr)^{n_h}
  \Bigl( \frac{12 \kappa_3}{g_3^2} \Bigr)^{n_\kappa}
  \Bigl( \frac{D_i}{\mD} \Bigr)^{m}
  \Bigl( \frac{g_3 \phi}{\mD} \Bigr)^n \ ,
  \end{align}
  where
  $\ell$ is the considered loop order using
  $n_\lambda$ insertions of $\lambda_3$,
  $n_h$ insertions of $h_3$, and
  $n_\kappa$ insertions of $\kappa_3$.
}
\begin{align}
\label{eq:softer leading matching contribution}
\mathcal L_\text{softer} \supset \text{const.} \times \frac{\mD^3}{4\pi}
  \Bigl( \frac{D_i}{\mD} \Bigr)^m
  \Bigl( \frac{g_3 \phi}{\mD} \Bigr)^n
  \ ,
\end{align}
where the constant prefactor is expected to be of $\mathcal{O}(1)$.
At first glance,
this might suggest that the theory is ill-defined,
since higher-dimensional operators
are apparently not suppressed by powers of $g_3$ compared to super-renormalizable ones.
However, when computing vacuum diagrams in the symmetric phase, each additional vertex from
an operator~\eqref{eq:softer leading matching contribution} introduces a suppression
\begin{align}
\Bigl( \frac{g_3^2}{4\pi \bar\mu_3} \Bigr)^{\frac{n}2}
\Bigl( \frac{\bar\mu_3}{\mD} \Bigr)^{n+m-3} \propto
\Bigl( \frac1{16 \pi^2 y} \Bigr)^{\frac{n}4}
\Bigl( 3 g^2 y \Bigr)^{\frac{n+m-3}2}
\ ,
\end{align}
where
$\bar\mu_3 \ll \mD$ is the effective Higgs mass parameter and
$\bar{y} \equiv \bar\mu_3^2/\bar g_3^4$.
This means that the theory can be used to compute the effective potential in the symmetric phase if 
\begin{align}
\label{eq:symmetric phase constraint}
(3 g^2)^{-1} \gg \bar{y} \gg (4\pi)^{-2} \ .
\end{align}
The situation is more complex in the broken phase since
the Higgs condensate can generate contributions to
the spatial gauge boson and Higgs masses
that disrupt the validity of the power counting for strong phase transitions.
This imposes additional consistency constraints,
similar to those in eq.~\eqref{eq:symmetric phase constraint}.
These constraints are analyzed in detail in
secs.~\ref{sec:soft vs softer} and \ref{sec:EFT:validity}.

In practice, we construct the softer effective theory using the same process as
in sec.~\ref{sec:soft:EFT}.
By again including operators up to dimension six and using
field redefinitions to then eliminate redundant operators,
we obtain the Lagrangian
\begin{align}
\label{eq:lag:ultrasoft}
\mathcal L_\text{softer} =
  \frac{1}{4}F_{ij}F_{ij}
+ (\overline D_i\phi)^\dagger (\overline D_i\phi)
&
+ \bar\mu_3^2\phi^\dagger\phi
+ \bar\lambda_3 \, (\phi^\dagger\phi)^2
+ c_6 \, (\phi^\dagger\phi)^3
  \nn &
+ \alphaSFR_{D^2 \phi^4} \, \phi^\dagger\phi(\overline D_i\phi)^\dagger(\overline D_i\phi)
+ \alphaSFR_{\phi^2 F^2} \, F_{ij}F_{ij}\phi^\dagger\phi
\ ,
\end{align}
where $\overline D_i = \partial_i - i \bar g_3 B_i $ is
the effective covariant derivative,
$\bar{g}_3$ its associated gauge coupling, and
$\bar\lambda_3$ is the quartic Higgs self-interaction.
The corresponding effective masses
are the ones from eq.~\eqref{eq:masses:3d}
adapted to the softer Lagrangian~\eqref{eq:lag:ultrasoft}.
After matching at one loop,
the gauge coupling $\bar{g}_3 = g_3$ is the same as in
the soft scale theory,
while $\bar\lambda_3$ is given as~\cite{Farakos:1994kx}
\begin{align}
\label{eq:soft lam3 matching}
\bar\lambda_{3,\rmi{red}}^{ } &=
\frac{\bar\lambda_3}{4\pi\mD} =
\lambda_\rmi{red}^{ }
  - \Bigl( 1 + \frac{\mu_3^2}{6 \mD^2} \Bigr) \frac{h_\text{red}^2}2
  + \mathcal{O}(g^4)
  \ ,
\end{align}
where
\begin{align}
h_\text{red} &= \frac{h_3}{4\pi\mD}
\ , &
\kappa_{\rmi{red}} &= \frac{\kappa_3}{4\pi\mD}
\ , &
\lambda_\text{red} &= \frac{\lambda_3}{4\pi\mD}
\,,
\end{align}
are the relevant effective expansion parameters of the soft theory.
Analogously to the soft theory,
the mass parameter $\bar\mu_3^2$ is
the only UV divergent Wilson coefficient in the super-renormalizable theory
without marginal operators.
To ensure that its running is consistent with that of $\mu_3^2$,
a two-loop matching is required,
leading to the expression
\begin{align}
\label{eq:soft mu3 matching}
\bar\mu_3^2 &= \mu^2_3
  - \mD^2 h_\rmi{red}^{ }
  - \mD^2 h_\rmi{red}^{2} \Bigl[ 1 + 2 \ln\frac{\Lamd}{2 \mD} \Bigr]
  + 6\mD^2 h_\rmi{red}^{ } \kappa_{\rmi{red}}^{ }
  + \mathcal{O}(g^5)
  \ ,
\end{align}
where the last term is of $\mathcal{O}(g^6)$,
which is already of the order of hard three-loop contributions to
the scalar mass, while
the soft three-loop contribution is of $\mathcal{O}(g^5)$.
The result in eq.~\eqref{eq:soft mu3 matching} is readily obtained
using {\tt DRalgo}~\cite{Ekstedt:2022bff}
or partially available in~\cite{Farakos:1994kx}.
The dimension-six Wilson coefficients at one loop,
\begin{subequations}
\label{eq:softer marginal matching}
\begin{align}
\alphaSFR_{\phi^2 F^2} &=
    \alphaFR_{\phi^2 F^2}
  + \mathcal{O}(g^2)
  \,, \\
\alphaSFR_{D^2 \phi^4} &=
    \alphaFR_{D^2 \phi^4}
  - 4\pi \frac{h_\text{red}^2}{12\mD}
  + \mathcal{O}(g^2)
  \,, \\
\alphaSFR_{\phi^6} &=
    \alphaFR_{\phi^6}
  + (4\pi)^2 \frac{h_\text{red}^2}6 \left( h_\text{red} - \lambda_\text{red} \right) 
  + \mathcal{O}(g^4)
  \,,
\end{align}\end{subequations}
were determined in appendix~\ref{sec:dimred:soft-softer}.
Written in terms of $g$ and $\lambda$, one has
\begin{align}
\label{eq:super soft beta}
\bar{c}_6 &=
  \alphaFR_{\phi^6}
  + \frac{\sqrt3 \, g^3}{8 \pi}
    \Bigl( \frac{\mDLO}{\mD^{ }} \Bigr)^3
    \bigl( 1 - x_\rmii{LO} \bigr)
  + \mathcal O(g^4)
  \ ,
\end{align}
where
we used
the leading-order (LO)
results of
$x_\rmii{LO}= \frac{\lambda}{g^2}$ and
$\mDLO = \frac{g T}{\sqrt3}$.
As generally expected from eq.~\eqref{eq:softer leading matching contribution},
the leading non-vanishing contribution to
$\alphaSFR_{\phi^6}$ appears at $\mathcal O(g^3)$,
while higher-order corrections to the Debye mass $\mD$ become important at $\mathcal O(g^5)$.
In the following sections,
when discussing the phase transition dynamics, we also require the combination
\begin{align}
\label{eq:x matching}
\bar{x} \equiv \frac{\bar\lambda_3}{\bar g_3^2} &=
    x_\rmii{LO}
  - \frac{\sqrt3 \, g}{8\pi}
    \Bigl( \frac{\mDLO}{\mD} \Bigr)
    \Bigl( 1 + \frac{\mu_3^2}{6 \mD^2} \Bigr)
  + \mathcal O (g^2)
  \ .
\end{align}
It is this quantity,
the ratio of thermal scalar self-coupling to thermal gauge coupling,
that controls the strength of a phase transition.

\subsection{Supersoft scale effective potential}
\label{sec:soft vs softer}

Before focusing on the thermodynamics,
we first detail the regimes of validity for
the two EFT setups~\eqref{eq:DR:A} and~\eqref{eq:DR:B}.
To this end, we consider the effective potential;
see also~\cite{Farakos:1994kx,Hirvonen:2021zej} for a more detailed derivation
or~\cite{Gould:2023ovu} for further qualitative details.

In gauge-Higgs theories,
and in particular in the Abelian Higgs model,
radiative corrections generically induce a barrier between
the local minima of the effective potential.
Therefore, it is necessary to inspect these corrections to correctly describe first-order transitions.
In general, the leading contributions are generated by spatial and temporal gauge boson modes, while the Higgs contributions are subdominant.
Working within the soft scale EFT, and
expressing the effective potential in terms of
the 3d background field $\phid$, one has
\begin{subequations}\begin{align}
\Delta V_{\text{tree}} &=
    \frac{1}{2} \mu_3^2 \phid^2
  + \frac{1}{4} \lambda_{3}^{ } \phid^4
  + \frac{1}{8} c_6^{ } \phid^6
  \, , \\
\label{eq:Vloop1-gauge}
\Delta V_{\text{1loop,gauge}} &=
  - \frac{1}{12\pi} \Bigl( 2 \mB^3 + m_\rmii{$B_0$}^3  \Bigr) =
  - \frac{1}{12\pi} \Bigl( 2 g^3_3 \phid^3 + (\mD^2 + h_3^{ } \phid^{2})^{\frac{3}{2}} \Bigr)
  \, ,
\end{align}\end{subequations}
where $\Delta V (\phid) = V(\phid) - V(0)$, while
$\mB$ is the spatial and
$m_\rmii{$B_0$}$ the temporal gauge boson mass defined in eq.~\eqref{eq:masses:3d}.
Notably,
$c_6$ is the only marginal operator coefficient that contributes to the tree-level potential,
while all other coefficients only contribute at one-loop.
Their effect remains subdominant compared to
the leading contributions from gauge fields and is therefore omitted here.
Additionally,
since $\lambda_3$ is positive, the tree-level potential does not exhibit
a barrier between its minima;
cf.~\cite{Camargo-Molina:2021zgz}.

If there is no hierarchy between
$\mD^2$ and Higgs contribution to the temporal gauge boson mass $h_3^{ } \phid^2$,
one has to use the~\eqref{eq:DR:B} setup.
The leading-order effective potential is given by the sum of
tree-level and one-loop gauge field contributions,
\begin{align}
\label{eq:VLO-soft}
\Delta V_\text{soft} &=
  \Delta V_{\text{tree}}
  + \Delta V_{\text{1loop,gauge}}
  \nn &=
    \frac{1}{2} \mu_3^2 \phid^2
  + \frac{1}{4} \lambda_3^{ } \phid^4
  + \frac{1}{8} c_6^{ } \phid^6
   -\frac{1}{12\pi} \Bigl( 2 g^3_3 \phid^3 + (\mD^2 + h_3^{ } \phid^2)^{\frac{3}{2}}  \Bigr)
   \,.
\end{align}

A hierarchy
$\mD^2 \gg h_3^{ } \phid^2$
allows for constructing an EFT at
the softer scale
by integrating out the temporal mode
using the~\eqref{eq:DR:A} setup.
In this approach,
soft contributions are perturbatively absorbed
into the Wilson coefficients of the softer theory.
This corresponds to an expansion in $h_3^{ } \phid^2/\mD^2$, or equivalently,
using leading-order (hard-scale) matching relations, an expansion in $3\, \phid^2 /T$,
where the factor 3 results from the definition of the Debye mass.
Consequently,
the softer EFT is only valid for relatively weak transitions,
where $\phid^2 /T \sim v_4^2 / T^2  \ll 1/3$.
On the other hand, the softer EFT setup is not suitable for describing strong phase transitions
with large field-induced mass contributions.%
\footnote{\label{fn:general debye mass}
  For a generic ${\rm SU}(N)$ gauge-scalar theory with 
  $N_t$ scalars in the adjoint and
  $N_d$ scalars in the fundamental representation, as well as
  $\Nf$ fermionic degrees of freedom,
  a similar argument leads to
  $\phid^2/T \ll \frac{2}{3}N + \frac{2}{3}N_t + \frac{1}{3} N_d + \frac{1}{6}\Nf $.
  Due to the large number of fermions with a high degree of internal symmetry,
  the fermionic contribution can render the Debye mass large and hence increase
  the range of validity for a softer EFT.
  For example, in
  the SM $\Nf = 3(1+\Nc)$ for 3 generations of both lepton and
  $\Nc = 3$ quark doublets, which leads to
   $\phid^2/T \ll \frac{11}{3}$
  using $N=2$, $N_d =1$, $N_t=0$.
  For non-perturbative studies that explicitly retain the temporal mode in the lattice action,
  see e.g.~\cite{Kajantie:1993ag,Jakovac:1994xg}.
}
In sec.~\ref{sec:thermodynamics},
we specifically demonstrate this by showing that
for very strong transitions,
higher-dimensional operators become dominant when contributions from
the soft temporal sector are properly accounted for.

The soft to softer EFT
matching relations~\eqref{eq:soft lam3 matching}, \eqref{eq:soft mu3 matching}, and~\eqref{eq:softer marginal matching}
follow from the series expansion of last term in eq.~\eqref{eq:Vloop1-gauge}.
However, the tree-level potential within the softer EFT does not have a barrier,
which is again generated by radiative corrections.
By including one-loop contributions generated by the spatial gauge boson modes,
one obtains the leading-order effective potential
\begin{align}
\label{eq:softer-to-supersoft}
\Delta  V_{\text{softer}} &=
      \frac{1}{2} \bar{\mu}_3^2 \phid^2
    + \frac{1}{4} \bar{\lambda}_3^{ } \phid^4
    + \frac{1}{8} \bar{c}_6^{ } \phid^6
    - \frac{1}{6\pi} \bar{g}^3_3 \phid^3
  \;.
\end{align}

Conversely,
while
there is in principle no reason to presume
a hierarchy between $\mD^2$ and $h_3^{ } \phid^{2}$ in the soft theory setup,
a reasonable estimate
of the effective potential for strong transitions, where
$h_3 \, \phid^2 \gtrsim \mD$, can be obtained by setting
$\mD^2 \to 0$ in eq.~\eqref{eq:VLO-soft}.
In this limit, one obtains
\begin{align}
\label{eq:soft-to-supersoft}
\Delta V^{\mathcal{E}}_\rmi{soft} =
\Delta V_\rmi{soft}^{ }\bigr|_{\mD^2 \ll h_3^{ } \phid^2} &=
      \frac{1}{2} \mu_3^2 \phid^2
    + \frac{1}{4} \lambda_3^{ } \phid^4
    + \frac{1}{8} c_6^{ } \phid^6
    - \frac{\phid^3}{6\pi} \Big( g^3_3 + \frac{1}{2} h^{3/2}_3 \Big)
    \,,
\end{align}
where the so-called enhancement factor $\mathcal{E}$
is defined in eq.~\eqref{eq:enhancement factor definition}.
While this approximation is relatively crude,
it works reasonably well in the regime where $\phid$ becomes large enough
for the high-temperature expansion to start breaking down.
For a more detailed discussion,
see sec.~\ref{subsec:higher dimensional operator importance}.
One key advantage of using the approximation is that it allows for
an analytical estimate of
the latent heat at
the critical temperature in the~\eqref{eq:DR:B} setup.
Inspecting eq.~\eqref{eq:soft-to-supersoft},
we see that the temporal mode effectively enhances the cubic term.
Indeed, utilizing leading-order (hard-scale) matching relations~\cite{Ekstedt:2024etx},
one obtains
\begin{align}
\label{eq:effective-barrier}
  - T \frac{\phid^3}{6\pi} \Big( g^3_3 + \frac{1}{2} h^{3/2}_3 \Big) =
  - T v^3_4 \frac{g^3}{12\pi} \Big(
      (2)_{\text{spatial}}
    + (1)_{\text{temporal}}
  \Big)
  + \mathcal{O}(g^4)
  \,.
\end{align}
In the limit where the field-dependent contribution dominates over the Debye mass,
the temporal mode gives a 50\% {\em soft-scale enhancement} to the cubic term.
Therefore, temporal modes are expected to further increase
the strength of already strong transitions
that push the high-temperature expansion to its limits.

Since the effective potential, due to its mass dimension being $[\Delta V]=3$,
exhibits a trivial $T^3$ scaling,
it is practical to define dimensionless potentials
when computing thermodynamic quantities
\begin{align}
  \label{eq:tildeV:omega}
  \Delta \widetilde V(\varphi, \omega_q) &\equiv \frac{\Delta V(\phid)}{\omega_q^3} \ , &
\varphi^2 &= \frac{\phid^2}{\omega_q} \ , 
\end{align}
where $\omega_q \propto T$ is some characteristic energy scale.
One convenient choice is to identify $\omega_q$ with the square of the gauge coupling that sets the size of the cubic $\phid^3$ term (i.e.\ the size of the potential barrier separating the local minima), so that
\begin{align}\label{eq:reference scale definition}
\omega_q &=
\begin{cases}
g_3^2 & \text{full soft potential~\eqref{eq:VLO-soft}} \\
\bar{g}_3^2 & \text{full softer potential~\eqref{eq:softer-to-supersoft}} \\
g_\text{eff}^2 & \text{$\mD^2 \ll h_3 \phid^2$ potential~\eqref{eq:soft-to-supersoft}}
\end{cases}
\ , &
g_\text{eff}^3 &= g_3^3 + \frac12 h_3^{\frac32} \approx \frac32 g_3^3
\ .
\end{align}
In this convention,
the corresponding rescaled effective potentials associated with
at leading order effective potentials of~\eqref{eq:VLO-soft}--\eqref{eq:soft-to-supersoft}
share the same overall shape
\begin{equation}
    \Delta \widetilde{V}(\varphi,A,B,C,D)\equiv
        \frac{A}{2} \varphi^2
      + \frac{B}{4} \varphi^4
      + \frac{C}{8} \varphi^6
      - \frac1{6\pi} \varphi^3
      - \frac1{12\pi} \left( D + \varphi^2 \right)^{\frac32}
      \, ,
\end{equation}
where the size of the $\varphi^3$ term is not a free parameter but fixed by the normalization.
Explicitly, they are given as
\begin{subequations}
\begin{align}\label{eq:softer-to-supersoft rescaled}
\Delta  \widetilde V_\text{softer} &=
\frac{\bar{y}}2 \varphi^2
+ \frac{\bar{x}}4 \varphi^4
+ \frac{\bar{c}_6}8 \varphi^6
- \frac1{6\pi} \varphi^3
\ , \\[2mm]\label{eq:VLO-soft rescaled}
\Delta \widetilde V_\text{soft} &=
\frac{y}2 \varphi^2
+ \frac{x}4 \varphi^4
+ \frac{c_6}8 \varphi^6
- \frac1{6\pi} \varphi^3
- \frac1{12\pi}  \bigl(
  \yD^{ } +\varphi^2 
  \bigr)^{\frac32}
\ , \\[2mm]\label{eq:soft-to-supersoft rescaled}
\Delta \widetilde{V}^{\mathcal{E}}_\rmi{soft} =
\Delta  \widetilde V_\rmi{soft}^{ }\bigr|_{\yD^{ } \ll \varphi^2} &=
\frac{y}2 \varphi^2
+ \frac{x}4 \varphi^4
+ \frac{c_6}8 \varphi^6
- \frac1{6\pi} \varphi^3
\ ,
\end{align}
\end{subequations}
where we have defined the dimensionless quantities
\begin{subequations}\begin{align}
x &= \frac{\lambda_3}{\omega_q} \ , &
\bar{x} &= \frac{\overline \lambda_3}{\omega_q} \ ,
\\
y &= \frac{\mu_3^2}{\omega_q^2} \ , &
\bar{y} &= \frac{\bar{\mu}_3^2}{\omega_q^2} \ , &
\yD &= \frac{\mD^2}{\omega_q^2} \ .
\end{align}\end{subequations}

Notice, that the quantities $x$ and $y$ in the rescaled
full soft potential~\eqref{eq:VLO-soft rescaled} are not the same as
the $x$ and $y$ in the approximated potential~\eqref{eq:soft-to-supersoft rescaled},
as the two potentials are defined using different choices of $\omega_q$
to reflect the effective enhancement of the cubic term.
In general, if the size of the cubic term in
the \emph{original} potential $\Delta V(\phid)$ is rescaled by some factor $\mathcal E$,
this leads to a corresponding rescaling
\begin{align}
\label{eq:temporal enhancement}
\omega_q &\to \mathcal E^{\frac23} \omega_q \ , &
  x &\to \mathcal E^{-\frac{2}{3}} x \ , &
  y &\to \mathcal E^{-\frac{4}{3}} y \ .
\end{align}
In our case,
the effective height of the barrier in the original approximated
potential~\eqref{eq:soft-to-supersoft} is enhanced by a factor
\begin{align}
\label{eq:enhancement factor definition}
\mathcal E &= \frac{g_\text{eff}^3}{g_3^3} \approx \frac32
\,,
\end{align}
compared to the barrier in the softer potential~\eqref{eq:softer-to-supersoft}
due to the impact of temporal gauge modes.
We expect that this enhancement is of the same order for general ${\rm SU}(N)$ theories with one fundamental scalar.

To summarize, within our perturbative description of the Abelian Higgs model effective potential, the supersoft scale is always the characteristic scale associated with first-order phase transitions.
We have constructed two approximations
eqs.~\eqref{eq:softer-to-supersoft} and~\eqref{eq:soft-to-supersoft} for
the LO potential eq.~\eqref{eq:VLO-soft},
corresponding to two limits, for relatively weak and strong transitions, respectively.
These approximations arise from how one treats the temporal mode.
We remark that the latter construction for strong transitions, where the temporal mode is not integrated out to construct a softer EFT,
has indeed been applied in recent work~\cite{Kierkla:2023von,Gould:2024jjt} in the context of other models,
and analogous constructions can be obtained using the
Higgs effective field theory (HEFT) functionalities in
{\tt HEFT.m} of {\tt DRalgo}~\cite{Ekstedt:2022bff}
for generic models.
See also~\cite{Bertenstam:2025jvd} for a recent application of {\tt DRalgo}.

We emphasize that constructing the supersoft scale effective potential is qualitatively different from
the matching computations
used to compute the Wilson coefficients in the soft and softer EFTs, which are performed in the symmetric phase with $\phid = 0$.
In contrast,
the supersoft scale effective potential is constructed only
in the Higgs phase around a non-vanishing scalar background.
In the Higgs phase,
in addition to
the temporal gauge field mode, also
the spatial gauge field modes are soft
and hence integrated out to construct
the effective potential~\cite{Ekstedt:2022zro,Gould:2023ovu,Ekstedt:2024etx}.

\section{Phase transition thermodynamics}
\label{sec:thermodynamics}

The free energy,
or equivalently the pressure $p(T)$, governs the behavior of systems in thermal equilibrium.
In particular, the pressure determines the strength of a phase transition,
which is often characterized in terms of the parameter
\begin{align}\label{eq:alpha def}
\alpha &\equiv \frac{\Delta \theta}{3\omega_+}
\ ,
\end{align}
where $\Delta \theta = \theta_+ - \theta_-$ is the discontinuity of the pseudo trace anomaly
$\theta_\pm =  e - p/\cs^2$~\cite{Hindmarsh:2017gnf,Giese:2020rtr} across the transition,
$\omega = e + p$ the enthalpy,
$e = T \partial_\T p - p$ the energy density, and
$\cs$ the speed of sound in the plasma.
The subscript ${}_\pm$ specifies that a quantity is evaluated in either the high- or low-temperature phase,
while the normalization factor 3 ensures that $\alpha$ is consistent with the definition in the bag model~\cite{Giese:2020rtr}.
In general, the speed of sound $\cs$ is also determined by the pressure,
but for our purposes below it is sufficient to use the leading-order result
$\cs^2 = 1/3$~\cite{Tenkanen:2022tly}.
We likewise use the leading result for the pressure in the high-temperature phase,
\begin{align}
\label{eq:p:plus}
  p_{+} = g_\star \frac{\pi^2}{90} T^4 + \mathcal{O}(g^2 T^4)
  \ ,
\end{align}
where
$g_\star = g_{\rmii{DS},\star} = 4$ is
the effective number of relativistic degrees of freedom in
the dark sector (DS) Abelian Higgs model.
This represents a simplification, as we do not include
the full number of relativistic degrees of freedom in the SM.
In principle, the total effective number should account for both
the SM and the dark sector~\cite{Gould:2019qek}, leading to
$g_{\rmi{tot},\star} = g_{\rmii{SM},\star} + g_{\rmii{DS},\star} \, \xi_\rmii{D}^4$,
where $\xi_\rmii{D} \equiv T_\rmii{DS}/T_\rmii{SM}$ accounts for
a possible temperature difference between the
two sectors~\cite{Husdal:2016haj,Bringmann:2023iuz}.

The pressure in the low-temperature or Higgs phase is
\begin{align}
  p(T) &= p_{+} - T \Delta V(\phid_{\text{min}}^{ })
  \ , &
  \Delta V (\phid_{\text{min}}^{ }) &= \omega_q^3 \Delta \widetilde V (\varphi_{\text{min}}^{ })
  \ , &
  \phid_{\text{min}}^2 &= \omega_q^{ } \varphi_{\text{min}}^2
  \ ,
\end{align} 
where $\Delta \widetilde V(\varphi)$ is
the rescaled supersoft scale effective potential defined in
sec.~\ref{sec:soft vs softer},
$\omega_q$ the corresponding normalization scale,
and $\phid_{\text{min}}(T)$ the value of the background field that minimizes the potential.
Above the critical temperature $\Tc$, the origin
$\phid_{\text{min}} = 0$ is the global minimum of the effective potential.
For smaller temperatures, the global minimum shifts to a finite value
$\phid_{\text{min}} \neq 0$ that is separated from the now local minimum at the origin by a potential barrier.
Directly at the critical temperature, the two minima are degenerate, so that
$\Delta V (\phid_{\text{min}}) = 0$.
Using the definition~\eqref{eq:alpha def}, one obtains the relation
\begin{align}\label{eq:alpha intermediate}
\alpha &\approx
    \frac{1}{3 \, \partial_\T p_+} \Bigl( T \frac{{\rm d}}{{\rm d} T} - 3 \Bigr) \Delta V
  =
  \frac{30}{(2\pi)^2 g_\star} T \frac{{\rm d}}{{\rm d} T} \Bigl( \frac{\omega_q^3}{T^3} \Delta \widetilde V\Bigr)
  \ ,
\end{align}
which we use going forward.

\subsection{Leading-order effective potentials}

To proceed, one must provide a suitable expression for the effective potential.
Following the discussion in sec.~\ref{sec:soft vs softer},
this can be achieved through different approaches,
resulting in at least two potentially viable EFT constructions:
the~\eqref{eq:DR:A} or
\eqref{eq:DR:B} setups.
We now discuss each setup separately.

\subsubsection*{Softer-induced effective potential}

In the~\eqref{eq:DR:A} EFT setup,
the effective potential is constructed by first integrating out
the temporal scalar before addressing the spatial gauge boson modes.
At leading order in the super-renormalizable theory,
one obtains the expression given in~\cite{Ekstedt:2024etx}.
Following~\cite{Ekstedt:2024etx}, it is straightforward to add higher-order perturbative corrections to the effective potential.
We do not consider them in this work, but instead investigate the impact of the leading correction due to the marginal operator $(\phi^\dagger \phi)^3$,
which yields the effective potential~\eqref{eq:softer-to-supersoft} and the associated rescaled potential 
\begin{align}
\label{eq:softer-to-supersoft rescaled reprise}
\Delta  \widetilde V_\text{softer} &=
\frac{\bar{y}}2 \varphi^2
+ \frac{\bar{x}}4 \varphi^4
+ \frac{\bar{c}_6}8 \varphi^6
- \frac1{6\pi} \varphi^3 \ , &
\omega_q &= \bar{g}_3^2 \ ,
\end{align}
which we have already defined in~\eqref{eq:softer-to-supersoft rescaled}.
The field value $\varphi_\text{min}$ that minimizes the rescaled potential
is a function of the dimensionless parameters
$\bar{y}$, $\bar{x}$, and $\bar{c}_6$.
Using that
$\bar{g}_3^2/T$, $\bar{x}$, and $\bar c_6$, are temperature independent at leading order~\cite{Gould:2019qek},
cf.\ also
eqs.~\eqref{eq:soft eft couplings}, \eqref{eq:super soft beta}, and \eqref{eq:x matching},
one can simplify
the expression for the transition strength of
eq.~\eqref{eq:alpha intermediate} by using the chain rule
\begin{align}
\label{eq:v chain rule}
  T \frac{{\rm d}}{{\rm d}T}
  \Bigl( \frac{\Delta V (\varphi_\text{min})}{T^3} \Bigr)
  &\approx \frac{\bar g_3^6 \eta_{\bar{y}}}{T^3} \partial_{\bar{y}} \Delta \widetilde V (\varphi_\text{min})
  \ , &
  \eta_{\bar{y}} &\equiv T \frac{\text{d} \bar{y}}{\text{d} T}
  \, .
\end{align}
At leading order
(without including higher-order loop corrections),
the derivative $\partial_{\bar{y}} \Delta \widetilde V$ is just
$\varphi_\text{min}^2$, which yields the final result
\begin{align}
\label{eq:alpha}
\alpha(T) \approx \frac{15}{(2\pi)^2 g_\star} \frac{\bar g^6_3 \eta_{\bar{y}}}{T^3} \varphi_\text{min}^2 \ .
\end{align}
We emphasize that this expression is only valid at leading order.
At higher orders, it is the jump in
the 3d scalar condensate~\cite{Farakos:1994xh}
\begin{align}
  \Delta \langle \phi^\dagger \phi \rangle \equiv \frac{{\rm d}}{{\rm d} \bar \mu_3^2} \Delta V
  =
  \bar g_3^2 \partial_{\bar{y}} \Delta \widetilde V (\varphi_\text{min})
\,,
\end{align}
that contributes to $\alpha$.%
\footnote{%
  Calculation in terms of the condensate can be made manifestly gauge-invariant order by order,
  by applying strict expansion around the leading-order minimum.
  The leading-order result in eq.~\eqref{eq:alpha} is also gauge invariant,
  since the minimum of the leading-order potential does not depend on
  the gauge fixing parameter.
}
At the critical temperature, where $\Delta \widetilde V = 0$ must hold,
it is possible to give an analytic expression for the Higgs phase minimum,
which is located at
\begin{align}
\label{eq:phimin:F:r:softer}
  \varphi_\text{min} &= \frac{
     \overline{\mathcal F}_+
   - \overline{\mathcal F}_-}{2 \pi \bar{x} \sqrt{\bar{r}}}
  \,, &
 \overline{\mathcal F}_\pm &\equiv \left( \sqrt{1 + \bar{r}} \pm \sqrt{\bar{r}} \right)^{\frac13}
  \,, &
  \bar{r} &= \frac{3}{(2\pi)^2}\frac{\bar{c}_6}{\bar{x}^3}
  \,.
\end{align}
For $\byc$, one likewise obtains
\begin{align}
\label{eq:y:softer:EFT:induced}
    \byc = \frac{
      (\overline{\mathcal F}_+^{ } - \overline{\mathcal F}_-^{ })^2
      (\overline{\mathcal F}_+^{2} + \overline{\mathcal F}_-^{2})}{
    (4\pi)^2 \bar{x} \, \bar{r}} =
    \frac12 \bar{x} \, \varphi_\text{min}^2 \frac{(\overline{\mathcal F}_+^2
      + \overline{\mathcal F}_-^2)}2
  \,.
\end{align}
When neglecting the impact of the sextic operator,
these expressions reduce to the well-known limit
\begin{align}
\varphi_\text{min} &= \frac1{3\pi \bar{x}} \ , &
\byc &= \frac1{18\pi^2 \bar{x}}
\ .
\end{align}
This result implies that the strength of
the phase transition scales as $\varphi_\text{min}^2 \propto \bar{x}^{-2}$,
while including higher-dimensional operators is
most relevant for large values of $\bar{r}$, which scales as $\bar{x}^{-3}$.
In other words, small values of $\bar{x}$ characterize the region of parameter space with both the strongest transitions and the largest impact of higher-dimensional operators.

\subsubsection*{Soft-induced effective potential}

Conversely, if the effective potential is computed by treating
temporal scalars and spatial
gauge boson modes on equal footing,
following the~\eqref{eq:DR:B} setup,
the leading-order potential\textemdash including the sextic operator\textemdash
is given by eq.~\eqref{eq:VLO-soft}.
The associated rescaled potential is
\begin{align}
\label{eq:VLO-soft rescaled reprise}
\Delta \widetilde V_\text{soft} &\equiv
    \frac{y}2 \varphi^2
    + \frac{x}4 \varphi^4
    + \frac{c_6}8 \varphi^6
    - \frac{1}{12\pi} \Bigl( 2 \varphi^3 + \bigl( \yD^{ } + \varphi^2 \bigr)^{\frac32} \Bigr)
\ , &
\omega_q &= g_3^2
\ ,
\end{align}
which we have already defined in~\eqref{eq:VLO-soft rescaled}.
In analogy to the~\eqref{eq:DR:A} setup,
the minimum of the rescaled potential
$\varphi_{\text{min}}$
depends on
$\yD$, $y$, $x$, and $c_6$, where $y$ is the only parameter that is temperature dependent at leading order.
This gives
\begin{align}
\label{eq:alpha_soft}
\alpha(T) &\approx \frac{15}{(2\pi)^2 g_\star} \frac{g^6_3 \eta_y}{T^3} \varphi_\text{min}^2 \ , &
\eta_y &\equiv T \frac{\text{d} y}{\text{d} T} \approx \eta_{\bar{y}} \ .
\end{align}
This expression is the same as
eq.~\eqref{eq:alpha},
except that it contains
$g_3$ in place of $\bar{g}_3$ and that
$\varphi_\text{min}$
now minimizes soft-induced potential~\eqref{eq:VLO-soft rescaled reprise} instead of
the softer-induced one given in eq.~\eqref{eq:softer-to-supersoft rescaled reprise}.
Hence, the main difference between the~\eqref{eq:DR:A} and~\eqref{eq:DR:B} approaches is the predicted location of the minimum.

Due to the non-analytic structure of the potential~\eqref{eq:VLO-soft rescaled reprise},
it is quite challenging to find an analytic expression for $\varphi_\text{min}$.
However, as discussed in sec.~\ref{sec:soft vs softer},
we may estimate the potential for strong transitions by setting $\yD \to 0$,
which yields the rescaled effective potential~\eqref{eq:soft-to-supersoft rescaled}.
This potential is formally of the same shape as
the softer-induced effective potential~\eqref{eq:softer-to-supersoft rescaled},
allowing us to derive an analytic estimate for the minimum at $\Tc$.
To obtain the correct result, it is however important to remember that the potential~\eqref{eq:soft-to-supersoft rescaled} is defined using a normalization scale $\omega_q = g_\text{eff}^2$
that differs from the scale $\omega_q = g_3^2$ of the full soft potential~\eqref{eq:VLO-soft rescaled reprise}.
This impacts the definition of $x$ and $y$ as well as the location of the minimum $\varphi_\text{min}$.
Using the replacements~\eqref{eq:temporal enhancement} and
$\varphi_\text{min} \to \varphi_\text{min}/\mathcal E^{\frac13}$
to account for the change in $\omega_q$, we find the minimum
\begin{align}
  \label{eq:softer:}
\varphi_\text{min} &\approx \mathcal E \times \frac{\mathcal F_+ - \mathcal F_-}{2\pi x \sqrt{r}}
  \,, &
\mathcal F_\pm &\equiv \left( \sqrt{1 + r} \pm \sqrt{r} \right)^{\frac13}
  \,, &
  r &= \mathcal E^2 \times \frac3{(2\pi)^2}\frac{c_6}{x^3} \ ,
\end{align}
where
$x = \lambda_3 / g_3^2$ is defined as usual.
The quantity $\mathcal E = 3/2$
was already defined in~\eqref{eq:enhancement factor definition} and
parametrizes the enhancement of the transition strength due to temporal gauge boson modes.
The corresponding soft-enhanced critical $\yc$ is
\begin{align}
\label{eq:y:soft:EFT:enhanced}
    \yc =
    \mathcal{E}^2
    \frac{
      (\mathcal F_+^{ } - \mathcal F_-^{ })^2
      (\mathcal F_+^{2} + \mathcal F_-^{2})}{
    (4\pi)^2 x \,r} =
    \frac12 x \, \varphi_\text{min}^2 \frac{(\mathcal F_+^2
      + \mathcal F_-^2)}2
  \,.
\end{align}
When neglecting the impact of the sextic operator,
one finds the limit
\begin{align}
  \varphi_\text{min} &= \frac{\mathcal E}{3\pi x} \ , &
  \yc &= \frac{\mathcal E^2}{18 \pi^2 x}
  \ .
\end{align}
As in the~\eqref{eq:DR:A} setup,
small values of $x$ characterize the region with both
strong transitions and the largest impact of higher dimensional operators.
Therefore, our analysis will primarily focus on the regime of small $x$-values.

\subsection{EFT consistency at the critical temperature}
\label{sec:EFT:validity}

Qualitatively,
small values of $x$ and $\bar{x} \sim x$ yield strong phase transitions because they correspond to large values of the background field $\varphi_\text{min} \sim x^{-1}$.
In sec.~\ref{sec:soft vs softer},
we already discussed how the high-temperature expansion and with it consistency of the EFT description breaks down for sufficiently large values of $\varphi_\text{min}$
(or equivalently, small values of $x$).
On the other hand, it is well known that
the thermal EFTs remain consistent but become non-perturbative
for large values of $x$~\cite{Kajantie:1995kf,Kajantie:1996mn,Kajantie:1998bg}.
In the softer theory, this non-perturbative behavior is often related to
the existence of a critical endpoint for the transition,
which is characterized by a divergent correlation length that is
said to signal the associated breakdown of perturbation theory.
For the Abelian Higgs model it is located at $\bxc \approx 0.28$~\cite{Mo:2001fi}.
The existence of this endpoint is not readily apparent in perturbation theory and is
thus typically investigated using lattice or other non-perturbative methods.

In this section, we quantitatively examine
constraints on the consistency and perturbativity of the soft and softer EFTs.
One particular focus is the
impact of higher-dimensional operators in
the regime close to the eventual breakdown of the high-temperature expansion.
For each EFT introduced in sec.~\ref{sec:3d:eft},
we examine
(i) the power counting that is used to construct the EFT
by truncating the infinite tower of higher-dimensional operators that contribute to it in principle,
and
(ii) the perturbative power counting that underlies the loop expansion within the EFT.
It is important to emphasize that these are two independent considerations:
\begin{itemize}
  \item[(i)]
    \hyperref[sec:EFT:consistency]{\em Validity of power counting}:\\
    If the truncation of higher-dimensional operators is invalid,
    the resulting effective theory becomes inconsistent,
    as it would require accounting for the entire tower of
    effective interactions even at leading order.
    In the soft and softer EFTs, this signals the breakdown of the high-temperature expansion and
    the EFT cannot be applied to compute observables.
  \item[(ii)]
    \hyperref[sec:EFT:perturbativity]{\em Perturbativity of the loop expansion}:\\
    If the power counting holds but the loop expansion fails,
    the EFT remains consistent but becomes non-perturbative.
    In this case, non-perturbative methods like lattice simulations are needed
    to compute thermodynamic quantities such as $\alpha(T)$.
\end{itemize}
One important factor in the context of the soft and softer EFTs is that
non-perturbative behavior propagates upwards through the chain of effective theories.
Therefore,
if the softer EFT is non-perturbative, then so is the soft EFT; and if
the soft EFT is non-perturbative, then so is the full theory.

\subsubsection*{(i) Consistency}
\label{sec:EFT:consistency}

In the symmetric phase,
the power counting~\eqref{eq:soft power counting} ensures that
the soft-scale EFT is consistent as long as the coupling constants in
the full theory are perturbative,
\begin{align}
\lambda, g^2 &\ll (4\pi)^2 \ .
\end{align}
The power counting for the softer theory is more subtle, and only
consistent if condition~\eqref{eq:symmetric phase constraint} is satisfied.
Working at the critical temperature $\Tc$ and using $\yc \approx \frac1{18\pi^2 \bar{x}}$,
this gives the further constraint
\begin{align}
\label{eq:softer consistency symmetric phase}
\frac{g^2}{6 \pi^2} \approx 0.017 g^2 &\ll \bar{x} \ll \frac89 \ .
\end{align}
In the broken phase, the Higgs mechanism generates additional mass contributions to
the effective masses of the Higgs boson, the spatial gauge boson modes, and the temporal gauge boson modes
which can spoil the consistency of the EFT power counting.
The most stringent constraint turns out to be associated with the mass of the spatial gauge boson modes,
\begin{align}
\mB = g_3 \phid = g v_4 \big( 1 + \order{g^2} \big) \ .
\end{align}
To ensure consistency of the soft theory in the broken phase, this mass has to be small compared to
the hard-scale $\pi T$.
Working at the critical temperature $\Tc$ and using
$\phid \approx \frac{\mathcal E g_3}{3\pi x}$,
this yields the constraint
\begin{align}
\frac{\mathcal E g^2}{3 \pi^2} \approx 0.051 \, g^2 \ll x \ .
\end{align}
Since the leading contribution to the effective potential from higher-dimensional operators in the soft theory scales as
\begin{align}
V_\text{soft} \supset \pi^2 T^3 \Bigl( \frac{g \phid}{2 \pi T} \Bigr)^{2n}
\, ,
\end{align}
this condition also ensures that contributions from higher-dimensional operators are suppressed compared to those from super-renormalizable operators.

To also ensure consistency of the softer theory,
$\mB$ has to be small compared to the
Debye mass $\mD \approx \frac{g T}{\sqrt 3}$.
Working at the critical temperature $\Tc$ and using $\phid \approx \frac{\bar{g}_3}{3\pi \bar{x}}$,
this gives the constraint
\begin{align}
\label{eq:softer eft validity}
\frac{\sqrt3}{3\pi} g \approx 0.18 \, g \ll \bar{x}
\ .
\end{align}
The leading contribution to the effective potential from higher-dimensional operators in the softer theory scales as
\begin{align}
\Delta V_\text{softer} \supset \frac{\mD^3}{4\pi} \Bigl( \frac{g_3 \phid}{\mD} \Bigr)^{2n} \ .
\end{align}
Thus, as in the soft case,
eq.~\eqref{eq:softer eft validity} also ensures that
higher-dimensional operator contributions remain subdominant,
provided the effective theory is consistent from the outset.

\subsubsection*{(ii) Perturbativity}
\label{sec:EFT:perturbativity}

In general, thermal field theories are non-perturbative if they contain light bosons with masses that are not much larger than the relevant non-perturbative scale associated with
the Linde IR problem.
Since particles are generally heavier in the broken phase than in the symmetric phase, the resulting constraints are more stringent in the symmetric phase.
Furthermore, non-perturbative behavior in the softer theory implies that
the soft theory is also non-perturbative.
Therefore,
we only consider the behavior of the softer theory in the symmetric phase,
as that is sufficient to test the onset of non-perturbativity.

As a starting point,
we consider diagrams that contain only $\bar{g}_3^2 B_i^2 \phi^\dagger \phi$ interactions.
In the symmetric phase, the effective expansion parameter for this subset of diagrams is
\begin{align}
\alpha_\text{eff} \equiv \frac{2 \bar{g}_3^2}{4\pi \bar\mu_3} =
\frac1{2 \pi \byc^{1/2}} \ .
\end{align}
Both the soft and softer theories can only be perturbative if $\alpha_\text{eff} < 1$.
Again working at the critical temperature $\Tc$ and using
$\byc \approx \frac1{18 \pi^2 \bar{x}}$, this yields the condition
\begin{align}
\label{eq:expansion parameter constraint}
\bar{x} \ll \frac29 \approx 0.22
\,.
\end{align}
This constraint is equivalent to demanding that the mass of the Higgs
$m_h\sim \bar{\mu}_3$
is larger than the non-perturbative scale
$\Lambda_\rmii{NP} \equiv \frac{\bar{g}_3^2}{4\pi}$,
thereby avoiding the Linde IR problem.
A very similar bound can be obtained by noting that perturbation theory is expected to break down in the vicinity of the critical point at
$\bxc \approx 0.28$ due to the associated divergence of the correlation length
$\xi = 1 / \bar \mu_3$.
On a quantitative level, the relevant statement is that perturbation theory is expected to break down if the Ginzburg-criterion
\cite{Cardy:1996xt}
\begin{align}
\frac12 \bar{g}_3^2 \varphi_\text{min}^2 \gg \xi^{2-d}
\,,
\end{align}
is violated in $d=3$ dimensions.
Again working at the critical temperature $\Tc$ and using
$\varphi_\text{min} = \frac1{3\pi \bar{x}}$ and
$\byc = \frac1{18 \pi^2 \bar{x}}$,
this gives the constraint
\begin{align}
\bar{x} \ll \Bigl( \frac1{18 \pi^2} \Bigr)^{1/3} \approx 0.18
\,,
\end{align}
which is very similar but somewhat more stringent than
eq.~\eqref{eq:expansion parameter constraint}.

Finally, since the parameter $x$ receives loop corrections from integrating out both the hard and soft scales,
one might also be worried that sufficiently small values of $x$ are fine-tuned.
This concern is more immediate in case of the softer theory, since the matching contribution from integrating out the soft scale is larger than the hard scale contribution.
Indeed, choosing the matching scale such that
thermal logarithms vanish ({\em viz.}~$\Lb = 0$), one has
\begin{align}
\bar{x} &= x_\rmii{LO} - \frac{\sqrt3 \, g}{8\pi} + \frac{g^2}{8 \pi^2} +\mathcal{O}(g^3) \ ,
\end{align}
which implies that values of $x$ much smaller than
\begin{align}
\delta \bar{x}_\text{1loop} = \frac{\sqrt3}{8\pi} g \approx 0.07 \, g
\,,
\end{align}
require large cancellations between loop-orders,
leading to fine-tuning and the associated instability of the perturbative expansion.
However, since this constraint is less stringent than condition~\eqref{eq:softer eft validity},
it is fortunately not relevant in practice.

\subsubsection*{Summary}

Combining the aforementioned constraints and working at the critical temperature,
we find that the softer EFT remains consistent and perturbative in both
the symmetric and the broken phase if
\begin{align}
\label{eq:x constraints}
0.18 \, g \ll \bar{x} \ll 0.18 \ .
\end{align}
This narrow region suggests that the softer EFT should be applied with caution,
particularly in the broken phase,
where it becomes inconsistent for $\bar{x} < 0.18 \, g$.
In contrast, eq.~\eqref{eq:softer consistency symmetric phase} implies that
in the symmetric phase, the softer EFT remains consistent for much smaller values,
breaking down only at $\bar{x} < 0.017 \, g^2$.
Therefore, one viable approach for the description of
strong first-order transitions may be to use the softer EFT only for
computing the effective potential in the symmetric phase,
while using complementary techniques~\cite{Laine:2017hdk,Curtin:2022ovx} 
for the computation in the broken phase.

The soft theory remains consistent and perturbative at the critical temperature $\Tc$
in a larger regime
\begin{align}
0.051 g^2 \ll x \ll 0.18 \ .
\end{align}
For even smaller values of $x$, higher-dimensional operators start to contribute at leading order to the soft theory and eventually dominate the effective potential.
In this regime, the high-temperature expansion breaks down, which also means that thermodynamics cannot be studied using non-perturbative lattice simulations
that are performed in the EFT with only super-renormalizable operators.
Any accurate study in the small-$x$ region then necessitates
a computation without resorting to
high-temperature expansions;
see e.g.~\cite{Laine:2000kv,Laine:2017hdk}.

We note that our findings can readily be generalized to more general gauge groups.
In particular,
${\rm SU}(N)$ theories typically exhibit larger Debye masses than
the U(1) theory
due to group factors (see also footnote~\ref{fn:general debye mass}),
which would extend the regime of consistency for
the soft and softer theories to smaller values of $x$.
However,
relatively large gauge couplings, similar to
the SM SU(2) gauge coupling $g_2^2 \sim 0.4$,
restrict the regime of consistency to larger values of $x$.
Furthermore, for a SU(2) theory with a scalar in the fundamental representation,
the critical endpoint is located well below that of the U(1) case.
We therefore expect perturbativity to break down for smaller $x$ values
compared to the Abelian Higgs model,
further limiting the regime in which perturbative techniques can be applied.

\subsection{Impact of higher-dimensional operators}
\label{subsec:higher dimensional operator importance}

To investigate the impact of the $(\phi^\dagger \phi)^3$ operator on
the strength of the phase transition in a more quantitative way,
we consider the ratio
\begin{align}
  R &\equiv \frac{\alpha(\Tc)}{\alpha_{c_6 = 0}(\Tc)}
  \ .
\end{align}
Working with the~\eqref{eq:DR:A} setup,
 we immediately find the analytic expression
\begin{align}
R &= \frac{9}{4\bar{r}} \bigl(
       \overline{\mathcal F}_+
     - \overline{\mathcal F}_-
   \bigr)^2 =
   1 - \frac{8\bar{r}}{27} +\mathcal{O}(\bar{r}^2)
   \ .
\end{align}
Finding an exact analytic expression or $R$ in the~\eqref{eq:DR:B} setup
is challenging due to the non-analytic structure of
the effective potential~\eqref{eq:VLO-soft rescaled reprise}.
Instead, we determine the ratio numerically.
However,
an analytic estimate for $R$ can still be obtained by approximating
the effective potential with eq.~\eqref{eq:soft-to-supersoft},
where the Debye mass $\mD$ is set to zero.
This yields the ratio
\begin{align}
\label{eq:R soft analytic estimate}
  R\approx \frac{9}{4 r}\bigl(
         \mathcal F_+
       - \mathcal F_-)^2
     =
     1-\frac{8 r}{27}
     +\mathcal{O}(r^2)
 \, ,
\end{align}
which we shall compare to
the full numerical result in the regime where the effects of $c_6$ become relevant.
\begin{figure}[t]
  \centering
  \includegraphics[width=0.5\textwidth]{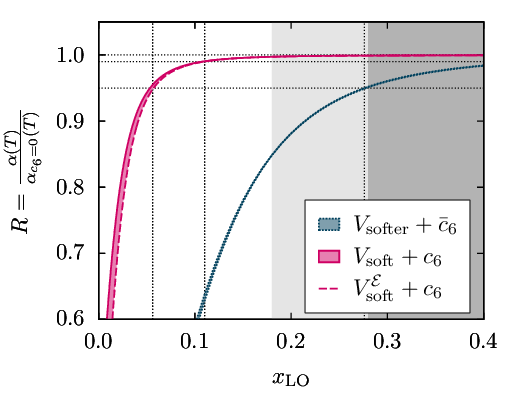}%
  \includegraphics[width=0.5\textwidth]{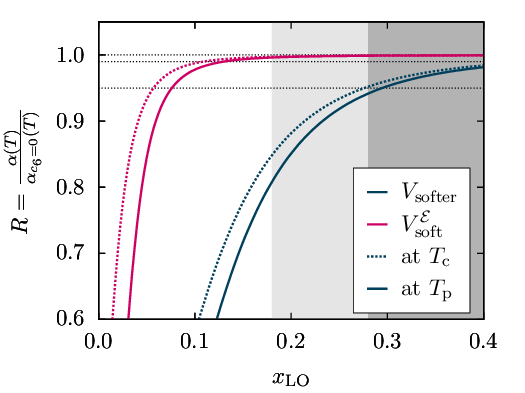}
  \caption{%
    Impact of the $(\phi^\dagger \phi)^3$ operator on
    the transition strength $\alpha$ for
    $g_0\equiv g(\LamD=\overline{T}) = 1$
    and
    various $x_\rmii{LO} \approx x \approx \bar{x}$.
    Left:
    Blue lines are obtained from
    the softer-induced potential~\eqref{eq:softer-to-supersoft} and
    magenta lines from
    the soft-induced potential~\eqref{eq:VLO-soft}.
    Uncertainty bands arise from varying
    the hard-to-soft matching renormalization scale within
    $\LamD \in [\overline{T}, 5\overline{T}]$ where $\overline{T}\equiv 4\pi e^{-\gammaE} T$
    and
    $x_{\rmii{LO}} \equiv x_{\rmii{LO}}(\LamD =\overline{T})$.
    The dashed magenta line depicts the $\mD \to 0$ limit for
    $\LamD = \overline{T}$ using eq.~\eqref{eq:soft-to-supersoft}.
    Horizontal dotted lines mark relative differences of
    5\% and 1\% compared to the result without the sextic operator.
    The dark gray region is excluded by
    the critical endpoint
    $\bxc \approx 0.28$~\cite{Mo:2001fi}
    and
    the light gray region is excluded by perturbativity constraints~\eqref{eq:x constraints}.
    The soft-scale contribution to $\alpha$ has a significantly greater impact
    than that of the hard scale alone.
    For the soft potential (cf.\ sec.~\ref{sec:soft vs softer}),
    and strong transitions (small~$x$),
    the numerical evaluation of eq.~\eqref{eq:VLO-soft} agrees with
    its dashed analytical estimate~\eqref{eq:soft-to-supersoft}.
    Right:
    the same ratio as on the left but at
    $\Tc$ (dotted) and
    $\Tp$ (solid) using $\LamD = \overline{T}$.
}
\label{fig:condensate-c6-effect}
\end{figure}

Figure~\ref{fig:condensate-c6-effect} (right) depicts
the various expressions of the ratio $R$ as a function of the reference value
$g = 1$ and
$x_\rmii{LO} \equiv \lambda / g^2$
at $\Tc$.
The blue lines are obtained in the~\eqref{eq:DR:A} setup, while
the magenta lines are obtained in the~\eqref{eq:DR:B} setup.
The shading indicates the region with
$x_\rmii{LO} > 0.2$, where our current perturbative approach is not applicable due the Linde IR problem.
The softer theory is expected to be inconsistent for $\bar{x} < 0.2$ since higher-dimensional operators start to contribute to the effective potential at leading order.
Indeed, fig.~\ref{fig:condensate-c6-effect} shows
that the $(\phi^\dagger \phi)^3$ operators rapidly becomes dominant for smaller values of
$x_\rmii{LO} \approx \bar{x}$, and that correctly accounting for it
significantly decreases the transition strength.
For sufficiently low $\bar{x}_\rmii{LO}$,
the value of $\alpha$ becomes negative since it is dominated by $\bar{c}_6$
through eq.~\eqref{eq:phimin:F:r:softer},
which further indicates that this regime is unphysical.

On the other hand,
the soft theory is expected to remain consistent for significantly smaller values of $x$.
In the plot, the $(\phi^\dagger \phi)^3$ operator becomes
dominant for $x_\rmii{LO} \lesssim 0.05$.
In the intermediate regime, $0.05 < x_\rmii{LO} < 0.2$, where
we expect the soft theory to be applicable,
it decreases the strength of the transition by less than 10\%.
We also observe that the analytical estimate~\eqref{eq:R soft analytic estimate} is in excellent agreement with the full numerical result for $R$ in the soft theory.

Overall,
both the soft and softer theories become increasingly unreliable
for small values of $x_\rmii{LO}$,
where the high-temperature expansion is expected to break down.
While our analysis has focused on the critical temperature,
there is no reason to expect that neglecting higher-dimensional operators would
yield more accurate results for e.g.\ $\alpha$ at lower temperatures, particularly below $\Tc$,
where bubble nucleation occurs.
Indeed, we have already plotted the ratio $R$ at the percolation temperature $\Tp$, where the transition completes,
as shown on the right-hand side of figure~\ref{fig:condensate-c6-effect}.
In sec.~\ref{sec:GW}, we explore how $\Tp$ is determined by estimating
the statistical contribution to the bubble nucleation rate and
relating it to cosmological evolution.

\subsection{Comparing higher-dimensional operators and loop corrections}

Given the stringent consistency and perturbativity constraints on the softer EFT,
one natural question that arises is what role the higher-dimensional operators actually play in a perturbative computation of the effective potential.
Our previous discussion provides a clear answer for both limiting cases of $\bar{x}$:
\begin{itemize}
  \item
    For $\bar{x} > 0.2$,
    the softer theory is consistent but non-perturbative,
    which means that loop corrections dominate the effective potential,
    rendering a perturbative computation unfeasible,
    while higher-dimensional operators are essentially irrelevant.
  \item
    For $\bar{x} < 0.2 \, g$,
    the softer theory is formally perturbative, so that
    loop corrections remain subdominant.
    However, higher-dimensional operators start contributing at leading order and
    ultimately lead to the breakdown of the high-temperature expansion.
\end{itemize}
But what about the intermediate regime $0.2 \, g < \bar{x} < 0.2$?
In the symmetric phase, spatial gauge boson modes are approximately massless,
leading to a Linde problem at the ultrasoft scale.
The resulting non-perturbative contributions to the effective potential are expected to scale as
\begin{align}
V_\rmii{NP} &\sim \frac{\Lambda_\rmii{NP}^3}{4\pi} = \frac{\bar \mu_3^3}{4\pi}
  \Bigl( \frac1{(4\pi)^2 y} \Bigr)^{\frac32}
\ , &
\Lambda_\rmii{NP} &= \frac{\bar g_3^2}{4\pi}
 \ .
\end{align}
In comparison, the leading contribution due to higher-dimensional operators scales as
\begin{align}
V &\supset \frac{\bar \mu_3^3}{4\pi}
  \Bigl( \frac1{(4\pi)^2 y} \Bigr)^2
  \Bigl( \frac{\bar \mu_3}{\mD} \Bigr)^3 
  \ ,
\end{align}
which implies that higher-dimensional operator contributions are expected to be small compared to non-perturbative corrections,
and therefore largely unimportant for the computation for the effective potential in the symmetric phase.

A priori,
we expect higher-dimensional operators to be relatively more important in the broken phase,
since they can contribute to the effective potential at tree level.
Since we already include the leading
$\frac{g_3^3 \phi^3}{6\pi}$ term from one-loop gauge boson diagrams,
the main question is whether the
$(\phi^\dagger \phi)^3$ contribution is substantial compared to
the size of the neglected higher-loop corrections.
Adapting the power counting from~\cite{Ekstedt:2024etx},
at the critical temperature $\Tc$, and using $\phid \approx \frac{\bar g_3}{3\pi \bar{x}}$
as well as $\byc \approx \frac1{18 \pi^2 \bar x}$,
higher-order loop corrections to
the effective potential in the broken phase scale as
\begin{align}
\label{eq:loop diagram scaling}
\frac{(\bar g_3 \phid)^3 x^{\frac{n_\lambda + m}2} }{4\pi} \Bigl( \frac{\bar g_3}{4\pi \phid} \Bigr)^{\ell - 1}
&\approx \frac{\bar g_3^6}{108 \pi^4} \Bigl( \frac34 \Bigr)^{\ell - 1} x^{- 4 + \ell + \frac{n_\lambda + m}2} \ ,
\end{align}
where $\ell$ denotes the loop order and $n_\lambda$ the number of Higgs self-interaction vertices.
The parameter $0 \leq m \leq 3$ depends on the effective mass scale of the diagram.
In diagrams without Higgs self-interactions,
$m = 0$ since the effective mass scale is determined by the gauge boson mass,
and in pure scalar diagrams without vector boson lines,
$m = 3$ since the effective mass scale is determined by the Higgs mass.
Using~\eqref{eq:super soft beta}, \eqref{eq:loop diagram scaling}, and
$\bar g_3^2 = g^2 T + \mathcal O(g^4)$, this implies the next-to-leading (NLO) loop correction,
generated by two-loop gauge boson diagrams with $n_\lambda = m = 0$,
is expected to be subdominant compared to the $(\phi^\dagger \phi)^3$ operator if
\begin{align}
  \bar{x} < \biggl( \frac{2}{\sqrt3 \, (3\pi)^3} \biggr)^{\frac14} g^{\frac34}
  \approx 0.19 \, g^{\frac34}
  \ .
\end{align}
The analysis in~\cite{Ekstedt:2024etx} includes diagrams
up to
N$^4$LO, which scale as $x^{-1/2}$,
but neglects
N$^5$LO diagrams with $\ell = 2$ and $n_\lambda + m = \ell + 2$.
These neglected diagrams are expected to be subdominant compared to
$(\phi^\dagger \phi)^3$ contributions if
\begin{align}
  \bar{x} < \biggl( \frac{2}{3^{\frac12} (3 \pi)^3} \biggr)^{\frac16} g^{\frac12} \approx
  0.33 \, g^{\frac12}
  \ .
\end{align}
These estimates show that for small $x$,
uncertainties from neglecting higher-dimensional operators
outweigh those from neglecting higher-loop corrections.
Specifically, for $\bar{x} < 0.19 \, g^{\frac34}$,
the contributions due to higher-dimensional operators dominate even over
the leading two-loop
contributions to the effective potential.

\section{Gravitational wave prospects}
\label{sec:GW}

In the previous section, we discussed the impact of higher-dimensional operators on the phase transition strength at the critical temperature.
To relate these findings to gravitational wave predictions, we need to consider the impact below the critical temperature, and in particular at the percolation temperature $\Tp$, where the phase transition is completed.

To analyze the leading effect of the sextic operator,
we consider the leading-order potentials of
eqs.~\eqref{eq:softer-to-supersoft rescaled} and~\eqref{eq:soft-to-supersoft rescaled} 
and use {\tt FindBounce}~\cite{Guada:2020xnz}
to construct a semi-classical bounce solution $\phib$ that extremizes the effective action
\begin{align}
\label{eq:nucleation action}
  S_3[\phid] &\equiv \int_{\vec{x}} \biggl[
    \frac{1}{2} (\partial_i \phid)^2 + \widetilde V(\phid)
  \biggr]
  \,, &
  \widetilde V(\phid) &\equiv
    \frac{y}{2} \phid^2
  + \frac{x}{4} \phid^4
  + \frac{c_6}{8} \phid^6
  - \frac1{6\pi} \phid^3
  \,,
\end{align}
and thus solves the classical equation of motion $\delta S_3[\phib] =0$.
This stationary solution corresponds to the physical profile of the critical bubble,
and the resulting extremal value of the effective action $S_3[\phid]=S_3(x,y,c_6)$
is the main ingredient needed to determine the thermal bubble nucleation rate at leading order in the semi-classical approximation
or nucleation EFT~\cite{Gould:2021ccf}.
This formulation guarantees theoretical self-consistency,
including the gauge-invariance~\cite{Hirvonen:2021zej,Lofgren:2021ogg}, and
avoids double counting contributions of different energy scales \cite{Gould:2021ccf}.
In particular, it would allow for a consistent inclusion of all large thermal corrections from
the hard scale, including two-loop thermal masses essential for
the renormalization scale independence~\cite{Croon:2020cgk,Gould:2021oba}.
We emphasize that resolving these aforementioned issues is essential for theoretical consistency and ensuring physically meaningful results, but commonly not achieved,
cf.\ e.g.~\cite{Grojean:2006bp,Delaunay:2007wb,Dorsch:2016nrg,Ellis:2018mja}.

The form of the potential in~\eqref{eq:nucleation action} encompasses both
the softer-induced potential~\eqref{eq:softer-to-supersoft rescaled}
and the approximated soft-induced potential~\eqref{eq:soft-to-supersoft rescaled}.
Higher-dimensional kinetic (or derivative) effects
from the scalar and vector sectors of
the corresponding EFT also contribute to the effective action.
These effects can arise both
from a derivative expansion of the prefactor $A$ in the nucleation rate and
from the high-temperature expansion.
Since the derivative expansion does not necessarily converge
at the tail of the bounce~\cite{Kierkla:2023von,Kierkla:2025qyz},
we do not include its higher-order derivative terms explicitly
in eq.~\eqref{eq:nucleation action}.
Instead, we argue that their effects can be more systematically included through
the corresponding fluctuation determinants~\cite{Ekstedt:2021kyx,Ekstedt:2023sqc,Kierkla:2025qyz}.

Using the same approach as~\cite{Croon:2020cgk,Ekstedt:2022ceo,Ekstedt:2024etx}
(also cf.~\cite{Athron:2023rfq}),
we follow~\cite{Enqvist:1991xw} and define the percolation time $\tp$ as the time at which volume fraction of space in the metastable phase is $f(\tp) = 1/e$.
While it should be possible, in principle,
to capture the time-evolution and higher-orders of this volume fraction 
from first principles using
perturbative real-time computations~\cite{%
  Cornwall:1974vz,Berges:2004yj,Kainulainen:2021eki} or
non-perturbative, numerical
simulations~\cite{Hindmarsh:2013xza,Moore:2000jw,Ajmi:2022nmq,Guo:2023gwv,Gould:2024chm},
the standard approach in the literature is to follow~\cite{Guth:1979bh,Guth:1981uk}
and utilize the phenomenological expression
\begin{align}
\label{eq:volume-fraction}
  f(t) &= e^{-I(t)} \ , &
  I(t) &\equiv \int\displaylimits_{\tc}^{t} \hspace{-3pt} \text{d} t' \,
  \Gamma(T(t^\prime)) V(t,t^\prime) \ , &
  V(t, t^\prime) &= \frac{4\pi}3 \vw^3 (t - t^\prime)^3
  \,.
\end{align} 
Here, $\tc$ is the time at which the temperature of the early universe plasma is equal to the critical temperature $\Tc$,
$\Gamma (t) = A(t) \, e^{-S_3(t)}$ the bubble nucleation rate
(i.e.\ the probability of nucleation per unit time per unit volume),
$V(t,t')$ the volume of a (spherically symmetric) bubble that nucleated at $t^\prime$ at some later time $t$, 
$\vw$ the terminal bubble wall velocity, 
and the exponentiation in eq.~\eqref{eq:volume-fraction} accounts for the fact that bubbles can overlap.
Formula~\eqref{eq:volume-fraction} implicitly assumes the thin-wall approximation,
where the volume occupied by the bubble walls is negligible,
and that all bubbles grow at the same rate.

The explicit expression for $V(t,t’)$ in terms of $\vw(t)$ also assumes
that the bubble wall velocity rapidly reaches its terminal value,
and that this value remains approximately constant over time.
Further, assuming that the non-equilibrium prefactor $A(t)$ is approximately constant,
the volume fraction integral becomes
\begin{align}
\label{eq:volume fraction integral}
I(\tp) &\approx \frac{4\pi \vw^3}3 A \int\displaylimits_{\tc}^{\tp} {\rm d}t' e^{-S_3(t')} (\tp - t')^3
\approx  8 \pi \vw^3 \frac{A \, e^{- S_3(\tp)}}{\beta^4} 
\begin{cases}
1  & \gamma \ll 1 \\
\gamma^3 e^\gamma  & \gamma \gg 1
\end{cases}
\ ,
\end{align}
where
\begin{align}
  \beta &\equiv - \frac{{\rm d} S_3}{{\rm d}\tp} \ , &
\gamma &\equiv \beta (\tp - \tc)
\ .
\end{align} 
The quantity $\beta$ defines the inverse duration of the phase transition.
The integral in eq.~\eqref{eq:volume fraction integral} has been evaluated by expanding the action to linear order in $t^\prime - \tp$,
\begin{align}
  S_3(t) =
      S_3(\tp)
    + \frac{{\rm d}S_3}{{\rm d}\tp} (t - \tp)
    + \order{(t-\tp)^2}
    \ .
\end{align}
This approximation captures the correct leading-order dynamics of bubble
nucleation because the integrand in eq.~\eqref{eq:volume fraction integral} is
exponentially suppressed and therefore rapidly decreases for smaller values of
$t^\prime$. This also further justifies our assumption that both $A$ and $\vw$
can be treated as approximately constant.
Nevertheless, higher-order corrections can be systematically included by
evaluating eq.~\eqref{eq:volume fraction integral} numerically, while tracking
the full temperature dependence of
each quantity~\cite{Hindmarsh:2013xza,Ajmi:2022nmq,Guo:2023gwv}.

Using
the approximate expression on the right-hand side of
eq.~\eqref{eq:volume fraction integral} and assuming that $\gamma \ll 1$,
the percolation condition
$I(\Tp) = 1$
for maximally colliding bubbles%
\footnote{%
  The percolation condition used here
  is more conservative than the standard one~\cite{Ellis:2018mja,Turner:1992tz},
  which requires
  the probability of a point in space remaining in
  the false vacuum to be
  $P(\Tp)=e^{-I(\Tp)} \simeq 71\%$
  corresponding to
  $I(\Tp) \approx 0.34$.
}
thus yields the relation
\begin{align}
\label{eq:Tp-exact}
S_3(\Tp) =
  \ln (8 \pi)
  + 3 \ln \vw
  + \ln\frac{A}{H^4}
  - 4 \ln \frac{\beta}{H}
  \,.
\end{align}
In a radiation-dominated FRW cosmology with an ideal gas equation of state,
the Hubble rate is given by%
\footnote{%
  The effective number of radiation degrees of freedom, $g_\star$,
  explicitly enters this relation.
  If our setup is treated as a dark sector decoupled from
  the SM,
  the numerical values of eq.~\eqref{eq:H:T} differ
  from those in the SM,
  as discussed below eq.~\eqref{eq:p:plus}.
}
\begin{align}
\label{eq:H:T}
H(T) &= \sqrt{\frac{4 g_\star \pi^3}{45}} \frac{T^2}{M_{\rmii{Pl}}} \ , &
M_{\rmii{Pl}} &= 1.22 \times 10^{19}~{\rm GeV}
\ .
\end{align}
At this stage, we have not yet specified explicit expressions for the prefactor
$A$ or the bubble wall velocity $\vw$.
However, since these quantities appear in
eq.~\eqref{eq:Tp-exact} only logarithmically,
their exact values are unimportant at leading order;
it suffices to estimate their order of magnitude.
Accordingly, we approximate $A \approx T^4$ by dimensional analysis and take
$\vw \sim 1$, and then impose
the approximate condition~\cite{Gould:2022ran,Ekstedt:2022ceo,Ekstedt:2024etx,Kierkla:2025qyz}
\begin{align}
\label{eq:Tp-approx}
\left. S_3[\phib] \right|_{T = \Tp} \approx 126
  \,,
\end{align}
to find the percolation temperature.
Inverting this condition, one obtains a relationship
$y(\Tp) = \yp(x,c_6)$
that makes it possible to find the percolation temperature $\Tp$
purely in
terms of the 3d quantities $y,x,c_6$,
allowing for efficient scans in the $(x,y)$-plane (while also varying $c_6$).
However, it is also straightforward to relax the
assumptions underlying
eq.~\eqref{eq:Tp-approx}
for a given parameter point of the parent theory,
and to determine $\Tp$ directly from eq.~\eqref{eq:Tp-exact}.
For vanishing $c_6$,
one could find a fit function for $S_3(x,y,c_6=0)$
in analogy to
the one in~\cite{Ekstedt:2022ceo}.

For non-zero $c_6$,
deriving a general fit becomes more challenging.
Instead, we evaluate $S_3$ using eq.~\eqref{eq:nucleation action},
including the soft enhancement with
the coefficient $c_6 = c_6(g)$ at fixed $g$.
This approach allows us to illustrate our key findings
during the parameter scan outlined below.
The corresponding fit takes the form
\begin{align}
\label{eq:S3:fit}
  S_3= \kappa\biggl[A+B\gamma+\frac{C}{1-\gamma}+\frac{D}{(1-\gamma)^2}\biggr]
  \,,
\end{align}
where
$\kappa = 64 \pi^2 y^{3/2}$,
$\gamma = y/\yc(x,c_6)$
with $\yc(x,c_6)$ given in eq.~\eqref{eq:y:soft:EFT:enhanced}.
In the soft theory,
the action fitting parameters are given in table~\ref{tab:fit:soft:softer} (left)
\begin{table}[]
    \centering
    \begin{tabular}[t]{|c|c|c|c|}
         \hline
         $A$ & $B$ & $C$ & $D$
         \\
         \hline
         \hline
         0.751 & -0.413 & 0.704 & 0.075\\
         \hline 
    \end{tabular}
    \hspace{1cm}
    \begin{tabular}[t]{|c|c|c|c|c|}
         \hline
         $g$ & $A$ & $B$ & $C$ & $D$
         \\
         \hline
         \hline
         0.9 &  -30.418 & 56.244 & -2.310 & 0.267\\
         0.7 & 3.353 & -2.030 & 0.484 & 0.206\\
         0.5 & 4.178 & -4.383 & 0.945 & 0.196\\
         \hline 
    \end{tabular}
    \caption{%
      Soft (left) and
      softer (right) theory fit parameters for
      the 3d action~\eqref{eq:S3:fit}.
    }
    \label{tab:fit:soft:softer}
\end{table}
while in the softer theory,
they depend on the value of $g$
as given in table~\ref{tab:fit:soft:softer} (right).
In the softer theory,
the fit is also more sensitive
to values of $g$ as the fitted $S_{3}(x,y,\bar{c}_{6})$
utilizes
$\bar{c}_{6}\sim g^3 $ from eq.~\eqref{eq:super soft beta}
instead of
$c_{6}\sim g^6 $ from eq.~\eqref{eq:c6:soft:EFT}.

The right panel in
fig.~\ref{fig:condensate-c6-effect} compares the ratio of the transition strength
$R = \alpha(T) / \alpha_{c_6 = 0}(T)$
at the critical temperature (dotted lines) and the percolation temperature (solid lines).
As expected, for each fixed $x_{\rmii{LO}}$ the sextic operator is slightly more important at $\Tp < \Tc$,
and reduces the transition strength by $\mathcal{O}(1\%)$
more than at the critical temperature in the regime
that is still within reach of the high-temperature expansion.
For the inverse duration, we find a smaller $\sim1 - 4\%$ reduction
due to the sextic operator in the range $x \in [0.02,0.1]$,
concluding that this quantity appears less sensitive compared to $\alpha$,
based on our simple leading-order estimates.

Besides the transition strength $\alpha$, gravitational wave predictions also depend on
average bubble separation $R_* H_*$ (relative to the Hubble parameter in radiation dominated epoch),
which can be related in our setup to
the inverse duration of the transition $\beta$,
via~\cite{Caprini:2019egz} (cf.~\cite{Enqvist:1991xw})

\begin{align}
  R_* H_* \approx (8\pi)^{\frac13} \max\{\vw,\cs\} \Bigl(\frac{\beta}{H_*}\Bigr)^{-1}
  \,,
\end{align}
where for the strong transitions in our case $\vw > \cs$.
We compute the inverse duration of the transition by using the relation~\cite{Gould:2019qek}
\begin{align}
\label{eq:beta}
\frac{\beta}{H} &=
    T \frac{{\rm d} S_3[\phib]}{{\rm d}T} \approx
    \eta_y \partial_y S_3[\phib] \Bigr|_{T = \Tp}
    \ .
\end{align}
To connect our computation to observational prospects of LISA and DECIGO detectors,
we scan over the model parameters of the parent theory, and
compute thermal parameters $\alpha$ and $\beta/H$ at $\Tp$.

\begin{figure}[t!]
  \centering
  \includegraphics[width=0.5\textwidth]{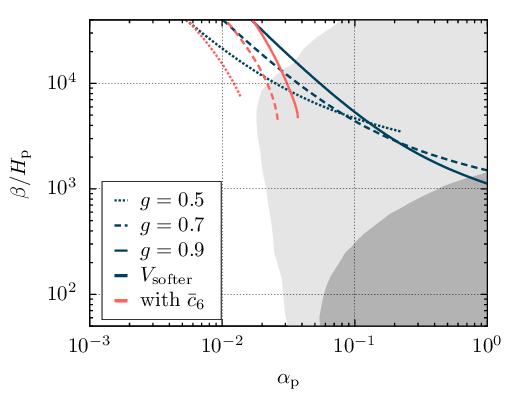}%
  \phantom{\includegraphics[width=0.5\textwidth]{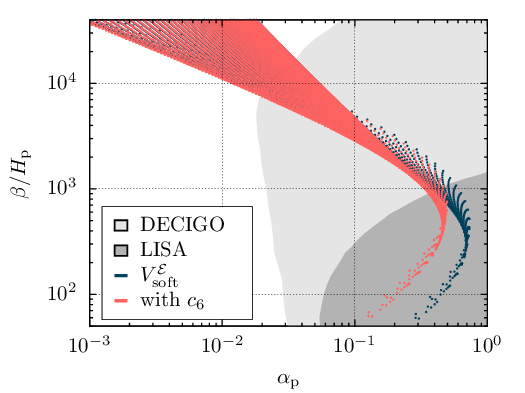}}
  \\
  \includegraphics[width=0.5\textwidth]{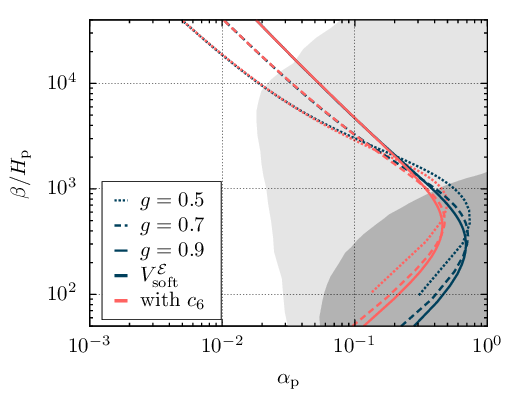}%
  \includegraphics[width=0.5\textwidth]{fig//alpha.beta.scan.soft}%
  \caption{%
  Top left: parameter scan over $x$
  of the softer EFT potential~\eqref{eq:softer-to-supersoft}
  at fixed values
  $g=0.5$ (dotted),
  $g=0.7$ (dashed), and
  $g=0.9$ (solid), where $x$ decreases from left to right.
  Bottom row:
  as the top row but for
  the soft EFT potential~\eqref{eq:VLO-soft}
  which exhibits soft enhancement due to the included temporal mode.
  Bottom right:
  parameter scan using the grid~\eqref{eq:scan:grid}
  in the $(\alphap,\beta/\Hp)$-plane
  with (red) and without (blue) LO effect from $c_6$.
  Also shown are the tentative sensitivity regions for
  LISA~\cite{Caprini:2019egz} and
  DECIGO~\cite{Kawamura:2011zz};
  see footnote~\ref{footnote:LISA}.%
  }
  \label{fig:alpha-beta}
\end{figure}

First, we work in the softer EFT~\eqref{eq:DR:A}.
To employ eqs.~\eqref{eq:alpha} and~\eqref{eq:beta},
we insert the leading-order temperature dependence of $y$
via its $\eta$-function
\begin{align}
  \eta_y \equiv T \frac{{\rm d}y}{{\rm d}T} \approx
  - \frac{2}{g^4} \frac{\mu_{ }^2}{T^2}
  = \frac{1}{g^4} \frac{M^2}{T^2}
  \,,
\end{align}
where we have used leading-order matching relations, as well as tree-level relation
$\mu_{ }^2 = -\frac{1}{2} M^2$ between
the 4d scalar mass parameter and physical scalar mass $M$.%
\footnote{%
  To relate the scalar mass parameter to its pole mass $M$
  at higher orders
  as well as
  relating other \MSbar{} and physical parameters,
  one needs to employ zero-temperature one-loop vacuum renormalization,
  cf.\ e.g.~\cite{Kajantie:1995dw,Niemi:2021qvp}.
}
Next, we could directly compute thermal parameters
$\Tp$,
$\alpha$ and
$\beta/H$ in terms of three input parameters
$g$,
$\lambda = x_\rmii{LO} g^2$ (using the LO relation) and $M$.
However, we can simplify the parameter scan considerably by eliminating the ratio $M/T$.
To this end, we note that at the critical temperature
$y(\Tc) = \yc(x) = 1/(18\pi^2 x)$ and by using the leading-order matching relation,
we find
\begin{align}
\label{eq:eliminate-MT}
  \frac{M^2}{\Tc^2} \approx
      \frac{g^2}{2}
    - \frac{g^4}{9\pi^2 x_\rmii{LO}}
    + \frac{2 g^2 x_\rmii{LO}}{3}
    - \tau \frac{g^3}{2 \sqrt{3}\pi}
\equiv F(g,x_\rmii{LO},\tau)
  \approx
  \frac{M^2}{\Tp^2}
\,. 
\end{align}
For convenience,
we defined an auxiliary parameter
$\tau = 1$ for the softer theory, while for
the soft enhancement $\tau = 0$.
We know that for a radiatively generated barrier, the amount of supercooling,
i.e.\ the difference $\Delta T \equiv \Tc - \Tp$, is small~\cite{Ekstedt:2024etx}.
Therefore, we also approximate
$M^2/\Tp^2 \approx F(g,x)$ using eq.~\eqref{eq:eliminate-MT}.
We have thus eliminated the ratio $M^2/\Tp^2$ in terms of $g$ and $x$,
and can hence express
\begin{align}
\frac{\beta}{H} &\approx \frac{1}{g^4}F(g,x,1) \partial_y S_3(\yp,x,c_6)
  \,, \\
\alpha &\approx \frac{30}{(2\pi)^2 g_\star} g^2 F(g,x,1) \partial_y \Delta \widetilde V_\text{softer}(\yp,x,c_6)
  \,,
\end{align}
at leading order.
Within these approximations, we can readily scan the entire parameter space,
i.e.\ the $(x,g)$-plane.
For this, we perform a uniform scan with
\begin{align}
\label{eq:scan:grid}
  x &\in [0.01,0.3] &\text{with step size } &&\Delta x &= 0.002
  \,,\nn
  g &\in [0.1, 0.9] &\text{with step size } &&\Delta g &= 0.02
  \,,
\end{align}
and recast our result in
the $(\alphap,\beta/\Hp)$-plane in
fig.~\ref{fig:alpha-beta} (top row).

Before discussing the result, we note that our
discussion so far has considered only the softer EFT~\eqref{eq:DR:A}.
Since we have already concluded
that the soft-to-softer EFT construction is unreliable
for strong transitions (at small $x$),
we also consider
the soft-to-supersoft EFT
utilizing the approximated soft-induced effective potential~\eqref{eq:soft-to-supersoft}.
In this case, the formulae for $\alpha$ and $\beta/H$ need to be adjusted to include the effects of soft enhancement factor $\mathcal{E} = \frac{3}{2}$.
This yields 
\begin{align}
  \frac{\beta_{\rmii{$\mathcal{E}$}}}{H} &\approx
      \frac{\mathcal{E}^{-\frac{4}{3}}}{g^4}F(g,x,0)\,
      \partial_\rmii{$Y$} S_3(Y_{\rm p},X,c_6)
  \,, \\
  \alpha_{\rmii{$\mathcal{E}$}} &\approx
      \frac{30}{(2\pi)^2 g_\star}
      \mathcal{E}^{\frac{2}{3}} g^2 F(g,x,0)\,
      \partial_y \Delta\widetilde V_{\mD^2 \ll h_3^{ } \phid^2}(\yp,x,c_6)
  \,,
\end{align}
where we defined the notation
$X \equiv \mathcal{E}^{-\frac{2}{3}} x$ and
$Y \equiv \mathcal{E}^{-\frac{4}{3}} y$.
The subscript $\{\alpha,\beta\}_\rmii{$\mathcal{E}$}$ indicates
the inclusion of the soft enhancement.

Figure~\ref{fig:alpha-beta} depicts the predicted values of
$\alpha$ and $\beta$ for various values of $g$ and $x$.
The top row is obtained using
the soft-to-softer EFT setup~\eqref{eq:DR:A}, while the bottom row is obtained using
the soft-to-supersoft EFT~\eqref{eq:DR:B}.
In each plot, the value of $x$ decreases from left to right, with
the smallest $x$ values mapping into the bottom right corner.
The points approach the tentative LISA
sensitivity window%
\footnote{%
  \label{footnote:LISA}
  The sensitivity regions are taken purely tentative
  at a wall velocity $\vw=0.95$ for
  LISA at ${\rm SNR} = 5$~\cite{LISA:2017pwj} with
  $\mathcal{T}=4$~year mission duration~\cite{Caprini:2019egz} and
  DECIGO with the {\em Correlation} design~\cite{Kawamura:2011zz} were
  taken from~\cite{Friedrich:2022cak}.
  As discussed below eq.~\eqref{eq:p:plus},
  $g_\star = 4$ which differs from the SM one used in the sensitivity regions.
}
for
large values of $\alpha$ and small values of $\beta/H$.%
\footnote{%
  For a signal to be visible for LISA and DECIGO,
  the temperature $\Tp$ has to be in a $\sim 100$~GeV range to produce
  a GW power spectrum
  in the appropriate frequency range.
  To ensure this, we have implicitly assumed here that the scalar mass $M$ is
  $\mathcal{O}(100~{\rm GeV})$.
}
To showcase the dependence on $g$,
the plots of the left-hand side display
results for three fixed values of
$g=0.5, 0.7, 0.9$ while varying $x$.
In each case, including the leading effect of the dimension-six operator reduces $\alpha$.
The branching between the curves with and without higher-operator effect
increases as $\alpha$ and $\beta$ approach
the LISA sensitivity region,
which indicates the increasing uncertainty
and an eventual breakdown of the high-temperature expansion.
The same trend is evident in
the plot on the bottom right-hand side
of fig.~\ref{fig:alpha-beta},
where the tail containing the strongest and longest transitions shifts to
lower values of $\alpha$.

Comparing the top and bottom
plots on the left-hand side,
we observe that it is crucially important to treat the temporal mode correctly
for strong transitions.
In the top row,
including sextic operator effects shifts
all points within
the reach of LISA and DECIGO
toward undetectably small values of $\alpha$.
This signals the limitation of
the softer EFT at small $x$,
and is consistent with the behavior in
fig.~\ref{fig:condensate-c6-effect}.
Numerically, in the regime of even larger $x$,
the curves for the softer EFT with $\bar{c}_6 \neq 0$ and $c_6 = 0$ do not fully converge.
This is due to the limitations of our simplified scanning setup,
described by the fit in eq.~\eqref{eq:S3:fit},
which is more sensitive in this region to
large derivatives of $\frac{{\rm d}}{{\rm d} \ln T} S_3(x, y, \bar{c}_6)$.
While this regime lies outside the focus of our analysis,
a more accurate approach would be a direct computation of the bounce action.

The bottom row of fig.~\ref{fig:alpha-beta} shows
that the temporal mode strengthens the transition in the small-$x$ regime due
to soft enhancement,
shifting points toward larger $\alpha$.
However, even with this improved treatment of the temporal mode,
the impact of the sextic operator remains significant in
the tentative LISA window and
is still non-negligible within the tentative sensitivity range of DECIGO.
This suggests that even if the temporal modes are treated correctly,
the high-temperature expansion may become unreliable for sufficiently strong transitions
such as those targeted by next-generation GW detectors.

In this regard, our work builds on and extends
the findings of~\cite{Gould:2019qek},
which concluded that, in the absence of higher-dimensional operators,
radiatively-generated one-step transitions strong enough to be detectable by LISA
cannot be reliably described.
Our results suggest an even stronger conclusion.
The high-temperature expansion itself may break down entirely,
thereby invalidating the EFT description.
While we arrive at this result within a simplified toy model,
we expect the conclusion to hold more broadly.
This is due to the generic infrared structure of gauge-Higgs theories and
the qualitatively similar features of dimensional reduction,
including the emergence of sizable higher-dimensional operators in
the regime of strongest transitions.
To reach these conclusions regarding the relevance of higher-dimensional operators,
we adopted the following simplifying assumptions:
(i) leading-order in dimensional reduction,
(ii) leading-order in the nucleation effective theory, and
(iii) negligible supercooling, as noted below eq.~\eqref{eq:eliminate-MT}.
Each of these approximations is individually robust when including higher-order corrections.
Accordingly, we anticipate that our main conclusions will remain valid
when such corrections are taken into account.

\section{Conclusions and outlook}
\label{sec:outlook}

Comprehensively understanding the effects of higher-dimensional operators
in thermal EFTs for the electroweak phase transition has been a long-standing challenge.
In this article, we take a significant step toward this goal by studying
the Abelian Higgs model as a simplified toy model.
This model
shares many important features with more complex and
realistic gauge-Higgs theories relevant for electroweak phase transitions,
allowing us to probe the impact of higher-dimensional operators in a controlled setting.

In particular,
we have achieved a complete matching of the marginal sextic operators
by constructing a complete, minimal, and fully gauge-invariant operator basis in both
the soft and softer EFTs.
This step is crucial for ensuring the theoretical consistency of the EFT framework and
accurately capturing the physical properties of the full theory.
As is often the case in advancing the understanding of hot electroweak-like theories,
we have drawn inspiration from analogies in
hot QCD~\cite{Chapman:1994vk,Laine:2018lgj} and
weaponized existing tools for general EFT matching,
such as {\tt Matchete}~\cite{Fuentes-Martin:2022jrf}.
We anticipate that these developments will enhance
the capabilities of {\tt DRalgo}\cite{Ekstedt:2022bff},
which, in turn, utilizes functionalities from
{\tt GroupMath}~\cite{Fonseca:2020vke}.

We have also clarified the role of
temporal gauge boson modes
by demonstrating that treating them within the framework of
soft-to-softer EFT matching\textemdash
i.e., integrating out temporal modes and absorbing their effects into
the scalar sector parameters~\cite{Kajantie:1995dw,Farakos:1994kx}\textemdash
induces large contributions from marginal operators in the regime of strong transitions.
As a result, soft-to-softer EFT matching proves unreliable for strong transitions.%
\footnote{%
  This conclusion is softened in theories with larger Debye masses,
  such as ${\rm SU}(N)$ theories with many charged fermions,
  where the softer EFT is expected to perform better.
  This could be tested by repeating
  a similar study within SU(2) gauge-Higgs theory
  and by computing higher-loop corrections akin to~\cite{Ekstedt:2024etx},
  while treating temporal gauge fields on equal footing with spatial ones.
  In addition, a direct comparison would require lattice studies
  that explicitly incorporate
  dynamical temporal gauge fields,
  cf.~\cite{%
    Kajantie:1993ag,Jakovac:1994xg,Kajantie:1997pd,%
    Kajantie:2000iz,Hart:2000ha,Gould:2022ran}.
}
Instead, temporal modes should be treated within the framework of
soft-to-supersoft matching~\cite{Gould:2023ovu}, as implemented
in~\cite{Kierkla:2023von,Lewicki:2024xan,Gould:2024jjt,Kierkla:2025qyz}.
This refinement leads to an enhanced potential barrier compared
to including only spatial mode contributions.

For radiatively induced phase transitions in the Abelian Higgs model,
we demonstrated that including the leading effect of the sextic $(\phi^\dagger \phi)^3$-operator
in the effective potential leads to a significant
$\mathcal{O}(5\%)$ reduction in the phase transition strength
in the regime of strong transitions that are still within reach of the high-temperature expansion.
For even stronger transitions, the sextic operator contribution quickly increases.
This signals a potential breakdown of
the EFT description due to a lack of thermal scale hierarchies,
as the large background field value induces masses for
the Matsubara zero-modes that are comparable to the hard scale.
Such a breakdown would not only limit the applicability of state-of-the-art perturbative techniques,
but also of non-perturbative lattice studies.
Overcoming these limitations would require the development of new techniques to compute
the phase transition thermodynamics entirely without using
high-temperature expansions~\cite{Laine:2000kv,Laine:2017hdk},
while still being able to resum infrared-sensitive contributions at small field values.
Our findings highlight the importance of such endeavors,
and in particular the need for direct lattice simulations of the full theory in
four dimensions~\cite{Bunk:1992kf,Fodor:1994sj,Csikor:1995jj,Csikor:1998eu,Aoki:1999fi}.

Extending the present study to more complicated BSM theories could help to comprehensively
investigate the reliability of current state-of-the-art non-perturbative studies,
such as~\cite{Kainulainen:2019kyp,Niemi:2020hto,Niemi:2024axp},
which rely on the high-temperature expansion and
thermal EFTs that are truncated to not include marginal operators.
Scrutinizing the accuracy and the limits of such analyses is of key importance
to reliably predict the thermodynamics of cosmological phase transitions,
which, in turn, affect open problems such as the matter-antimatter asymmetry of
the universe in the framework of electroweak baryogenesis~\cite{Morrissey:2012db},
as well as predictions for a gravitational wave background generated by such
transitions~\cite{Caprini:2015zlo,Caprini:2019egz,LISACosmologyWorkingGroup:2022jok}.

For gravitational wave predictions,
we found that including the sextic operator in
the small-$x$ regime decreases both $\alpha$ and $\beta/H$.
While the reduction in $\beta/H$ improves sensitivity for LISA,
the overall weakening of the transition is generally unfavorable for observational prospects,
which benefit from stronger phase transitions.
We have found that it is indeed the strongest transitions that
are most affected by higher-order operators and
the associated possibility of a breakdown of the high-temperature expansion.
While this
puts serious limitations on
the predictability of primordial remnants from thermal phase transitions at LISA and
similar experiments,
it also poses an opportunity to refine theoretical approaches.

\section*{Acknowledgments}

We generously thank
Andreas Ekstedt,
Lauri Niemi and
Juuso {\"O}sterman for their contributions and insights at the early stages of this work.
In addition, we thank 
Jaakko Annala,
Oliver Gould, 
Joonas Hirvonen,
Pablo Navarrete,
Risto Paatelainen,
Kari Rummukainen,
Riikka Sepp{\"a},
Kaapo Sepp{\"a}nen, and
Jorinde van de Vis
for illuminating discussions.
We also thank Cristina Puchades-Ibáñez
for pointing out a typo in eq.~\eqref{eq:c6:soft:EFT} compared
to the original version of the manuscript.
FB and PS were supported by
the Swiss National Science Foundation (SNSF) under grant
\href{https://data.snf.ch/grants/grant/215997}{{\tt PZ00P2-215997}}.
PK was supported by the SNF under grant
\href{https://data.snf.ch/grants/grant/217885}{{\tt P500PT-217885}}.
TT was supported by the European Union (ERC, CoCoS, 101142449).
\\

\noindent
{\bf Data availability statement.}
The datasets supporting
figs.~\ref{fig:condensate-c6-effect} and \ref{fig:alpha-beta}
are publicly available at~\cite{ZenodoData}.

\appendix
\renewcommand{\thesection}{\Alph{section}}
\renewcommand{\thesubsection}{\Alph{section}.\arabic{subsection}}
\renewcommand{\theequation}{\Alph{section}.\arabic{equation}}

\section{Field redefinitions}
\label{sec:field:redefinitions}

To construct an effective field theory, one has to specify a power counting in terms of some small parameter $\epsilon$
such that, at any given order in $\epsilon$, only
a finite number of operators contribute to its action
\begin{align}\label{eq:generic eft action}
S_\rmii{EFT} \left[\left\{ \Phi_i \right\} \right] &= \sum_{n=0}^{\infty} \epsilon^n S_n \left[ \left\{ \Phi_i \right\} \right] \ ,
\end{align}
where $\left\{\Phi_i\right\}$ collectively denotes the various fields of the theory.
If the action contains a term
\begin{align}
S_\rmii{EFT} \left[\left\{ \Phi_i \right\} \right] &\supset \epsilon^n \vspace{-5pt} \int \text{d}^d x \, \frac{\delta S_0}{\delta \Phi_i (x)} \mathcal O[\left\{\Phi_i\right\}](x) \ , &
n &\geq 1 \ , 
\end{align}
where $\mathcal O[\left\{\Phi_i\right\}](x)$ is a composite operator that depends only on fields evaluated at coordinate $x$,
this term can be removed from the action up to corrections of order $\epsilon^{n+1}$ by means of the field redefinition
\begin{align}\label{eq:field redefinitions}
\Phi_i &\to \widetilde \Phi_i - \epsilon^n \mathcal O[\{\widetilde \Phi_i\}] \ .
\end{align}
Since the action~\eqref{eq:generic eft action} must already contain all operators that are consistent with the symmetries of the effective theory,
the net effect of~\eqref{eq:field redefinitions} is to
rescale the Wilson coefficients of higher-order operators.%
\footnote{%
  Note that if $\Phi_i$ is a gauge field, the gauge-fixing and ghost Lagrangians should be considered a part of the zeroth-order effective action $S_0$,
  so that the field redefinition~\eqref{eq:field redefinitions} implicitly changes the gauge-fixing.
  The original gauge-fixing can then be restored by using a normal gauge transformation.
}
Furthermore, working in dimensional regularization and excepting fermions
that are subject to a chiral anomaly, it can be shown~\cite{Arzt:1993gz}%
\footnote{%
  While~\cite{Georgi:1991ch,Arzt:1993gz} only consider theories in
  Minkowski space, our primary focus is on Euclidean theories.
  However, the proof does not rely on any information about the signature of
  the metric tensor and hence translates to Euclidean geometry without modification.
}
that field redefinitions of the form~\eqref{eq:field redefinitions} do not alter the path integral measure $\mathcal D \{ \Phi_i \}$ in the generating functional
\begin{align}
Z[J_i] &= e^{i W[J_i]} = \frac1{\mathcal N} \int \vspace{-5pt} \mathcal D \{ \Phi_i \} \, e^{- S_\text{EFT} \left[\left\{ \Phi_i \right\} \right] + \int \text{d}^d x J_i (x) \Phi_i(x) } \ ,
\end{align}
which is normalized such that $Z[0] = 1$.
The source term $J_i (x) \Phi_i(x)$, and with it $Z[J_i]$ as a whole, is \emph{not} in general invariant under~\eqref{eq:field redefinitions},
but this can be shown to not affect
on-shell $S$-matrix elements~\cite{Georgi:1991ch,Arzt:1993gz}.
We focus on the Euclidean 1PI effective action $\mathrm\Gamma[\phi_i]$,
which obeys the recursive relation
\begin{align}
e^{- \mathrm\Gamma[\phi_i]} &= \int \vspace{-5pt} \mathcal D \{ \Phi_i \} \, e^{- S_\text{EFT} \left[\left\{ \Phi_i \right\} \right] + \int \text{d}^d x J_i (x) (\Phi_i(x) - \phi_i)  } \ , &
J_i(x) &= \frac{\delta \mathrm\Gamma}{\delta \phi_i(x)} \ .
\end{align}
For field configurations that obey the equations of motion $J_i(x) = 0$,
this relation, together with the invariance of the path integral measure, implies that the effective action is not affected by the field redefinition~\eqref{eq:field redefinitions}.
Since the effective potential $V(\phi)$ is just the effective action of a constant field configuration divided by the volume of spacetime,
this result implies that both the location of the minima of the effective potential as well as the value of the potential at those minima are likewise unaffected by the field redefinition.

To utilize the field redefinition~\eqref{eq:field redefinitions} in practice, one has to identify the leading-order non-vanishing contribution to the effective action $S_\rmii{EFT}$.
In our setup, the first non-vanishing contributions are of $\mathcal{O}(g^2)$,
so that the corresponding effective Lagrangian is
\begin{align}
\mathcal L &=
D_i \phi^\dagger D_i \phi
+ \mu_3^2 \phi^\dagger\phi
+ \lambda_3 (\phi^\dagger \phi)^2
\nn &
+ \frac14 F_{ij}F_{ij}
+ \frac12 (\partial_i B_0)^2
+ g_3^2 \phi^\dagger\phi B_0^2
+ \frac12 \mD^2 B_0^2 \ .
\end{align}
This yields the equations of motion
\begin{subequations}
\label{eq:loeom}
\begin{align}
  D^2 \phi &=
  \left( \mu_3^2 + 2 \lambda_3^{ } (\phi^\dagger \phi) + g_3^2 B_0^2 \right) \phi
  \ , &
\partial^2 B_0 &=
  \left( \mD^2 + 2 g_3^2 (\phi^\dagger \phi) \right) B_0
  \ , \\[2mm]
D^2 \phi^\dagger &=
  \left( \mu_3^2 + 2 \lambda_3^{ } (\phi^\dagger \phi) + g_3^2 B_0^2 \right) \phi^\dagger
  \ , &
  \partial_i F_{ij} &= i g_3 \, \phi^\dagger D_j \phi + \textit{h.c.}
  \ ,
\end{align}\end{subequations}
which we can use to eliminate any higher-order operator containing the terms
$D^2 \phi$,
$D^2 \phi^\dagger$,
$\partial^2 B_0$, and
$\partial_i F_{ij}$.

\section{Details of dimensional reduction} 
\label{sec:dimred:hard-soft}

This appendix collects
renormalization group equations (RGE),
the matching relations of
the fundamental, four-dimensional theory, defined in eq.~\eqref{eq:lag:4d} to its
dimensionally reduced, three-dimensional, effective theory in eq.~\eqref{eq:lag:3d},
and
the thermal effective potential computed within the EFT.

\subsection{Renormalization and $\beta$-functions at zero temperature}
\label{sec:rge}

The renormalization group equations listed below
are associated with the parameters of the model in eq.~\eqref{eq:lag:4d} and
encode their running with respect to
the \MSbar{} renormalization scale $\LamD$ via the $\beta$-functions.
To this end, we use
\begin{equation}
\label{eq:rge:g1}
t \equiv \ln\LamD^2
\;,
\end{equation}
where $\LamD^2 \equiv 4\pi e^{-\gammaE} \Lambda^2$,
and find
at one-loop level
\begin{align}
\partial_{t}^{ }
g^2 &=
  \frac{1}{(4\pi)^2} \Big( \frac{1}{3} g^{4} \Big)
  \;, \\
\partial_{t}^{ }
\mu_{ }^{2} &=
  \frac{1}{(4\pi)^2} \mu_{ }^{2} \Bigl( 4\lambda - 3 g^{2} \Bigr)
  \;, \\
\partial_{t}^{ }
\lambda_{ } &=
  \frac{1}{(4\pi)^2} \Bigl( 10\lambda^{2} - 6g^{2}\lambda + 3 g^{4} \Bigr)
  \;.
\end{align}

\subsection{One-loop dimension-six coefficients of the soft 3d EFT}
\label{sec:dim6:preShift}

Dimension-four vertices and matching relations for
the Abelian Higgs model are known~\cite{Farakos:1994kx,Hirvonen:2021zej}.
Below we collect the results for
the corresponding matching relations up to dimension-six operators
and before field redefinitions.
Results are given in dimensional regularization and
with dimension $d=3-2\epsilon$.

In tables~\ref{tab:softop:ren} and \ref{tab:softop:dim6},
all the operators up to dimension 6 along with
their respective Wilson coefficients are listed
($D_i\equiv \partial_i-i g_3 Z_{B_i}^{1/2}B_i$).
Operators containing an odd number of $B_0$ fields are
excluded because the matching conditions require them to vanish
due to the Abelian nature of the theory and
charge conjugation symmetry~\cite{Hart:2000ha}.
\begin{table}[t]
\centering
\begin{tabular}[t]{|l|l|}
\hline
\multicolumn{2}{|c|}{dimension~2}\\
\hline
$B_0^2$ & $Z_{B_0}\, \widehat m_\rmii{D}^2/2$
\\ \hline
$\phi^\dagger\phi$ & $Z_\phi\, \hat \mu^2_3$
\\ \hline
\end{tabular}
\hspace{1cm}
\begin{tabular}[t]{|l|l|}
\hline
\multicolumn{2}{|c|}{dimension~4}\\
\hline
$(\partial_i B_0)^2/2$ & $Z_{B_0}$
\\ \hline
$F_{ij}F_{ij}/4$ & $Z_{B_i}$
\\ \hline
$(D_i\phi)^\dagger(D_i\phi)$ & $Z_\phi$
\\ \hline
$(\phi^\dagger\phi)^2$ & $Z_\phi^2\alphaF_{\phi^4} \;$
\\ \hline
$(\phi^\dagger\phi) B_0^2$ & $Z_\phi Z_{B_0}\alphaF_{\phi^2 B_0^2}$
\\ \hline
$B_0^4$ & $Z_{B_0}^2\alphaF_{B_0^4}$
\\ \hline
\end{tabular}
\caption{%
  Super-renormalizable operators and their corresponding
  Wilson coefficients in the soft-scale 3d effective theory.
  Note that the wave-function renormalization constants
  $Z_\phi$, $Z_{B_0}$ and $Z_{B_i}$ are subsequently reabsorbed into field operators
  $\phi$, $B_0$, and $B_i$ to ensure canonical normalization.
}
\label{tab:softop:ren}
\end{table}
\begin{table}[t]
\centering
\begin{tabular}[t]{|l|l|}
\hline
\multicolumn{2}{|c|}{dimension-six operator basis}\\
\hline
$F_{ij}F_{ij} B_0^2$ &
$Z_{B_0} Z_{B_i}\alphaF_{B_0^2 F^2}$ \\ \hline
$F_{ij} F_{ij} \phi^\dagger \phi$ &
$Z_{B_i} Z_\phi\alphaF_{\phi^2 F^2}$ \\ \hline
$(D_i \phi^\dagger D_i \phi) (\phi^\dagger \phi)$ &
$Z_\phi^2  \alphaF_{D^2\phi^4, 1}$ \\ \hline
$(D_i\phi^\dagger D_i \phi) B_0^2$ &
$Z_\phi Z_{B_0}\alphaF_{D^2\phi^2 B_0^2, 3}$ \\ \hline
$B_0^6$ & $Z_{B_0}^3\alphaF_{B_0^6}$ \\ \hline
$B_0^4 (\phi^\dagger\phi) $ &
$Z_\phi Z_{B_0}^2\alphaF_{\phi^2 B_0^4}$ \\ \hline
$B_0^2 (\phi^\dagger\phi)^2 $ &
$Z_\phi^2 Z_{B_0}\alphaF_{\phi^4 B_0^2}$ \\ \hline
$(\phi^\dagger\phi)^3$ &
$Z_\phi^3\alphaF_{\phi^6 }$ \\ \hline
\end{tabular}
\hspace{1cm}
\begin{tabular}[t]{|l|l|}
\hline
\multicolumn{2}{|c|}{Redundant operators}\\
\hline
$(\partial_i F_{ij})^2$ &
$Z_{B_i}\alphaF_{D^2 F^2}$ \\ \hline
$B_0 \Box^2 B_0$ &
$Z_{B_0}\alphaF_{D^4B_0^2}$ \\ \hline
$B_0^3 \Box B_0$ &
$Z_{B_0}^2\alphaF_{D^2B_0^4}$ \\ \hline
$(D^2 \phi^\dagger) (D^2 \phi)$ &
$Z_\phi\alphaF_{D^4\phi^2}$ \\ \hline
$(\phi^\dagger \phi) (\phi^\dagger D^2 \phi + \textit{h.c.})$ &
$Z_\phi^2 \alphaF_{D^2\phi^4,2}$ \\ \hline
$(\partial_i F_{ij}) i \phi^\dagger (D_j \phi)$ &
$Z_\phi Z_{B_i}^{1/2}\alphaF_{D^2\phi^2 F}$ \\ \hline
$(\phi^\dagger \phi) B_0 \Box B_0$ &
$Z_\phi Z_{B_0}\alphaF_{D^2\phi^2 B_0^2, 1}$ \\ \hline
$\phi^\dagger (D_i ^2\phi) B_0^2 + \textit{h.c.}$ &
$Z_\phi Z_{B_0}\alphaF_{D^2\phi^2 B_0^2, 2}$ \\ \hline
\end{tabular}
\caption{%
  All possible dimension-six operators in the soft-scale 3d effective theory.
  The operators of all the redundant operators can be eliminated using
  the leading-order equations of motion~\eqref{eq:loeom}.
  In the final Lagrangian, we also combine partial integration with
  the equations of motion to remove the operator
  $(D_i \phi^\dagger D_i \phi) B_0^2$ in favor of
  $(\phi^\dagger \phi) (\partial_i B_0 \partial_i B_0)$.
  }
  \label{tab:softop:dim6}
\end{table}

Below, we present the result of the one-loop matching between the two theories.
The fundamental theory
lives
in $D=d+1$ spacetime
and
the EFT
in $d=3$ spatial dimensions.
Together with the field normalizations
\begin{eqnarray}
    Z_{B_i} &=& 1+\frac{1}{3}g^{2}\mathcal{I}_{2}^{ }
    \;, \\[2mm]
    Z_{B_0} &=& 1+\frac{1}{3}(4-d)g^{2}\mathcal{I}_{2}^{ }
    \;, \\[2mm]
    Z_\phi &=& 1+(\xi-3)g^{2}\mathcal{I}_{2}^{ }
    \;,
\end{eqnarray}
we display 
the coefficients up to dimension-four operators
{\em viz.}
\begin{eqnarray}
    \hat g_3^2 &=&
        g^2T^{ }
      \Bigl[
          1
        - \frac{1}{3}g^{2}\mathcal{I}_{2}^{ }
      \Bigr]
    \;, \\[2mm]
    \hat\mu_3^2 &=&
    \mu^2+(dg^{2}+4\lambda)\mathcal{I}_{1}^{ }
    \;, \\[2mm]
    \widehat m_{\rmii{D}}^2 &=& 2(d-1)g^{2}\mathcal{I}_{1}^{ }
    \;, \\[2mm]
    \hat \lambda_3 &=&
      \lambda\,T
    -\Bigl[dg^{4}-6g^{2}\lambda+10\lambda^2\Bigr] \mathcal{I}_{2}^{ }\,T
    \;, \\[2mm]
    \hat h_3 &=&
      g^{2}\,T-\frac{1}{3}\Bigl[(2d-5)g^{4}+12(d-3)g^{2}\lambda\Bigl]\mathcal{I}_{2}^{ }\,T
    \;, \\[2mm]
    \hat\kappa_{3} &=& -\frac{1}{6}(d-3)(d-1)g^{4}\mathcal{I}_{2}^{ }\,T
    \;.
\end{eqnarray}
Here and henceforth,
we employ
hatted coefficients
(e.g.\ $\alphaF_i$) for quantities before field redefinitions and
un-hatted coefficients
(e.g.\ $\alphaFR_i$) for quantities after field redefinitions.

The novel dimension-six operator coefficients read
\begin{eqnarray}
    \alphaF_{D^2F^2} &=& -\frac{1}{30}g^{2}\mathcal{I}_{3}^{ }
    \;, \\[2mm]  
    \alphaF_{D^4B_0^2} &=&
      \frac{1}{30}(d-6)g^{2}\mathcal{I}_{3}^{ }
    \;, \\[2mm]  
    \alphaF_{D^2B_0^4} &=&
    -\frac{1}{9}(d-5)(d-4)g^{4}\mathcal{I}_{3}^{ }\,T
    \;, \\[2mm] 
    \alphaF_{B_0^2F^2} &=&
    -\frac{1}{6}(d-5)g^{4}\mathcal{I}_{3}^{ }\,T
    \;, \\[2mm] 
    \alphaF_{B_0^6} &=& \frac{1}{45}(d-5)(d-3)(d-1)g^{6}\mathcal{I}_{3}^{ }\,T^2
    \;, \\[2mm] 
    \alphaF_{D^4\phi^2} &=&
    -\frac{1}{3}(3\xi-5)g^{2}\mathcal{I}_{3}^{ }
    \;, \\[2mm] 
    \alphaF_{D^2\phi^4,1} &=&
    \frac{1}{3}\Bigl[(15-2d)g^{4}+80g^{2}\lambda-8\lambda^2\Bigr]\mathcal{I}_{3}^{ }\,T
    \;, \\[2mm] 
    \alphaF_{D^2\phi^4,2} &=&
      - \frac{1}{6}\Bigl[
          (2d+5)g^{4}-4(3\xi+5)g^{2}\lambda+16\lambda^2
      \Bigr]\mathcal{I}_{3}^{ }\,T
    \;, \hspace{1.2cm} \\[2mm] 
    \alphaF_{D^2\phi^2F} &=&
    \frac{4}{3}g^{3}\mathcal{I}_{3}^{ }\,T^{1/2}
    \;, \\[2mm] 
    \alphaF_{\phi^2F^2} &=&
    \frac{1}{6}\Bigl[7g^{4}-4g^2\lambda\Bigr]\mathcal{I}_{3}^{ }\,T^{}
    \;, \\[2mm] 
    \alphaF_{D^2\phi^2B_0^2,1} &=&
    -\frac{1}{3}\Bigl[(d-2)g^{4}+4(d-6)g^{2}\lambda\Bigr]\mathcal{I}_{3}^{ }\,T^{}
    \;, \\[2mm] 
    \alphaF_{D^2\phi^2B_0^2,2} &=&
    \frac{1}{6}\Bigl[3(d+2\xi-6)g^{4}-4(d-4)g^{2}\lambda\Bigr]\mathcal{I}_{3}^{ }\,T^{}
    \;, \\[2mm] 
    \alphaF_{D^2\phi^2B_0^2,3} &=&
    \frac{1}{3}(d-4)\Bigl[7g^{4}-4g^{2}\lambda\Bigr]\mathcal{I}_{3}^{ }\,T^{}
    \;, \\[2mm] 
    \alphaF_{\phi^4B_0^2} &=&
    \frac{4}{3}\Bigl[d g^{6}+3(d-\xi)g^{4}\lambda+(13d-63)g^{2}\lambda^2\Bigr]\mathcal{I}_{3}^{ }\,T^{2}
    \;, \\[2mm] 
    \alphaF_{\phi^2B_0^4} &=&
    \frac{1}{3}\Bigl[(d(d-4)-3\xi)g^{6}+4(d-3)(d-5)g^{4}\lambda\Bigr]\mathcal{I}_{3}^{ }\,T^{2}
    \;, \\[2mm] 
    \alphaF_{\phi^6} &=&
    \frac{4}{3}\Bigl[dg^{6}-3\xi g^{2}\lambda^2+28\lambda^3\Bigr]\mathcal{I}_{3}^{ }\,T^{2}
    \;.
\end{eqnarray}
Some of these 6-point vertices exhibit
evanescent operator terms, whose coefficients vanish in $d = 3$,
akin to the behavior observed in QCD~\cite{Laine:2018lgj}.
Such operators can in general be eliminated through a shift of
the operator basis~\cite{Aebischer:2022tvz}.

The thermal master integrals used in this context have the form
\begin{eqnarray}
  \mathcal{I}_{s}^{\alpha} &=&
  \Tint{P} \frac{p_0^{\alpha}}{[P^2]^{s}}
  =
  2T \frac{[2\pi T]^{d-2s+\alpha}}{(4\pi)^\frac{d}{2}}
  \frac{\Gamma\bigl(s-\frac{d}{2}\bigr)}{\Gamma(s)}
  \zeta_{2s-\alpha-d}
  \;,\\[2mm]
\label{eq:Z:3}
  \mathcal{I}_{3}^{ }
  &\stackrel{d=3-2\epsilon}{=}&
 \frac{\zeta^{ }_3 \, \mu^{-2\epsilon} }{128 \pi^4 T^2}
 \biggl\{
  1 + 2 \epsilon \biggl[
   \ln \biggl( \frac{\LamD e^{\gammaE}}{4\pi T} \biggr)
 + 1 - \gammaE
 + \frac{\zeta'_3}{\zeta^{ }_3}
 \biggr]
  + \mathcal{O}(\epsilon^2)
 \biggr\}
 \;,
\end{eqnarray}
where $\zeta^{ }_n \equiv \zeta(n)$ and
$\mathcal{I}_{s}^{ } = \mathcal{I}_{s}^{0}$.
Along the way, we utilized the
one-loop integration-by-parts (IBP) identity
\begin{align}
    \mathcal{I}_{s+1}^{\alpha+2} &=\Big(1-\frac{d}{2s}\Big)\mathcal{I}_{s}^{\alpha}
    \,.
\end{align}

\subsection{Dimension-six vertices of the soft 3d EFT in the $S/T$ basis}
\label{sec:eft:soft:dim6}

Below, we collect the results for dimension-six vertices
and the corresponding mapping
to the operator coefficients defined in the previous section.

Because the presence of a heat bath breaks Lorentz invariance,
we need to introduce separate notation for spatial and zero
spacetime indices akin to~\cite{Laine:2018lgj}.
The full Kronecker symbol is denoted by
\begin{equation}
 \delta^{ }_{\mu\nu} \; \equiv \; T^{ }_{\mu\nu} + S^{ }_{\mu\nu}
 \;,
 \quad
 T^{ }_{\mu\nu} \; \equiv \; \delta^{ }_{\mu 0}\delta^{ }_{\nu 0}
 \;,
 \quad
 S^{ }_{\mu\nu} \; \equiv \;  \delta^{ }_{\mu i}\delta^{ }_{\nu i}
 \;. \label{Smunu}
\end{equation}
We also introduce the totally symmetric tensors
\begin{eqnarray}
 T^{ }_{\mu\nu\rho\sigma} & \equiv &
\delta^{ }_{\mu 0} \delta^{ }_{\nu 0}
\delta^{ }_{\rho 0} \delta^{ }_{\sigma 0}
 \;, \\
 T^{ }_{\mu\nu\rho\sigma\alpha\beta} & \equiv &
 \delta^{ }_{\mu 0} \delta^{ }_{\nu 0}
 \delta^{ }_{\rho 0} \delta^{ }_{\sigma 0}
 \delta^{ }_{\alpha 0} \delta^{ }_{\beta 0}
 \;, \\
 \delta^{ }_{\mu\nu\rho\sigma} & \equiv &
 \delta^{ }_{\mu \nu} \delta^{ }_{\rho \sigma}
 + \mbox{2 permutations}
 \;, \\
 \delta^{ }_{\mu\nu\rho\sigma\alpha\beta} & \equiv &
 \delta^{ }_{\mu \nu}
 \delta^{ }_{\rho \sigma}
 \delta^{ }_{\alpha \beta}
 + \mbox{14 permutations}
 \;. \label{perm14}
\end{eqnarray}
It is advantageous to employ a basis in which the spatial and temporal indices are
strictly separated from each other.

\begin{figure}[t]
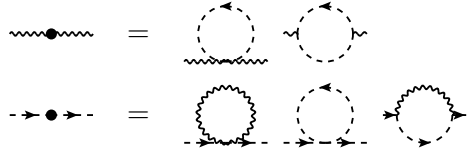

  \centering
  \begin{eqnarray*}
    \Vtxt(\Lglx,\Lglx) &=&
    \TopoST(\Lglx,\Asc1)
    \TopoSB(\Lglx,\Asc1,\Axx)
    \nn[2mm]
    \Vtxt(\Lsc1,\Lcs1) &=&
    \TopoST(\Lsc1,\Aglx)
    \TopoST(\Lsc1,\Asc1)
    \TopoSB(\Lsc1,\Aglx,\Asc1)
  \end{eqnarray*}
  \caption{%
    One-loop contributions to the
    Abelian Higgs model 2-point function at the soft EFT.
    Dashed directed lines denote soft scalars and
    wiggly lines soft spatial gauge fields.
    Diagrams are generated with {\tt Axodraw}~\cite{Collins:2016aya}.
  }
  \label{fig:2pt:diags}
\end{figure}
The 2-point vertex can be expressed as
\begin{align}
\label{2pt_ops}
  \delta S^{(2)}_\rmi{soft} =
  B^{ }_\mu(q)\,
  B^{ }_\nu(-q)\,
  \Bigl\{
      \eta^{ }_{1} \, q^2\bigl(q^2 S^{ }_{\mu\nu} - q_\mu^{ } q_\nu^{ } \bigr)
    + \eta^{ }_{2} \, q^4 T^{ }_{\mu\nu}
 \Bigr\}
 +
  \eta^{ }_{3} \, q^4\,
  \phi(q)\,
  \phi^\dagger(-q)\,
 \;.
\end{align}
Here and henceforth, for the matching coefficients in the action,
field normalization is accounted for
such that for a general correlator
e.g.\
$\Gamma_{\phi_1\dots,\phi_n}
= \eta_{\phi_1 \dots \phi_n} Z_{\phi_1}^\frac{1}{2}\dots Z_{\phi_n}^\frac{1}{2}$.
The contributing 2-point diagrams are collected in
fig.~\ref{fig:2pt:diags}.
A representation of the coefficients was chosen
to map directly into the $\alphaF$ coefficients
of sec.~\ref{sec:dim6:preShift},
\begin{align}
\label{eq:2pt:mapping}
  \eta_{1} &=
     \alphaF_{D^2F^2}
  \;,&
  \eta_{2} &=
    \alphaF_{D^4B_0^2}
  \;,&
  \eta_{3} &=
    \alphaF_{D^4\phi^2}
    \;.
\end{align}

\begin{figure}[t]
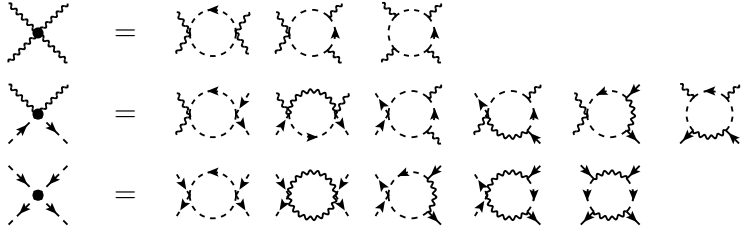

  \centering
  \begin{eqnarray*}
    \Vtxv(\Lglx,\Lglx,\Lglx,\Lglx) &=&
    \TopoVBlr(fex(\Lglx,\Lglx,\Lglx,\Lglx),\Asc1,\Axx)
    \TopoVCrl(fex(\Lglx,\Lglx,\Lglx,\Lglx),\Axx,\Asc1,\Axx)
    \TopoVD(fex(\Lglx,\Lglx,\Lglx,\Lglx),\Asc1,\Axx,\Axx,\Axx)
    \nn[2mm]
    \Vtxv(\Lglx,\Lglx,\Lsc1,\Lcs1) &=&
    \TopoVBlr(fex(\Lglx,\Lglx,\Lsc1,\Lcs1),\Asc1,\Axx)
    \TopoVBlr(fex(\Lglx,\Lsc1,\Lglx,\Lcs1),\Aglx,\Asc1)
    \TopoVCrl(fex(\Lsc1,\Lcs1,\Lglx,\Lglx),\Axx,\Asc1,\Axx)
    \TopoVC(fex(\Lglx,\Lcs1,\Lglx,\Lsc1),\Axx,\Asc1,\Aglx)
    \TopoVClr(fex(\Lglx,\Lglx,\Lsc1,\Lcs1),\Asc1,\Aglx,\Axx)
    \TopoVD(fex(\Lglx,\Lsc1,\Lcs1,\Lglx),\Axx,\Aglx,\Axx,\Asc1)
    \nn[2mm]
    \Vtxv(\Lsc1,\Lsc1,\Lcs1,\Lcs1)
    &=&
    \TopoVBlr(fex(\Lsc1,\Lcs1,\Lsc1,\Lcs1),\Asc1,\Axx)
    \TopoVBlr(fex(\Lsc1,\Lcs1,\Lsc1,\Lcs1),\Aglx,\Aglx)
    \TopoVCrl(fex(\Lsc1,\Lcs1,\Lsc1,\Lcs1),\Asc1,\Aglx,\Axx)
    \TopoVC(fex(\Lsc1,\Lcs1,\Lsc1,\Lcs1),\Aglx,\Acs1,\Aglx)
    \TopoVD(fex(\Lsc1,\Lcs1,\Lcs1,\Lsc1),\Acs1,\Aglx,\Asc1,\Aglx)
  \end{eqnarray*}
  \caption{%
    One-loop contributions to the
    Abelian Higgs model 4-point function at the soft EFT.
    Lines are defined in fig.~\ref{fig:2pt:diags}.
    Additional scalar line orientations are omitted for compactness.
  }
  \label{fig:4pt:diags}
\end{figure}
The 4-point vertex can be expressed as
\begin{eqnarray}
\label{4pt_ops}
\delta S^{(4)}_\rmi{soft} &=&
  \delta(q+r+s+t)
  \biggl[
  \nn  &+&
  B^{ }_\mu(q)\,
  B^{ }_\nu(r)\,
  B^{ }_\rho(s)\,
  B^{ }_\sigma(t)\,
  \Bigl\{
      \psi_{1}\bigl(
        S^{ }_{\mu\nu} q\cdot r
      + q_\mu r_\nu
    \bigr) \, T^{ }_{\rho\sigma}
    + \psi^{ }_{2} \, q^2 \, T^{ }_{\mu\nu\rho\sigma}
 \Bigr\}
  \nn &+&
  B^{ }_\mu(q)\,
  B^{ }_\nu(r)\,
  \phi(s)\,
  \phi^\dagger(t)\,
  \Bigl\{
      \psi^{ }_{3} \, \bigl(S^{ }_{\mu\nu}q^2 - q_\mu q_\nu\bigr)
    + \psi^{ }_{4} \, \bigl(S^{ }_{\mu\nu} q\cdot r - q_\mu r_\nu\bigr)
    \nn &&
    \hphantom{B^{ }_\mu(q)\,B^{ }_\nu(r)\,\phi(s)\,\phi^\dagger(t)\,}
    + \psi^{ }_{5} \, \bigl(
          S^{ }_{\mu\nu}(s^2 + t^2)
        - (q_\mu+2t_\mu)(r_\mu+2s_\nu)
    \bigr)
    \nn &&
    \hphantom{B^{ }_\mu(q)\,B^{ }_\nu(r)\,\phi(s)\,\phi^\dagger(t)\,}
  + \bigl(
      \psi^{ }_{6} \, q^2
    + \psi^{ }_{7} \, (s^2 + t^2)
    + \psi^{ }_{8} \, s\cdot t
    \bigr) \, T^{ }_{\mu\nu}
 \Bigr\}
  \nn &+&
  \phi(q)\,
  \phi(r)\,
  \phi^\dagger(s)\,
  \phi^\dagger(t)\,
  \Bigl\{
      \psi_{9}\,(q^2 + s^2)
    + \psi_{10}\,q\cdot s
  \Big\}
  \biggr]
 \;,
\end{eqnarray}
where
the contributing 4-point diagrams are collected in
fig.~\ref{fig:4pt:diags}.
Purely spatial contributions to the $B_\mu^4$ correlator vanish identically
after symmetry shifts and momentum conservation
due to the Abelian nature of the theory.
A representation of the coefficients can be chosen as
\begin{align}
\label{eq:4pt:mapping}
  \psi_1 &=-2\alphaF_{B_0^2F^2}
  \,,&
  \psi_2 &=-\alphaF_{D^2B_0^4}
  \,, &
  \psi_3 &=-\hat{g}_3^{ }\,\alphaF_{D^2\phi^2F}
  \,,&
  \psi_4 &=-2\alphaF_{\phi^2F^2}
  \,,&
  \nn
  \psi_5 &=\hat{g}_3^{2}\,\alphaF_{D^4\phi^2}
  \,, &
  \psi_6 &=-\alphaF_{D^2\phi^2B_0^2,1}
  \,,&
  \psi_7 &=-\alphaF_{D^2\phi^2B_0^2,2}
  \,, &
  \psi_8 &=-\alphaF_{D^2\phi^2B_0^2,3}
  \,,&
  \nn
  \psi_9 &=-\alphaF_{D^2\phi^4,2}
  \,, &
  \psi_{10} &=-\alphaF_{D^2\phi^4,1}
  \,.&
\end{align}

\begin{figure}[t]
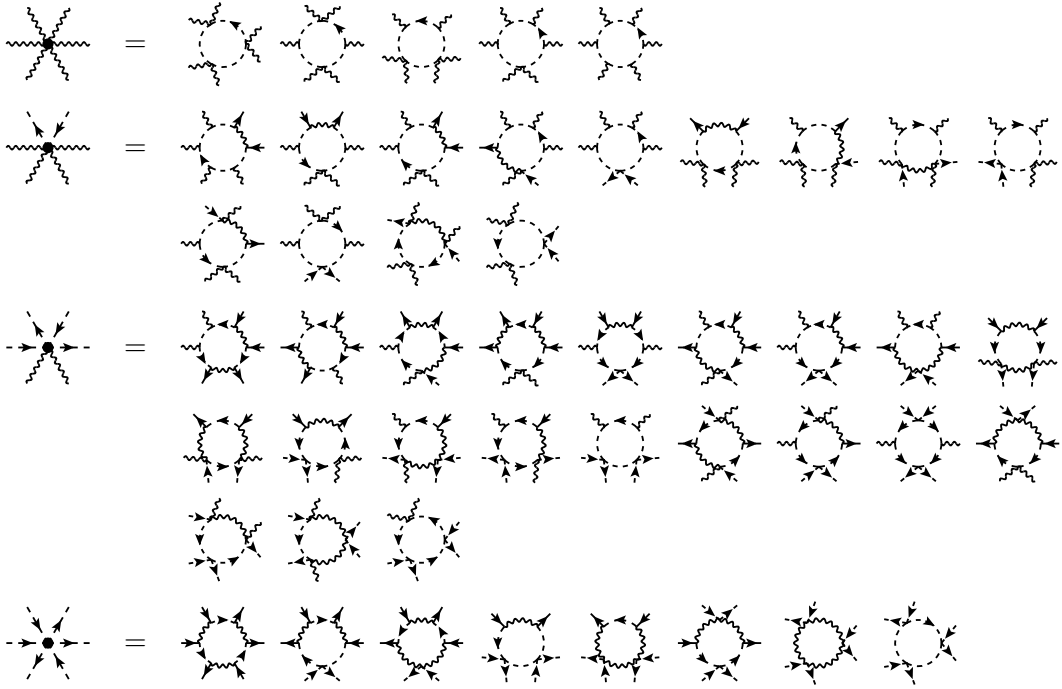

  \centering
  \begin{eqnarray*}
    \Vtxs(\Lglx,\Lglx,\Lglx,\Lglx,\Lglx,\Lglx) &=&
    \TopoSC(fex(\Lglx,\Lglx,\Lglx,\Lglx,\Lglx,\Lglx),\Asc1,\Axx,\Axx)
    \TopoSDo(fex(\Lglx,\Lglx,\Lglx,\Lglx,\Lglx,\Lglx),\Asc1,\Axx,\Axx,\Axx)
    \TopoSDa(fex(\Lglx,\Lglx,\Lglx,\Lglx,\Lglx,\Lglx),\Asc1,\Axx,\Axx,\Axx)
    \TopoSE(fex(\Lglx,\Lglx,\Lglx,\Lglx,\Lglx,\Lglx),\Asc1,\Axx,\Axx,\Axx,\Axx)
    \TopoSF(fex(\Lglx,\Lglx,\Lglx,\Lglx,\Lglx,\Lglx),\Asc1,\Axx,\Axx,\Axx,\Axx,\Axx)
    \nn[5mm]
    \Vtxs(\Lglx,\Lsc1,\Lcs1,\Lglx,\Lglx,\Lglx) &=&
    \TopoSF(fex(\Lsc1,\Lcs1,\Lglx,\Lglx,\Lglx,\Lglx),\Aglx,\Axx,\Axx,\Acs1,\Axx,\Axx)
    \TopoSE(fex(\Lglx,\Lcs1,\Lsc1,\Lglx,\Lglx,\Lglx),\Axx,\Aglx,\Axx,\Asc1,\Axx)
    \TopoSE(fex(\Lsc1,\Lcs1,\Lglx,\Lglx,\Lglx,\Lglx),\Aglx,\Axx,\Axx,\Acs1,\Axx)
    \TopoSE(fex(\Lglx,\Lglx,\Lglx,\Lcs1,\Lglx,\Lsc1),\Asc1,\Axx,\Axx,\Aglx,\Axx)
    \TopoSE(fex(\Lglx,\Lglx,\Lglx,\Lglx,\Lcs1,\Lsc1),\Asc1,\Axx,\Axx,\Axx,\Axx)
    \TopoSDa(fex(\Lsc1,\Lcs1,\Lglx,\Lglx,\Lglx,\Lglx),\Aglx,\Axx,\Acs1,\Axx)
    \TopoSDa(fex(\Lcs1,\Lglx,\Lglx,\Lglx,\Lglx,\Lsc1),\Axx,\Acs1,\Axx,\Aglx)
    \TopoSDa(fex(\Lglx,\Lglx,\Lglx,\Lsc1,\Lglx,\Lcs1),\Acs1,\Axx,\Aglx,\Axx)
    \TopoSDa(fex(\Lglx,\Lglx,\Lcs1,\Lsc1,\Lglx,\Lglx),\Acs1,\Axx,\Axx,\Axx)
    \nn[4mm] &&
    \TopoSDo(fex(\Lcs1,\Lglx,\Lsc1,\Lglx,\Lglx,\Lglx),\Aglx,\Axx,\Asc1,\Axx)
    \TopoSDo(fex(\Lglx,\Lglx,\Lglx,\Lglx,\Lsc1,\Lcs1),\Acs1,\Axx,\Axx,\Axx)
    \TopoSC(fex(\Lglx,\Lsc1,\Lglx,\Lcs1,\Lglx,\Lglx),\Aglx,\Acs1,\Acs1)
    \TopoSC(fex(\Lcs1,\Lsc1,\Lglx,\Lglx,\Lglx,\Lglx),\Axx,\Asc1,\Axx)
    \nn[5mm]
    \Vtxs(\Lsc1,\Lsc1,\Lcs1,\Lcs1,\Lglx,\Lglx) &=&
    \TopoSF(fex(\Lsc1,\Lsc1,\Lglx,\Lglx,\Lcs1,\Lcs1),\Aglx,\Asc1,\Axx,\Asc1,\Aglx,\Acs1)
    \TopoSF(fex(\Lsc1,\Lsc1,\Lglx,\Lcs1,\Lcs1,\Lglx),\Aglx,\Asc1,\Axx,\Aglx,\Axx,\Acs1)
    \TopoSE(fex(\Lsc1,\Lcs1,\Lcs1,\Lglx,\Lglx,\Lsc1),\Asc1,\Aglx,\Acs1,\Acs1,\Aglx)
    \TopoSE(fex(\Lsc1,\Lsc1,\Lcs1,\Lcs1,\Lglx,\Lglx),\Aglx,\Asc1,\Aglx,\Acs1,\Acs1)
    \TopoSE(fex(\Lglx,\Lsc1,\Lsc1,\Lglx,\Lcs1,\Lcs1),\Acs1,\Aglx,\Asc1,\Asc1,\Acs1)
    \TopoSE(fex(\Lsc1,\Lsc1,\Lglx,\Lcs1,\Lglx,\Lcs1),\Aglx,\Asc1,\Axx,\Aglx,\Acs1)
    \TopoSE(fex(\Lsc1,\Lsc1,\Lglx,\Lglx,\Lcs1,\Lcs1),\Aglx,\Asc1,\Axx,\Asc1,\Acs1)
    \TopoSE(fex(\Lsc1,\Lglx,\Lglx,\Lcs1,\Lcs1,\Lsc1),\Axx,\Asc1,\Axx,\Aglx,\Aglx)
    \TopoSDa(fex(\Lsc1,\Lsc1,\Lglx,\Lcs1,\Lcs1,\Lglx),\Aglx,\Asc1,\Aglx,\Acs1)
    \nn[4mm] &&
    \TopoSDa(fex(\Lsc1,\Lcs1,\Lglx,\Lsc1,\Lcs1,\Lglx),\Asc1,\Aglx,\Asc1,\Aglx)
    \TopoSDa(fex(\Lcs1,\Lsc1,\Lsc1,\Lcs1,\Lglx,\Lglx),\Aglx,\Asc1,\Asc1,\Asc1)
    \TopoSDa(fex(\Lsc1,\Lglx,\Lcs1,\Lglx,\Lcs1,\Lsc1),\Asc1,\Asc1,\Aglx,\Aglx)
    \TopoSDa(fex(\Lsc1,\Lglx,\Lcs1,\Lsc1,\Lglx,\Lcs1),\Asc1,\Asc1,\Asc1,\Aglx)
    \TopoSDa(fex(\Lglx,\Lglx,\Lsc1,\Lcs1,\Lsc1,\Lcs1),\Asc1,\Axx,\Axx,\Axx)
    \TopoSDo(fex(\Lcs1,\Lglx,\Lsc1,\Lcs1,\Lglx,\Lsc1),\Aglx,\Asc1,\Aglx,\Asc1)
    \TopoSDo(fex(\Lcs1,\Lglx,\Lsc1,\Lglx,\Lsc1,\Lcs1),\Aglx,\Asc1,\Asc1,\Asc1)
    \TopoSDo(fex(\Lglx,\Lsc1,\Lsc1,\Lglx,\Lcs1,\Lcs1),\Acs1,\Asc1,\Asc1,\Acs1)
    \TopoSDo(fex(\Lsc1,\Lcs1,\Lsc1,\Lcs1,\Lglx,\Lglx),\Aglx,\Aglx,\Acs1,\Acs1)
    \nn[4mm] &&
    \TopoSC(fex(\Lglx,\Lcs1,\Lglx,\Lsc1,\Lsc1,\Lcs1),\Aglx,\Asc1,\Asc1)
    \TopoSC(fex(\Lcs1,\Lsc1,\Lglx,\Lsc1,\Lcs1,\Lglx),\Aglx,\Asc1,\Aglx)
    \TopoSC(fex(\Lsc1,\Lcs1,\Lglx,\Lglx,\Lsc1,\Lcs1),\Asc1,\Asc1,\Asc1)
    \nn[5mm]
    \Vtxs(\Lsc1,\Lcs1,\Lsc1,\Lcs1,\Lcs1,\Lsc1) &=&
    \TopoSF(fex(\Lcs1,\Lcs1,\Lsc1,\Lsc1,\Lcs1,\Lsc1),\Aglx,\Acs1,\Aglx,\Asc1,\Aglx,\Asc1)
    \TopoSE(fex(\Lsc1,\Lcs1,\Lsc1,\Lcs1,\Lsc1,\Lcs1),\Aglx,\Acs1,\Aglx,\Acs1,\Axx)
    \TopoSE(fex(\Lsc1,\Lcs1,\Lsc1,\Lcs1,\Lcs1,\Lsc1),\Asc1,\Aglx,\Asc1,\Aglx,\Aglx)
    \TopoSD(fex(\Lcs1,\Lsc1,\Lsc1,\Lcs1,\Lsc1,\Lcs1),\Aglx,\Asc1,\Axx,\Axx)
    \TopoSDa(fex(\Lsc1,\Lcs1,\Lcs1,\Lsc1,\Lcs1,\Lsc1),\Asc1,\Aglx,\Aglx,\Aglx)
    \TopoSDo(fex(\Lcs1,\Lcs1,\Lsc1,\Lsc1,\Lsc1,\Lcs1),\Aglx,\Aglx,\Asc1,\Asc1)
    \TopoSC(fex(\Lsc1,\Lcs1,\Lsc1,\Lcs1,\Lsc1,\Lcs1),\Aglx,\Aglx,\Aglx)
    \TopoSC(fex(\Lsc1,\Lcs1,\Lsc1,\Lcs1,\Lsc1,\Lcs1),\Acs1,\Axx,\Axx)
  \end{eqnarray*}
  \caption{%
    One-loop contributions to the
    Abelian Higgs model 6-point function at the soft EFT.
    Lines are defined in fig.~\ref{fig:2pt:diags}.
    Additional scalar line orientations are omitted for compactness.
  }
  \label{fig:6pt:diags}
\end{figure}
The 6-point vertex can be expressed as
\begin{align}
\label{6pt_ops}
\delta S^{(6)}_\rmi{soft} &=
  \int_X
  B^{ }_\mu\,
  B^{ }_\nu\,
  B^{ }_\rho\,
  B^{ }_\sigma\,
  B^{ }_\alpha\,
  B^{ }_\beta\,
  \Bigl\{ 
     \chi^{ }_{1} \, S^{ }_{\mu\nu} S^{ }_{\rho\sigma} S^{ }_{\alpha\beta}
   + \chi^{ }_{2} \, S^{ }_{\mu\nu} S^{ }_{\rho\sigma} T^{ }_{\alpha\beta}
   + \chi^{ }_{3} \, S^{ }_{\mu\nu} T^{ }_{\rho\sigma\alpha\beta}
   + \chi^{ }_{4} \, T^{ }_{\mu\nu\rho\sigma\alpha\beta}
 \Bigr\}
  \nn[2mm] &+
  \int_X
  B^{ }_\mu\,
  B^{ }_\nu\,
  B^{ }_\rho\,
  B^{ }_\sigma\,
  \phi\,
  \phi^\dagger\,
  \Bigl\{
    \chi^{ }_{5} \, S^{ }_{\mu\nu} S^{ }_{\rho\sigma}
  + \chi^{ }_{6} \, S^{ }_{\mu\nu} T^{ }_{\rho\sigma}
  + \chi^{ }_{7} \, T^{ }_{\mu\nu\rho\sigma}
  \Bigr\}
  \nn[2mm] &+
  \int_X
  B^{ }_\mu\,
  B^{ }_\nu\,
  (
  \phi\,
  \phi^\dagger)^2\,
  \Bigl\{
   \chi^{ }_{8} \, S^{ }_{\mu\nu}
 + \chi^{ }_{9} \, T^{ }_{\mu\nu}
 \Bigr\}
 +
 \int_X
  (
  \phi\,
  \phi^\dagger)^3\,
  \chi_{10}
 \;,
\end{align}
where
the contributing 6-point diagrams are collected in
fig.~\ref{fig:6pt:diags}.
While purely spatial contributions vanish,
we present the relationship between the coefficients
$\chi_i$ and
$\alphaF_i$:
\begin{align}
\label{eq:6pt:mapping}
  \chi_{1} &= \chi_{2} = \chi_{3} = 0
  \,,&
  \chi_4 &=\alphaF_{B_0^6}
  \,,&
  \chi_5 &=\hat{g}_3^{4}\alphaF_{D^4\phi^2}
  \,, &
  \nn
  \chi_6 &=\hat{g}_3^{2}\bigl(
        \alphaF_{D^2 \phi^2 B_0^2,3}
      -2\alphaF_{D^2 \phi^2 B_0^2,2}
      \bigr)
  \,,&
  \chi_7 &=\alphaF_{B_0^4\phi^2}
  \,, &
  \chi_8 &=\hat{g}_3^2\bigl(
        \alphaF_{D^2\phi^4,1}
      -2\alphaF_{D^2\phi^4,2}
      \bigr)
  \,,&
  \nn
  \chi_9 &=\alphaF_{B_0^2\phi^4}
  \,, &
  \chi_{10} &=\alphaF_{\phi^6}
  \,.&
\end{align}

\subsection{Lagrangian of the soft 3d EFT after field redefinitions}
\label{sec:lag:final}

Below, we rewrite the most general 3d theory Lagrangian
from eqs.~\eqref{eq:lag:3d} and~\eqref{eq:lag:3d:dim6},
\begin{align}
\label{eq:lag:3d:full}
  \mathcal{L}^\rmi{3d}_{\rmi{soft}}&=
      \frac{1}{4}F_{ij}F_{ij}
    + (D_i\phi)^\dagger (D_i\phi)
    + \mu_{3}^2\phi^\dagger\phi
    + \lambda_3(\phi^\dagger\phi)^2
    \nn &
    + \frac{1}{2}(\partial_i B_0)(\partial_i B_0)
    + h_3(\phi^\dagger\phi)B_0^2
    + \frac{1}{2}\mD^2B_0^2
    + \kappa_3B_0^4
    \nn[2mm] &
    + \alphaFR_{\phi^2F^2}F_{ij}F_{ij}\phi^\dagger\phi
    + \alphaFR_{D^2\phi^4}\phi^\dagger\phi(D_i\phi)^\dagger(D_i\phi)
    + \alphaFR_{B_0^2F^2}B_0^2F_{ij}F_{ij}
    + \alphaFR_{D^2\phi^2B_0^2}\phi^\dagger\phi(\partial_iB_0)^2
    \nn[2mm] &
    + \alphaFR_{B_0^6}B_0^6
    + \alphaFR_{\phi^2B_0^4}(\phi^\dagger\phi)B_0^4
    + \alphaFR_{\phi^4B_0^2}(\phi^\dagger\phi)^2B_0^2
    + \alphaFR_{\phi^6}(\phi^\dagger\phi)^3
    \,.
\end{align}
Here, we included operators up to dimension six in the physical basis,
ensuring that redundant operators\textemdash
those removable via field redefinitions\textemdash
are excluded.
Utilizing {\tt Matchete}~\cite{Fuentes-Martin:2022jrf},
the coefficients of the physical basis up to dimension four are given by
\begin{align}
  \mu_{3}^2 &= \hat\mu_3^2\bigl(1+\alphaF_{D^4\phi^2}^{ }\hat\mu_3^2\bigr)
  \;, \\[2mm]
  \mD^2 &= \widehat m_\rmii{D}^2\bigl(1+2\alphaF_{D^4B_0^2}^{ } \widehat m_\rmii{D}^2\bigr)
  \;,\\[2mm]
    \kappa_3^{ } &=
      \hat\kappa_{3}^{ }\bigl(1+8\widehat m_\rmii{D}^2\alphaF_{D^4B_0^2}^{ }\bigr)
    + \widehat m_\rmii{D}^2 \alphaF_{D^2B_0^4}^{ }
  \;,\\[2mm]
    h_3^{ } &=
        \hat{h}_3^{ }\bigl(
          1
        + 4 \widehat m_\rmii{D}^2\alphaF_{D^4B_0^2}^{ }
        + 2 \hat\mu_3^2 \alphaF_{D^4\phi^2}^{ } 
        \bigr)
      + \hat\mu_3^2\bigl(
         2\alphaF_{D^2\phi^2B_0^2,2}^{ }
        - \alphaF_{D^2\phi^2B_0^2,3}^{ }
        \bigr)
    \nn&
      + \widehat m_\rmii{D}^2\bigl(
        \alphaF_{D^2\phi^2B_0^2,1}^{ }
      + \alphaF_{D^2\phi^2B_0^2,3}^{ }
      \bigr)
      \;,
      \\[2mm]
  \lambda_3^{ } &=
      \hat\lambda_{3}^{ }\bigl(1 + 4\hat\mu_3^2 \alphaF_{D^4\phi^2}^{ } \bigr)
      + \hat\mu_3^2\bigl(
        2\hat g_3^2\alphaF_{D^2F^2}^{ }
        +2\alphaF_{D^2\phi^4,2}^{ }
        +\hat g_3^{ }\alphaF_{D^2\phi^2F}^{ }
      \bigr)
    \;,
\end{align}
where $g_{3}^2 = \hat{g}_{3}^{2}$.
While these relations hold to all orders up to dimension-six operators,
we effectively truncate them at
NLO i.e.\
$\mathcal{O}(g^4)$ during the matching,
omitting terms of $\mathcal{O}(g^6)$.
Therefore dimension-six induced field redefinitions only
start to affect the dimension-four Wilson coefficients at
next-to-next-to-leading order (NNLO) which in particular corresponds to
three-loop level for the effective masses and
two-loop level for the effective couplings.

The non-renormalizable dimension-six derivative-operator coefficients
\begin{align}
  \alphaFR_{D^2\phi^2B_0^2}^{ }&=
      \alphaF_{D^2\phi^2B_0^2,3}^{ }
      \;, \\[2mm] 
  \alphaFR_{B_0^2F^2}^{ }&=
      \alphaF_{B_0^2F^2}^{ }
      \;, \\[2mm] 
  \alphaFR_{\phi^2F^2}^{ }&=
      \alphaF_{\phi^2F^2}^{ }
    \;, \\[2mm] 
  \alphaFR_{D^2\phi^4}^{ }&=
      \alphaF_{D^2\phi^4,1}^{ } 
    + 3 \hat g_3^{ }\, \alphaF_{D^2\phi^2F}^{ } 
    + 6 \hat g_3^{2}\, \alphaF_{D^2F^2}^{ }
    \,,
\end{align}
are truncated
at $\mathcal{O}(g^4)$.
The field-redefined marginal dimension-six coefficients,
\begin{align}
  \alphaFR_{B_0^6}^{ }&=
        \alphaF_{B_0^6}^{ }
      + 4 \hat\kappa_{3}^{ }\bigl(
        \alphaF_{D^2B_0^4}^{ }
      + 4 \hat\kappa_{3}^{ } \alphaF_{D^4B_0^2}^{ }
      \bigr)
      \;, \\[2mm]
  \alphaFR_{\phi^2B_0^4}^{ }&=
      \alphaF_{\phi^2B_0^4}^{ }
    +4\hat\kappa_{3}^{ }\bigl(
        \alphaF_{D^2\phi^2B_0^2,1}^{ }
      + \alphaF_{D^2\phi^2B_0^2,3}^{ }
      + 4\hat{h}_{3}^{ }\alphaF_{D^4B_0^2}^{ }
    \bigr)
    \nn&
    +\hat{h}_{3}^{ }\bigl(
        2\alphaF_{D^2\phi^2B_0^2,2}^{ }
      - \alphaF_{D^2\phi^2B_0^2,3}^{ }
      + 2\alphaF_{D^2B_0^4}^{ }
    \bigr)
    + \hat{h}_{3}^{2} \alphaF_{D^4\phi^2}^{ }
    \;, \\[2mm] 
  \alphaFR_{\phi^4B_0^2}^{ }&=
    \alphaF_{\phi^4B_0^2}^{ }
    +2 \hat\lambda_{3}^{ }\bigl(
       2\alphaF_{D^2\phi^2B_0^2,2}^{ }
      - \alphaF_{D^2\phi^2B_0^2,3}^{ }
    \bigr)
    \nn&
    +2\hat{h}_{3}^{ }\bigl(
        \alphaF_{D^2\phi^2B_0^2,1}^{ }
      + \alphaF_{D^2\phi^2B_0^2,3}^{ }
      + 2\hat\lambda_{3}^{ }\alphaF_{D^4\phi^2}^{ }
      + \widetilde\alphaFR_{D^2\phi^4}
    \bigr)
    +4 \hat{h}_{3}^{2} \alphaF_{D^4B_0^2}^{ }
    \;, \\[2mm] 
  \alphaFR_{\phi^6}^{ }&=
      \alphaF_{\phi^6}^{ }
      + 4 \hat\lambda_{3}^{ }\bigl(
        \widetilde\alphaFR_{D^2\phi^4}^{ }
        + \hat\lambda_{3}^{ }\alphaF_{D^4\phi^2}^{ }
      \bigr)
  \;, 
\end{align}
are truncated
at $\mathcal{O}(g^6)$,
using
\begin{align}
  \widetilde \alphaFR_{D^2\phi^4}^{ } &= 
      \alphaF_{D^2\phi^4,2}^{ }
      + \frac{1}{2} \hat{g}_{3}^{ } \, \alphaF_{D^2 \phi^2F}^{ }
      + \hat{g}_{3}^{2} \, \alphaF_{D^2F^2}^{ }
  \,.
\end{align}

Below, we present the expressions for the previously defined $\alpha$-coefficients
both in
general $d$ and $d=3-2\epsilon$ dimensions.
Notably, these coefficients are entirely independent of the gauge-fixing parameter $\xi$,
except for terms consistent with corrections arising from the matching of
the two theories at the three-loop level.
Therefore, in this basis,
the gauge independence of physical observables is manifest,
{\em viz.}
\begin{eqnarray}
    \alphaFR_{B_0^6} &=&\frac{1}{45}(d-5)(d-3)(d-1)g^{6}\mathcal{I}_{3}^{ }\,T^{2}
    + \mathcal{O}(g^8)
    \nn &\stackrel{d=3-2\epsilon}{=}&
    \mathcal{O}(g^8)
    \;, \\[2mm] 
    \alphaFR_{\phi^2B_0^4} &=&\frac{1}{9}\Bigl[(d-5)(d-1)g^{6}+12(d-5)(d-3)g^{4}\lambda\Bigr]\mathcal{I}_{3}^{ }\,T^{2}+\mathcal{O}(g^8)\nn &\stackrel{d=3-2\epsilon}{=}& -\frac{\zeta_3}{288\pi^4}g^{6}+\mathcal{O}(g^8)
    \;, \\[2mm] 
    \alphaFR_{\phi^4B_0^2} &=&\frac{2}{15}\Bigl[(36d-139)g^{6}+10(35-3d)g^{4}\lambda+10(13d-67)g^{2}\lambda^2\Bigr]\mathcal{I}_{3}^{ }\,T^{2}+\mathcal{O}(g^8)\nn &\stackrel{d=3-2\epsilon}{=}&-\frac{\zeta_3}{960\,\pi^4}\Big(31\,g^{6}\,-260\,g^{4}\,\lambda+280\,g^{2}\,\lambda^2\Big)+\mathcal{O}(g^8)
    \;, \\[2mm] 
    \alphaFR_{\phi^6} &=&\frac{4}{15}\Bigl[5d g^{6}-(5d+3)g^{4}\lambda+75g^{2}\lambda^2+100\lambda^3\Bigr]\mathcal{I}_{3}^{ }\,T^{2}+\mathcal{O}(g^8)\nn &\stackrel{d=3-2\epsilon}{=}&\frac{\zeta_3}{480\,\pi^4}\Big(15\,g^{6}-18\,g^{4}\,\lambda+75\,g^{2}\,\lambda^2+100\,\lambda^3\Big) +\mathcal{O}(g^8)
    \;, \\[2mm] 
    \alphaFR_{D^2\phi^2B_0^2} &=&\frac{1}{3}(d-4)\Bigl[7g^{4}-4g^{2}\,\lambda\Bigr]\mathcal{I}_{3}^{ }\,T^{}+\mathcal{O}(g^6)\nn &\stackrel{d=3-2\epsilon}{=}& -\frac{\zeta_3}{384\,\pi^4T}\Big(7\,g^{4}-4\,g^{2}\lambda\Big)+\mathcal{O}(g^6)
    \;, \\[2mm] 
    \alphaFR_{B_0^2F^2} &=&-\frac{1}{6}(d-5)g^{4}\,\mathcal{I}_{3}^{ }\,T^{}+\mathcal{O}(g^6)\nn &\stackrel{d=3-2\epsilon}{=}&\frac{\zeta_3}{384\pi^4T}g^{4} +\mathcal{O}(g^6)
    \;, \\[2mm] 
    \alphaFR_{\phi^2F^2} &=&\frac{1}{6}\Bigl[7g^{4}-4g^{2}\,\lambda\Bigr]\mathcal{I}_{3}^{ }\,T^{}+\mathcal{O}(g^6)\nn &\stackrel{d=3-2\epsilon}{=}& \frac{\zeta_3}{768\pi^4T}\Big(7\,g^{4}-4\,g^2\,\lambda\Big)+\mathcal{O}(g^6)
    \;, \\[2mm] 
    \alphaFR_{D^2\phi^4} &=&\frac{2}{15}\Bigl[(66-5d)g^{4}+200g^{2}\,\lambda-20\lambda^2\Bigr]\mathcal{I}_{3}^{ }\,T^{}+\mathcal{O}(g^6)\nn &\stackrel{d=3-2\epsilon}{=}& \frac{\zeta_3}{960\pi^4T}\Big(51\,g^{4}+200\,g^{2}\,\lambda-20\,\lambda^2\Big)+\mathcal{O}(g^6)
    \;,
\end{eqnarray}
where
\begin{align}
\label{eq:Lb}
    \Lb &\equiv
    2\ln\frac{\LamD}{T} - 2\bigl(\ln(4\pi) - \gammaE\bigr)
    \;.
\end{align}

After performing field redefinitions,
the Wilson coefficients are not completely gauge-independent, as
the dimension-four coefficients
$\mu_3^2$,
$h_3$,
and
$\lambda_3$,
still contain terms proportional to $\xi$.
However, these gauge-dependent terms are of the same order as
the corrections arising from three-loop dimensional reduction.%
\footnote{%
  This conclusion also holds for the
  three-loop dimensional reduction matching
  of the
  dimension-six
  operator coefficients
  $\alphaFR_{\phi^2B_0^4}$,
  $\alphaFR_{\phi^4B_0^2}$,
  and
  $\alphaFR_{\phi^6}$
  which is of $\mathcal{O}(g^{10})$.
}
This suggests that such higher-order corrections may further
cancel the residual $\xi$-dependence.
The matching relations in $d=3-2\epsilon$ up to $\mathcal{O}(g^6)$
for the
scalar and Debye masses,
are given by
\begin{align}
\mu_3^2 &=
    \Big[\mu^2\Big]_{\text{tree level}}
  + T^2\Big[\frac{1}{3}\lambda+\frac{1}{4}g^2\Big]_{\text{1loop}}
  - \Big[\Big(\mu^2+\frac{1}{4}g^2\,T^2+\frac{1}{3}\lambda T^2\Big)\delta Z_\phi\Big]_{\text{2loop}}
  \nn[1mm] &
  + \frac{T^2}{(4\pi)^2}\Big[
    - \frac{8 + 39\Lb}{36}g^4
    + \frac{2(1+3\Lb)}{3}\,g^2 \lambda
    - \frac{10}{3}\Lb\, \lambda^2
    + \bigl(
        3\,g^2
      - 4\,\lambda
      \bigr) \Lb\frac{\mu^2}{T^2}
  \nn&
  \hphantom{{}\frac{T^2}{(4\pi)^2}\Big[}
  + \bigl(-8\lambda^2-6g^4+8g^2\lambda\bigr)
    \Big(c+\ln\frac{3T}{\LamD}\Big)
  + %
    \frac{4\,g_3^4+8\,\lambda_3^2+2h_3^2-8g_3^2\,\lambda_3}{T^2}
    \ln\frac{\Lamd}{\LamD}
  \Big]_{\text{2loop}}
  \nn&
  + \frac{T^2}{(4\pi)^4}\Big[
      \frac{\Lb(216\,c_1-16+138\Lb^{ })+5\zeta_3}{24}g^6
      - \frac{\Lb(108\,c_1-18 + 54\Lb^{ })-5\,\zeta_3}{9}g^4\lambda
  \nn&
  \hphantom{{}+ \frac{T^2}{(4\pi)^4}\Big[}
    + \frac{2(162\,c_1\,\Lb+27\Lb^2+5\zeta_3)}{27}g^2\lambda^2
    + \frac{
      (27\Lb^2+5\zeta_3) g^2
    -4(27\Lb^2-5\zeta_3) \lambda
    }{3} g^2 \frac{\mu^2}{T^2}
  \nn&
  \hphantom{{}+ \frac{T^2}{(4\pi)^4}\Big[}
    + \frac{10\zeta_3}{3} g^2 \frac{\mu^4}{T^{4}}
    + \xi\,\mathcal{C}_{\mu_3}^{\xi}
  \Big]_{\rmii{FR}}
  + \Big[\delta \mu^2_3\Big]_{\text{3loop}}
  + \mathcal{O}(g^8)
    \;, \\[3mm]
\label{eq:mD:NNLO}
\mD^2 &=
    \Big[\frac{1}{3}g^2\,T^2\Big]_{\text{1loop}}
  + \frac{g^2 T^2}{(4\pi)^2}\Big[
        \frac{7-\Lb}{9}\,g^2
      + \frac{4}{3}\Bigl(\lambda + 3\frac{\mu^2}{T^2}\Bigr) 
    \Big]_{\text{2loop}}
  + \Big[\delta \mD^2\Big]_{\text{3loop}}
  \nn[1mm]&
  + \frac{g^4 T^2}{(4\pi)^4}\Big[
    \frac{5\Lb^2-25\Lb-70-6\zeta_3}{135}\,g^2
        - \frac{4(\Lb+2)}{9}\Bigl(
          \lambda
        + 3\frac{\mu^2}{T^2}
      \Bigr) 
    \Big]_{\rmii{FR}}
    + \mathcal{O}(g^8)
    \;,
\end{align}
where
terms arising through a field redefinition (FR)
are indicated explicitly.
From eq.~\eqref{eq:mD:NNLO},
one can directly infer the gauge-independence of the Debye mass at three-loop level;
see the QCD Debye mass~\cite{Ghisoiu:2015uza}.
The soft EFT couplings
are
\begin{align}
  g_3^{2} &=
    \Big[g^2\,T^{}\Big]_{\text{tree level}}
  - \frac{g^4 T}{(4\pi)^2}\Big[\frac{\Lb}{3}\Big]_{\text{1loop}}
  + \Big[\delta g_3^{2} - g^2 T\,\delta Z_{B_i}\Big]_{\text{2loop}}
  + \mathcal{O}(g^8)
  \;, \\[2mm] 
\kappa_3^{ } &=
    \frac{T}{(4\pi)^2}\Big[\frac{2}{3} g^4\Big]_{\text{1loop}}
  - \frac{T}{(4\pi)^4}\Big[\frac{4\zeta_3}{27}g^6\Big]_{\rmii{FR}}
  + \Big[\delta \kappa_3\Big]_{\text{2loop}}
  + \mathcal{O}(g^8)
    \;, \\[2mm] 
h_3^{ } &=
    \Big[g^2\,T^{}\Big]_{\text{tree level}}
    + \frac{T}{(4\pi)^2}\Big[\frac{4-\Lb}{3}g^4+8\,g^2\lambda\Big]_{\text{1loop}}
    + \Big[\delta h_3 - g^2 T\bigl(\delta Z_{B_0}+\delta Z_{\phi}\bigr)\Big]_{\text{2loop}}
  \nn&
  +\frac{\zeta_3 T}{(4\pi)^4} \frac{16}{45}\Big[
      - 2\,g^6
      + 15\,g^4\Bigl(\lambda + \frac{\mu^2}{T^{2}}\Bigr)\Big]_{\rmii{FR}}
  + \mathcal{O}(g^8)
    \;, \\[2mm]
\lambda_3 &=
    \Big[\lambda\,T^{}\Big]_{\text{tree level}}
  + \frac{T}{(4\pi)^2}\Big[\big(2-3\Lb\big)\,g^4+6\,\Lb\,g^2\lambda-10\,\Lb\,\lambda^2\Big]_{\text{1loop}}
  + \Big[\delta\lambda_3-2\lambda T^{}\,\delta Z_\phi\Big]_{\text{2loop}}
  \nn[1mm]&
  +\frac{\zeta_3 T}{(4\pi)^4}\Big[
      - \frac{6}{5}\Bigl(
            g^2
          + 4\frac{\mu^2}{T^{2}}
        \Bigr)\,g^4
      + \frac{76}{15} g^4\lambda
      + \frac{56}{9} g^2\lambda^2
      - \frac{32}{9} \Bigl(
          \lambda
        + 3\frac{\mu^2}{T^{2}}
        \Bigr)\lambda^2
      + \frac{80}{3} g^2 \lambda\,\frac{\mu^2}{T^{2}}
    \Big]_{\rmii{FR}}
    \nn[2mm]&
    + \mathcal{O}(g^8)
    \;,
\end{align}
which are gauge-independent at the given two-loop level.
Here,
$c\equiv\frac{1}{2}\big(\ln\frac{8\pi}{9}+(\ln\zeta_2)'-2\gammaE\big)$,
$c_1\equiv \ln({2\pi})-(\ln\zeta_2)'$.
We also introduced,
$\delta\mu^2_3$ as the scalar and
$\delta \mD^2$ as the Debye mass three-loop contribution
including their 3d running.
The coefficients
$\delta Z_\phi$,
$\delta Z_{B_0}$,
$\delta\kappa_3$,
$\delta h_3$,
$\delta\lambda_3$ are the two-loop contributions to
$Z_\phi$,
$Z_{B_0}$,
$\kappa_3$,
$h_3$,
$\lambda_3$
including their 3d running, while the residual gauge dependent part corresponds to 
\begin{align}
  \mathcal{C}_{\mu_3}^{\xi} &\equiv
    - \frac{\Lb(216\,c_1-16+138\Lb^{ })+9\zeta_3}{72} g^6
    + \frac{\Lb(12\,c_1-2+6\Lb^{ })-\zeta_3}{3} g^4\lambda
  \nn&
  - \frac{2(18\, c_1 \Lb+3\Lb^2+\zeta_3)}{9} g^2\lambda^2
  - \Bigl[
      (3\Lb^2+\zeta_3) g^2
    - \frac{4(3\Lb^2-\zeta_3)}{3} \lambda
  \Bigr]g^2\frac{\mu^2}{T^2}
  - 2\zeta_3 g^2 \frac{\mu^4}{T^{4}}
  \,.
\end{align}

\section{Softer-scale EFT}
\label{sec:dimred:soft-softer}

\begin{table}[t]
\centering
\begin{tabular}[t]{|l|l|}
\hline
\multicolumn{2}{|c|}{dimension~2}\\
\hline
$\phi^\dagger\phi$ & $\bar{Z}_\phi\, \hat{\bar\mu}^2_3$ \\ \hline
\end{tabular}
\begin{tabular}[t]{|l|l|}
\hline
\multicolumn{2}{|c|}{dimension~4}\\
\hline
$F_{ij}F_{ij}/4$ & $\bar{Z}_{B_i}$ \\ \hline
$(D_i\phi)^\dagger(D_i\phi)$ & $\bar{Z}_\phi$ \\ \hline
$(\phi^\dagger\phi)^2$ & $\bar{Z}_\phi^2\, \alphaSF_{\phi^4} \;$ \\ \hline
\end{tabular}
\begin{tabular}[t]{|l|l|}
\hline
\multicolumn{2}{|c|}{dimension~6}\\
\hline
$(\partial_i F_{ij})^2$ & $\bar{Z}_{B_i}\,\alphaSF_{D^2 F^2}$ \\ \hline
$(D^2 \phi^\dagger) (D^2 \phi)$ & $\bar{Z}_\phi\,\alphaSF_{D^4\phi^2}$ \\ \hline
$(\phi^\dagger \phi) (D_i \phi^\dagger D_i \phi)$ & $\bar{Z}_\phi^2  \,\alphaSF_{D^2\phi^4, 1}$ \\ \hline
$(\phi^\dagger \phi) (\phi^\dagger D^2 \phi + {\it h.c.})$ & $\bar{Z}_\phi^2 \,\alphaSF_{D^2\phi^4, 2}$ \\ \hline
$(\partial_i F_{ij}) i \phi^\dagger (D_j \phi)$ & $\bar{Z}_\phi \bar{Z}_{B_i}^{1/2}\,\alphaSF_{D^2\phi^2 F}$ \\ \hline
$F_{ij} F_{ij}  \phi^\dagger \phi$ & $\bar{Z}_{B_i}\bar{Z}_\phi\,\alphaSF_{\phi^2F^2}$ \\ \hline

$(\phi^\dagger\phi)^3$ & $\bar{Z}_\phi^3\,\alphaSF_{\phi^6}$ \\ \hline
\end{tabular}
\caption{%
  All possible dimension-six operators in the softer-scale 3d effective theory.
}
  \label{tab:ultrasoftop:dim6}
\end{table}
Our goal now is to integrate out the soft scale
and especially the massive field $B_0$.
The interactions between the Higgs and spatial gauge fields,
including all terms up to dimension-six operators,
are summarized in table~\ref{tab:ultrasoftop:dim6}.
Here, the covariant derivative is defined as
$D_i\equiv \partial_i-i \bar{g}_3 \bar{Z}_{B_i}^{1/2}B_i$.
\begin{figure}[t]
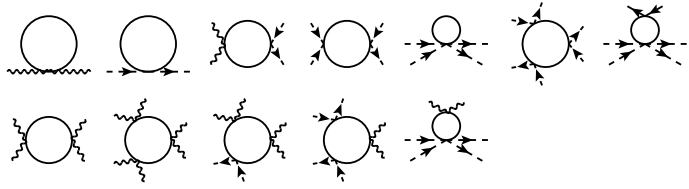

  \centering
  \begin{align*}
    &
    \TopoST(\Lglx,\Asa1)
    \TopoST(\Lsc1,\Asa1)
    \TopoVBlr(fex(\Lglx,\Lglx,\Lsc1,\Lcs1),\Asa1,\Asa1)
    \TopoVBlr(fex(\Lsc1,\Lsc1,\Lsc1,\Lcs1),\Asa1,\Asa1)
    \TopoVA(fex(\Lsc1,\Lsc1,\Lsc1,\Lcs1),\Asa1)
    \TopoSC(fex(\Lcs1,\Lsc1,\Lcs1,\Lsc1,\Lcs1,\Lsc1),\Asa1,\Asa1,\Asa1)
    \TopoSBx(fex(\Lsc1,\Lsc1,\Lsc1,\Lcs1,\Lsc1,\Lcs1),\Asa1,\Asa1)
    \nn[4mm]
    &
    \TopoVBlr(fex(\Lglx,\Lglx,\Lglx,\Lglx),\Asa1,\Asa1)
    \TopoSC(fex(\Lglx,\Lglx,\Lglx,\Lglx,\Lglx,\Lglx),\Asa1,\Asa1,\Asa1)
    \TopoSC(fex(\Lglx,\Lglx,\Lglx,\Lglx,\Lcs1,\Lsc1),\Asa1,\Asa1,\Asa1)
    \TopoSC(fex(\Lglx,\Lglx,\Lcs1,\Lsc1,\Lcs1,\Lsc1),\Asa1,\Asa1,\Asa1)
    \TopoSBx(fex(\Lsc1,\Lsc1,\Lsc1,\Lcs1,\Lglx,\Lglx),\Asa1,\Asa1)
  \end{align*}
  \caption{%
    One-loop contributions to the
    Abelian Higgs model 2-point, 4-point and 6-point functions at the softer EFT.
    Dashed directed lines denote softer scalars,
    wiggly lines softer spatial gauge fields, and
    solid lines temporal scalars $B_0$.
    Additional scalar line orientations are omitted for compactness.
    The diagrams listed in the second row
    contribute only to operators in the softer theory that are beyond dimension six.
  }
  \label{fig:softer:diags}
\end{figure}
To this end, we determine the softer EFT coefficients at the full one-loop level,
following the same approach as for the soft EFT
in sec.~\ref{sec:eft:soft:dim6}.
The contributing diagrams are collected in fig.~\ref{fig:softer:diags}.

The expressions for the
dimension-four coefficients of the one-loop matching read
\begin{align}
\label{eq:g3:softer:1l}
\hat{\bar{g}}_3^{2} &=
      g_3^{2}\Bigl[
        1
      -4 \alphaFR_{B_0^2 F^2}^{ } I_{1}
    \Bigr]
    \;, \\[2mm]
    \hat{\bar{\mu}}^2_3 &=
        \mu_3^2
        + \Bigl[
            h_3^{ }
          -\mD^{2}\alphaFR_{D^2\phi^2B_0^2}^{ }
        \Bigr]\,I_{1}^{ }
    \;, \\[2mm]
    \hat{\bar{\lambda}}_{3} &=
      \lambda_3^{ }
    + \Bigl[
        - h_3^2
          - \frac{2}{d-2} \mD^{2}\alphaFR_{\phi^4 B_0^2 }^{ }
        + \frac{2d}{d-2} h_3^{ } \mD^{2} \alphaFR_{D^2\phi^2B_0^2}^{ }
        - \frac{d+2}{d-2} \mD^{4} \alphaFR_{D^2\phi^2B_0^2}^{2}
      \Bigr]I_{2}^{ }
    \;.
\end{align}
In the corresponding
field normalizations, only
the spatial gauge bosons receive
a non-trivial one-loop contribution
at $\mathcal{O}(g^5)$,
\begin{align}
  \bar{Z}_{B_i}&=
    1 + 4\alphaFR_{B_0^2 F^2}^{ } I_{1}
  \;, &
    \bar{Z}_\phi &= 1
    \;.
\end{align}

In this context,
we also defined
the 3d one-loop master integrals
\begin{equation}
I^\alpha_\beta (m^2)
= \int_{\vec{p}}\frac{(p^2)^\alpha}{(p^2+m^2)^\beta}
= m^{d + 2 (\alpha - \beta)} \frac{
    \Gamma(\alpha + \frac{d}2)
    \Gamma(\beta - \alpha - \frac{d}2)}{\left( 4\pi \right)^{\frac{d}2} \Gamma(\beta) \Gamma(\frac{d}2)} \ ,
\end{equation}
where
$\int_{\vec{p}} = \int\frac{{\rm d}^dp}{(2\pi)^d}$,
$I_{\alpha}^{ } \equiv I_{\alpha}^{0} (\mD^2)$.
Explicitly, one has
\begin{align}
  I_1\, &\overset{d=3-2\epsilon}{=}\, - \frac{\mD}{4\pi} +\mathcal{O}(\epsilon)
   \,, \nn
  I_2\, &\overset{d=3-2\epsilon}{=}\,\, \frac{\mD^{-1}}{8 \pi}+\mathcal{O}(\epsilon)
  \,,\nn
  I_3\, &\overset{d=3-2\epsilon}{=}\,\, \frac{\mD^{-3}}{32 \pi}+\mathcal{O}(\epsilon)
  \,,
\end{align}
or can use the IBP identity
\begin{align}
   I_{s+1}(m^2) &= \Big(1-\frac{d}{2s}\Big) \frac{1}{m^2} I_{s}(m^2)
   \,.
\end{align}

For determining the dimension-six coefficients and
since the softer Wilson coefficients originate from the purely spatial part of the Lagrangian,
the mappings in
eqs.~\eqref{eq:2pt:mapping}, \eqref{eq:4pt:mapping}, and~\eqref{eq:6pt:mapping}
can be directly reused.
For the 2-point correction,
we obtain a vanishing result
\begin{align}
\label{eq:eta:softer}
  \bar\eta_{1} &= 0
  \,, &
  \bar\eta_{3} &= 0
  \,,
\end{align}
since
only for the non-Abelian case, cubic vertices between gauge fields
can generate a 2-point $\mathcal{O}(k^4)$ contribution,
and
since the soft-theory coefficients
$\alphaFR_{D^2F^2}$ and
$\alphaFR_{D^4\phi^2}$ vanish in their physical basis~\eqref{eq:lag:3d:full}
after field redefinitions.%
\footnote{%
  In the matching of EQCD and MQCD~\cite{Laine:2018lgj},
  these coefficients have been kept in the soft EFT and
  introduced effective 2-point vertices.
  By employing field redefinitions,
  however, the effect of these 2-point vertices
  can be absorbed by the other Wilson coefficients.
}

For the 4-point contributions to the action,
the result reads
\begin{align}
  \bar\psi_{1} &=
  \bar\psi_{3} =
  \bar\psi_{5} = 0
  \,,\\[2mm]
  \bar\psi_{4} &=
    - 2\alphaFR_{\phi^2 F^2}^{ }
    + 16\Bigl[
      - \frac{1}{d-4} h_{3}^{ } \mD^{2} \alphaFR_{B_0^2 F^2}^{ }
      + \frac{d}{(d-2)(d-4)} \mD^{4} \alphaFR_{B_0^2 F^2}^{ } \alphaFR_{D^2\phi^2B_0^2}^{ }
    \Bigr] I_{3}^{ }
  \,,\\[2mm]
  \bar\psi_{9} &=
      \Bigl[
        \frac{1}{3} h_3^{2}
      - \frac{2}{3}\frac{d+2}{d-4} h_{3}^{ } \mD^{2} \alphaFR_{D^2\phi^2B_0^2}^{ }
      + \frac{1}{3}\frac{d^2+6d-4}{(d-2)(d-4)} \mD^{4} \alphaFR_{D^2\phi^2B_0^2}^{2}
    \Bigr] I_{3}^{ }
  \,,\\[2mm]
  \bar\psi_{10} &=
    - \alphaFR_{D^2\phi^4}
    + 2\bar\psi_{9}
  \,.
\end{align}
For the 6-point contributions to the action,
the result reads
\begin{align}
  \bar\chi_{1} &=
  \bar\chi_{5} = 0
  \,,\\[2mm]
  \bar\chi_{8} &= g_{3}^{2} \alphaFR_{D^2\phi^4}^{ }
  \,,\\[2mm]
  \bar\chi_{10} &=
      \alphaFR_{\phi^6}
    + 4\Bigl[
        \frac{1}{3} h_3^{3}
      - \frac{2d\mD^{4} }{(d-2)(d-4)} \alphaFR_{\phi^4B_0^2}^{ } \alphaFR_{D^2\phi^2B_0^2}^{ }
      + \frac{2\mD^{2}}{d-4} h_{3}^{ } \alphaFR_{\phi^4B_0^2}^{ }
      - \frac{d\mD^{2}}{d-4} h_{3}^{2} \alphaFR_{D^2\phi^2B_0^2}^{ }
  \nn &
  \hphantom{{}=\alphaFR_{\phi^6}+4\Bigl[\frac{1}{3} h_3^{2}}
      + \frac{d(d+2)}{(d-2)(d-4)} h_{3}^{ } \mD^{4} \alphaFR_{D^2\phi^2B_0^2}^{2}
      - \frac{1}{3}\frac{(d+2)(d+4)}{(d-2)(d-4)} \mD^{6} \alphaFR_{D^2\phi^2B_0^2}^{3}
    \Bigr] I_{3}^{ }
  \,.
\end{align}
Here, the hard contributions are
$\delta\bar\eta_i = \eta_i$,
$\delta\bar\psi_i = \psi_i$,
$\delta\bar\chi_i = \chi_i$.

In the following,
we truncate the Wilson coefficients from the matching above at
the corresponding orders in a power counting in
the gauge coupling, $g$, of the fundamental 4d theory
using eq.~\eqref{eq:full softer power counting}.
This way, we exclude
orders of
\begin{align}
\label{eq:softer:EFT:trunctation}
  \mathcal{O}(g^4)
  \,,
  &&
  \text{for}
  &
  \text{ masses:}
  &&
  \hat{\bar{\mu}}_{3}^{2}
  \,,
  \nn
  \mathcal{O}(g^4)
  \,,
  &&
  \text{for}
  &
  \text{ relevant couplings:}
  &&
  \hat{\bar{\lambda}}_{3}^{2}
  \,,
  \hat{\bar{g}}_{3}^{2}
  \,,
  \nn
  \mathcal{O}(g^4)
  \,,
  &&
  \text{for}
  &
  \text{ marginal Wilson coefficients:}
  &&
  \alphaSF_{\phi^6}
  \,,
  \nn
  \mathcal{O}(g^2)
  \,,
  &&
  \text{for}
  &
  \text{ non-renormalizable Wilson coefficients:}
  &&
  \alphaSF_{D^2F^2}
  \,,
  \alphaSF_{D^4\phi^2}
  \,,
  \alphaSF_{D^2\phi^4,1}
  \,,
  \nn &&&&&
  \alphaSF_{D^2\phi^4,2}
  \,,
  \alphaSF_{D^2\phi^2F}
  \,,
  \alphaSF_{\phi^2F^2}
  \,.
\end{align}
The corresponding coefficients of
the one-loop matching
read,
for the field normalizations,
\begin{align}
  \bar{Z}_{B_i}&=
    1+\mathcal{O}(g^4)
  \;, &
    \bar{Z}_\phi &= 1
    \;,
\end{align}
and for the dimension-four coefficients,
\begin{eqnarray}
    \hat{\bar{\mu}}^2_3 &=&
        \mu_3^2
        + h_3^{ }\,I_{1}^{ }
      + \mathcal{O}(g^4)
    \;, \\[2mm]
  \hat{\bar{g}}_{3}^{2} &=&
        g_3^{2}
      + \mathcal{O}(g^4)
    \;, \\[2mm]
    \hat{\bar{\lambda}}_{3} &=&
      \lambda_3^{ }
    - h_3^2\,I_{2}^{ }
    + \mathcal{O}(g^4)
    \;.
\end{eqnarray}
For $\hat{\bar{\lambda}}_{3}$
soft one-loop corrections are parametrically larger than
hard one-loop correction.
By employing the
one-loop computations detailed below
eq.~\eqref{eq:g3:softer:1l}
together with the mappings,
eqs.~\eqref{eq:2pt:mapping}, \eqref{eq:4pt:mapping}, and~\eqref{eq:6pt:mapping},
the novel dimension-six operator coefficients, $\alphaSF$,
before
field redefinitions are
\begin{eqnarray}
    \alphaSF_{D^2F^2} &=& 0
    \;, \\[2mm]
    \alphaSF_{D^4\phi^2} &=& 0
    \;, \\[2mm]
    \alphaSF_{D^2\phi^4,1} &=&
      \alphaFR_{D^2\phi^4}
    - \frac{2}{3}\,h_3^2\,I_3^{ }
    + \mathcal{O}(g^2)
    \;, \\[2mm]
    \alphaSF_{D^2\phi^4,2} &=&
   -\frac{1}{3}\,h_3^2\,I_{3}^{ }
    + \mathcal{O}(g^2)
    \;, \\[2mm]
    \alphaSF_{D^2\phi^2F} &=& 0
    \;, \\[2mm]
    \alphaSF_{\phi^2F^2} &=&
      \alphaFR_{\phi^2F^2}
    + \mathcal{O}(g^2)
    \;, \\[2mm]
    \alphaSF_{\phi^6} &=&
      \alphaFR_{\phi^6}
    + \frac{4}{3}\,h_3^3\,I_{3}^{ }
    + \mathcal{O}(g^4)
    \;.
\end{eqnarray}

The Lagrangian in the physical basis is defined in eq.~\eqref{eq:lag:ultrasoft}.
From the redefinition of fields previously outlined in
the dimensional reduction procedure
(cf.\ sec.~\ref{sec:lag:final}),
we can deduce the expression for the coefficients in
the physical basis
for operators up to dimension four
{\em viz.}
\begin{eqnarray}
  \bar{\mu}_{3}^2 &=&
    \mu_{3}^2
  + h_3^{ }\,I_{1}^{ }
  + \mathcal{O}(g^4)
  \nn&\stackrel{d=3-2\epsilon}{=}&
  \mu_{3}^2
  - \frac{1}{4\pi}\,\mD^{ }\,h_3^{ }
  + \mathcal{O}(g^4)
  \nn&=&
      \mu^2
    + T^2\Big[\frac{1}{4}g^{2}
    + \frac{1}{3}\lambda \Big]_{\rmii{DR}}
    + T^2\Big[-\frac{1}{4\sqrt{3}\pi}g^3 \Big]_{\rmii{SS}}
    + \mathcal{O}(g^4)
    \;, \\[2mm]
\bar{\lambda}_3 &=&
    \lambda_3^{ }
  - h_3^2\,I_2^{ }
  - \frac{2}{3}\Big(
        h_3^{ } I_1^{ }
      + \mu^2_3\Big
    )h_3^2 I_3^{ }
  + \mathcal{O}(g^5)
  \nn &\stackrel{d=3-2\epsilon}{=}&
    \lambda_3
  - \frac{1}{8\pi}\,\frac{h_3^2}{\mD}
  + \frac{1}{192\pi^2}\frac{h_3^3}{\mD^2}
  - \frac{1}{48\pi}\frac{\mu^2_3\,h_3^2}{\mD^{3}}
  + \mathcal{O}(g^5)
  \nn &=&
  \lambda\,T
  + \frac{T}{(4\pi)^2}\Big[
      (2 - 3\Lb)g^4+6\Lb\,g^2\,\lambda-10\Lb\,\lambda^2\Big]_{\rmii{DR}}
  \nn&+&
  T\Big[-\frac{1}{64\sqrt{3}\pi}\Big(27\,g^3+4\,g\,\lambda+12\,g\, \frac{\mu^2}{T^{2}}\Big)+\frac{1}{64\pi^2}\,g^4\Big]_{\rmii{SS}}
  + \mathcal{O}(g^4)
    \;,
\end{eqnarray}
where
$\bar{g}_{3}^2 = \hat{\bar{g}}_{3}^2$ and
where, due to the truncation~\eqref{eq:softer:EFT:trunctation},
we do not include the two-loop correction to $\bar\mu_{3}$ from
eq.~\eqref{eq:soft mu3 matching}.
And for
operators of dimension six
\begin{eqnarray}
\alphaSFR_{D^2\phi^4} &=&
    \alphaFR_{D^2\phi^4}
  - \frac{2}{3}\,h_3^2\,I_3
  + \mathcal{O}(g^2)
  \nn &\stackrel{d=3-2\epsilon}{=}&
    \alphaFR_{D^2\phi^4}
  - \frac{1}{48\pi}\frac{h_3^2}{\mD^3}
  + \mathcal{O}(g^2)\nn&=&
    \frac{\zeta_3 T}{960\pi^4}\Big[51\,g^{4}+200\,g^{2}\lambda-20\,\lambda^2\Big]_{\rmii{DR}}
  \nn&+&
  \frac{1}{T}\Big[
    - \frac{\sqrt{3}}{16\pi}\,g
    + \frac{1}{128\sqrt{3}\,\pi^3}\Big((\Lb-4)g^3-24\,g\,\lambda\Big)
  \Big]_{\rmii{SS}}
  + \mathcal{O}(g^2)
    \;, \\[2mm] 
\alphaSFR_{\phi^2F^2} &=&
    \alphaFR_{\phi^2F^2}
  + \mathcal{O}(g^2)
    \nn &\stackrel{d=3-2\epsilon}{=}&
  \frac{\zeta_3}{768 \pi^4}\frac{1}{T}\Big[7\,g^{4}-4\,g^2\lambda\Big]_{\rmii{DR}}
    +\mathcal{O}(g^2)
    \;, \\[2mm] 
\alphaSFR_{\phi^6} &=&
    \alphaFR_{\phi^6}
  + \frac{4}{3}h_3^2\Big(h_3^2\,I_2-\lambda_3+h_3\Big)I_3
  + \mathcal{O}(g^4)
  \nn &\stackrel{d=3-2\epsilon}{=}&
    \alphaFR_{\phi^6}
  + \frac{1}{24\pi}\Big(h_3-\lambda_3\Big)\,\frac{h_3^2}{\mD^3}
  + \frac{1}{192\pi^2}\frac{h_3^4}{\mD^4}
  + \mathcal{O}(g^4)
  \nn &=&
    \frac{\zeta_3}{480\,\pi^4}\Big[
          15\,g^{6}
        - 18\,g^{4}\lambda
        + 75\,g^{2}\lambda^2
        + 100\,\lambda^3
  \Big]_{\rmii{DR}}
  \nn &+&
  \Big[
      \frac{\sqrt{3}g}{8\,\pi}\Big(g^2-\lambda\Big)
    + \frac{3}{64\pi^2}g^4
  \nn &&
    + \frac{g}{\sqrt{3}(4\pi)^3}\Big(3(\Lb+1)\,g^4-8(\Lb-4)g^2\lambda+(15\Lb-24)\lambda^2\Big)
  \nn &&
  -\frac{g^4}{(4\pi)^4}\Big((\Lb-4)g^2-24\lambda\Big)\Big]_{\rmii{SS}}
  + \mathcal{O}(g^4)
    \;.  
\end{eqnarray}
Contributions labelled
DR (dimensional reduction) come from the hard modes and
SS (soft scale) come from integrating out $B_0$.

{\small

}
\end{document}